\documentclass[a4paper,11pt]{article}
\pdfoutput=1 

\usepackage{jheppub} 
\usepackage[T1]{fontenc} 
\usepackage{lineno}
\usepackage{array,multirow,graphicx}
\usepackage{enumerate}

\vbadness=\maxdimen

\title{\boldmath Measurement of the charged-current electron (anti-)neutrino inclusive cross-sections at the T2K off-axis near detector ND280}

\newcommand{\noINSTHD}{1}
\newcommand{\noINSTEE}{2}
\newcommand{\noINSTFE}{3}
\newcommand{\noINSTD}{4}
\newcommand{\noINSTGA}{5}
\newcommand{\noINSTI}{6}
\newcommand{\noINSTGB}{7}
\newcommand{\noINSTFG}{8}
\newcommand{\noINSTFH}{9}
\newcommand{\noINSTBA}{10}
\newcommand{\noINSTEF}{11}
\newcommand{\noINSTIE}{12}
\newcommand{\noINSTEG}{13}
\newcommand{\noINSTHJ}{14}
\newcommand{\noINSTDG}{15}
\newcommand{\noINSTCB}{16}
\newcommand{\noINSTIB}{17}
\newcommand{\noINSTED}{18}
\newcommand{\noINSTEC}{19}
\newcommand{\noINSTHH}{20}
\newcommand{\noINSTEI}{21}
\newcommand{\noINSTGF}{22}
\newcommand{\noINSTBE}{23}
\newcommand{\noINSTBF}{24}
\newcommand{\noINSTBD}{25}
\newcommand{\noINSTEB}{26}
\newcommand{\noINSTHI}{27}
\newcommand{\noINSTHA}{28}
\newcommand{\noINSTID}{29}
\newcommand{\noINSTIF}{30}
\newcommand{\noINSTCC}{31}
\newcommand{\noINSTCD}{32}
\newcommand{\noINSTEJ}{33}
\newcommand{\noINSTFC}{34}
\newcommand{\noINSTFI}{35}
\newcommand{\noINSTHB}{36}
\newcommand{\noINSTCE}{37}
\newcommand{\noINSTDF}{38}
\newcommand{\noINSTFJ}{39}
\newcommand{\noINSTGJ}{40}
\newcommand{\noINSTCF}{41}
\newcommand{\noINSTGG}{42}
\newcommand{\noINSTIC}{43}
\newcommand{\noINSTGC}{44}
\newcommand{\noINSTFA}{45}
\newcommand{\noINSTE}{46}
\newcommand{\noINSTGD}{47}
\newcommand{\noINSTHC}{48}
\newcommand{\noINSTBC}{49}
\newcommand{\noINSTFB}{50}
\newcommand{\noINSTDI}{51}
\newcommand{\noINSTIA}{52}
\newcommand{\noINSTBB}{53}
\newcommand{\noINSTEH}{54}
\newcommand{\noINSTCH}{55}
\newcommand{\noINSTBJ}{56}
\newcommand{\noINSTCG}{57}
\newcommand{\noINSTHF}{58}
\newcommand{\noINSTGI}{59}
\newcommand{\noINSTHG}{60}
\newcommand{\noINSTF}{61}
\newcommand{\noINSTB}{62}
\newcommand{\noINSTG}{63}
\newcommand{\noINSTDJ}{64}
\newcommand{\noINSTDH}{65}
\newcommand{\noINSTFD}{66}
\newcommand{\noINSTGH}{67}
\newcommand{\noINSTEA}{68}
\newcommand{\noINSTHE}{69}
\newcommand{\noINSTH}{70}

\newcommand{\INSTHD}{\affiliation[\noINSTHD]{University Autonoma Madrid, Department of Theoretical Physics, 28049 Madrid, Spain}}
\newcommand{\INSTEE}{\affiliation[\noINSTEE]{University of Bern, Albert Einstein Center for Fundamental Physics, Laboratory for High Energy Physics (LHEP), Bern, Switzerland}}
\newcommand{\INSTFE}{\affiliation[\noINSTFE]{Boston University, Department of Physics, Boston, Massachusetts, U.S.A.}}
\newcommand{\INSTD}{\affiliation[\noINSTD]{University of British Columbia, Department of Physics and Astronomy, Vancouver, British Columbia, Canada}}
\newcommand{\INSTGA}{\affiliation[\noINSTGA]{University of California, Irvine, Department of Physics and Astronomy, Irvine, California, U.S.A.}}
\newcommand{\INSTI}{\affiliation[\noINSTI]{IRFU, CEA Saclay, Gif-sur-Yvette, France}}
\newcommand{\INSTGB}{\affiliation[\noINSTGB]{University of Colorado at Boulder, Department of Physics, Boulder, Colorado, U.S.A.}}
\newcommand{\INSTFG}{\affiliation[\noINSTFG]{Colorado State University, Department of Physics, Fort Collins, Colorado, U.S.A.}}
\newcommand{\INSTFH}{\affiliation[\noINSTFH]{Duke University, Department of Physics, Durham, North Carolina, U.S.A.}}
\newcommand{\INSTBA}{\affiliation[\noINSTBA]{Ecole Polytechnique, IN2P3-CNRS, Laboratoire Leprince-Ringuet, Palaiseau, France }}
\newcommand{\INSTEF}{\affiliation[\noINSTEF]{ETH Zurich, Institute for Particle Physics and Astrophysics, Zurich, Switzerland}}
\newcommand{\INSTIE}{\affiliation[\noINSTIE]{CERN European Organization for Nuclear Research, CH-1211 Gen\'eve 23, Switzerland}}
\newcommand{\INSTEG}{\affiliation[\noINSTEG]{University of Geneva, Section de Physique, DPNC, Geneva, Switzerland}}
\newcommand{\INSTHJ}{\affiliation[\noINSTHJ]{University of Glasgow, School of Physics and Astronomy, Glasgow, United Kingdom}}
\newcommand{\INSTDG}{\affiliation[\noINSTDG]{H. Niewodniczanski Institute of Nuclear Physics PAN, Cracow, Poland}}
\newcommand{\INSTCB}{\affiliation[\noINSTCB]{High Energy Accelerator Research Organization (KEK), Tsukuba, Ibaraki, Japan}}
\newcommand{\INSTIB}{\affiliation[\noINSTIB]{University of Houston, Department of Physics, Houston, Texas, U.S.A.}}
\newcommand{\INSTED}{\affiliation[\noINSTED]{Institut de Fisica d'Altes Energies (IFAE), The Barcelona Institute of Science and Technology, Campus UAB, Bellaterra (Barcelona) Spain}}
\newcommand{\INSTEC}{\affiliation[\noINSTEC]{IFIC (CSIC \& University of Valencia), Valencia, Spain}}
\newcommand{\INSTHH}{\affiliation[\noINSTHH]{Institute For Interdisciplinary Research in Science and Education (IFIRSE), ICISE, Quy Nhon, Vietnam}}
\newcommand{\INSTEI}{\affiliation[\noINSTEI]{Imperial College London, Department of Physics, London, United Kingdom}}
\newcommand{\INSTGF}{\affiliation[\noINSTGF]{INFN Sezione di Bari and Universit\`a e Politecnico di Bari, Dipartimento Interuniversitario di Fisica, Bari, Italy}}
\newcommand{\INSTBE}{\affiliation[\noINSTBE]{INFN Sezione di Napoli and Universit\`a di Napoli, Dipartimento di Fisica, Napoli, Italy}}
\newcommand{\INSTBF}{\affiliation[\noINSTBF]{INFN Sezione di Padova and Universit\`a di Padova, Dipartimento di Fisica, Padova, Italy}}
\newcommand{\INSTBD}{\affiliation[\noINSTBD]{INFN Sezione di Roma and Universit\`a di Roma ``La Sapienza'', Roma, Italy}}
\newcommand{\INSTEB}{\affiliation[\noINSTEB]{Institute for Nuclear Research of the Russian Academy of Sciences, Moscow, Russia}}
\newcommand{\INSTHI}{\affiliation[\noINSTHI]{International Centre of Physics, Institute of Physics (IOP), Vietnam Academy of Science and Technology (VAST), 10 Dao Tan, Ba Dinh, Hanoi, Vietnam}}
\newcommand{\INSTHA}{\affiliation[\noINSTHA]{Kavli Institute for the Physics and Mathematics of the Universe (WPI), The University of Tokyo Institutes for Advanced Study, University of Tokyo, Kashiwa, Chiba, Japan}}
\newcommand{\INSTID}{\affiliation[\noINSTID]{Keio University, Department of Physics, Kanagawa, Japan}}
\newcommand{\INSTIF}{\affiliation[\noINSTIF]{King's College London, Department of Physics, Strand, London WC2R 2LS, United Kingdom}}
\newcommand{\INSTCC}{\affiliation[\noINSTCC]{Kobe University, Kobe, Japan}}
\newcommand{\INSTCD}{\affiliation[\noINSTCD]{Kyoto University, Department of Physics, Kyoto, Japan}}
\newcommand{\INSTEJ}{\affiliation[\noINSTEJ]{Lancaster University, Physics Department, Lancaster, United Kingdom}}
\newcommand{\INSTFC}{\affiliation[\noINSTFC]{University of Liverpool, Department of Physics, Liverpool, United Kingdom}}
\newcommand{\INSTFI}{\affiliation[\noINSTFI]{Louisiana State University, Department of Physics and Astronomy, Baton Rouge, Louisiana, U.S.A.}}
\newcommand{\INSTHB}{\affiliation[\noINSTHB]{Michigan State University, Department of Physics and Astronomy,  East Lansing, Michigan, U.S.A.}}
\newcommand{\INSTCE}{\affiliation[\noINSTCE]{Miyagi University of Education, Department of Physics, Sendai, Japan}}
\newcommand{\INSTDF}{\affiliation[\noINSTDF]{National Centre for Nuclear Research, Warsaw, Poland}}
\newcommand{\INSTFJ}{\affiliation[\noINSTFJ]{State University of New York at Stony Brook, Department of Physics and Astronomy, Stony Brook, New York, U.S.A.}}
\newcommand{\INSTGJ}{\affiliation[\noINSTGJ]{Okayama University, Department of Physics, Okayama, Japan}}
\newcommand{\INSTCF}{\affiliation[\noINSTCF]{Osaka City University, Department of Physics, Osaka, Japan}}
\newcommand{\INSTGG}{\affiliation[\noINSTGG]{Oxford University, Department of Physics, Oxford, United Kingdom}}
\newcommand{\INSTIC}{\affiliation[\noINSTIC]{University of Pennsylvania, Department of Physics and Astronomy,  Philadelphia, PA, 19104, USA.}}
\newcommand{\INSTGC}{\affiliation[\noINSTGC]{University of Pittsburgh, Department of Physics and Astronomy, Pittsburgh, Pennsylvania, U.S.A.}}
\newcommand{\INSTFA}{\affiliation[\noINSTFA]{Queen Mary University of London, School of Physics and Astronomy, London, United Kingdom}}
\newcommand{\INSTE}{\affiliation[\noINSTE]{University of Regina, Department of Physics, Regina, Saskatchewan, Canada}}
\newcommand{\INSTGD}{\affiliation[\noINSTGD]{University of Rochester, Department of Physics and Astronomy, Rochester, New York, U.S.A.}}
\newcommand{\INSTHC}{\affiliation[\noINSTHC]{Royal Holloway University of London, Department of Physics, Egham, Surrey, United Kingdom}}
\newcommand{\INSTBC}{\affiliation[\noINSTBC]{RWTH Aachen University, III. Physikalisches Institut, Aachen, Germany}}
\newcommand{\INSTFB}{\affiliation[\noINSTFB]{University of Sheffield, Department of Physics and Astronomy, Sheffield, United Kingdom}}
\newcommand{\INSTDI}{\affiliation[\noINSTDI]{University of Silesia, Institute of Physics, Katowice, Poland}}
\newcommand{\INSTIA}{\affiliation[\noINSTIA]{SLAC National Accelerator Laboratory, Stanford University, Menlo Park, California, USA}}
\newcommand{\INSTBB}{\affiliation[\noINSTBB]{Sorbonne Universit\'e, Universit\'e Paris Diderot, CNRS/IN2P3, Laboratoire de Physique Nucl\'eaire et de Hautes Energies (LPNHE), Paris, France}}
\newcommand{\INSTEH}{\affiliation[\noINSTEH]{STFC, Rutherford Appleton Laboratory, Harwell Oxford,  and  Daresbury Laboratory, Warrington, United Kingdom}}
\newcommand{\INSTCH}{\affiliation[\noINSTCH]{University of Tokyo, Department of Physics, Tokyo, Japan}}
\newcommand{\INSTBJ}{\affiliation[\noINSTBJ]{University of Tokyo, Institute for Cosmic Ray Research, Kamioka Observatory, Kamioka, Japan}}
\newcommand{\INSTCG}{\affiliation[\noINSTCG]{University of Tokyo, Institute for Cosmic Ray Research, Research Center for Cosmic Neutrinos, Kashiwa, Japan}}
\newcommand{\INSTHF}{\affiliation[\noINSTHF]{Tokyo Institute of Technology, Department of Physics, Tokyo, Japan}}
\newcommand{\INSTGI}{\affiliation[\noINSTGI]{Tokyo Metropolitan University, Department of Physics, Tokyo, Japan}}
\newcommand{\INSTHG}{\affiliation[\noINSTHG]{Tokyo University of Science, Faculty of Science and Technology, Department of Physics, Noda, Chiba, Japan}}
\newcommand{\INSTF}{\affiliation[\noINSTF]{University of Toronto, Department of Physics, Toronto, Ontario, Canada}}
\newcommand{\INSTB}{\affiliation[\noINSTB]{TRIUMF, Vancouver, British Columbia, Canada}}
\newcommand{\INSTG}{\affiliation[\noINSTG]{University of Victoria, Department of Physics and Astronomy, Victoria, British Columbia, Canada}}
\newcommand{\INSTDJ}{\affiliation[\noINSTDJ]{University of Warsaw, Faculty of Physics, Warsaw, Poland}}
\newcommand{\INSTDH}{\affiliation[\noINSTDH]{Warsaw University of Technology, Institute of Radioelectronics and Multimedia Technology, Warsaw, Poland}}
\newcommand{\INSTFD}{\affiliation[\noINSTFD]{University of Warwick, Department of Physics, Coventry, United Kingdom}}
\newcommand{\INSTGH}{\affiliation[\noINSTGH]{University of Winnipeg, Department of Physics, Winnipeg, Manitoba, Canada}}
\newcommand{\INSTEA}{\affiliation[\noINSTEA]{Wroclaw University, Faculty of Physics and Astronomy, Wroclaw, Poland}}
\newcommand{\INSTHE}{\affiliation[\noINSTHE]{Yokohama National University, Faculty of Engineering, Yokohama, Japan}}
\newcommand{\INSTH}{\affiliation[\noINSTH]{York University, Department of Physics and Astronomy, Toronto, Ontario, Canada}}

\newcommand{\noINFN}{a}
\newcommand{\noJPARC}{b}
\newcommand{\noKAVLI}{c}
\newcommand{\noNRNU}{d}
\newcommand{\noVIETNAM}{e}
\newcommand{\noDUBNA}{f}
\newcommand{\noYOICHIRO}{g}
\newcommand{\noMBCC}{h}
\newcommand{\noDECEASED}{\ddag}

\newcommand{\NNINFN}{\affiliation[\noINFN]{Also at INFN-Laboratori Nazionali di Legnaro}}
\newcommand{\NNJPARC}{\affiliation[\noJPARC]{Also at J-PARC, Tokai, Japan}}
\newcommand{\NNKAVLI}{\affiliation[\noKAVLI]{Affiliated member at Kavli IPMU (WPI), the University of Tokyo, Japan}}
\newcommand{\NNNRNU}{\affiliation[\noNRNU]{Also at National Research Nuclear University "MEPhI" and Moscow Institute of Physics and Technology, Moscow, Russia}}
\newcommand{\NNDECEASED}{\affiliation[\noDECEASED]{Deceased}}
\newcommand{\NNDUBNA}{\affiliation[\noDUBNA]{Also at JINR, Dubna, Russia}}
\newcommand{\NNYOICHIRO}{\affiliation[\noYOICHIRO]{Also at Nambu Yoichiro Institute of Theoretical and Experimental Physics (NITEP)}}
\newcommand{\NNMBCC}{\affiliation[\noMBCC]{Also at BMCC/CUNY, Science Department, New York, New York, U.S.A.}}
\newcommand{\NNVIETNAM}{\affiliation[\noVIETNAM]{Also at the Graduate University of Science and Technology, Vietnam Academy of Science and Technology}}

\INSTHD
\INSTEE
\INSTFE
\INSTD
\INSTGA
\INSTI
\INSTGB
\INSTFG
\INSTFH
\INSTBA
\INSTEF
\INSTIE
\INSTEG
\INSTHJ
\INSTDG
\INSTCB
\INSTIB
\INSTED
\INSTEC
\INSTHH
\INSTEI
\INSTGF
\INSTBE
\INSTBF
\INSTBD
\INSTEB
\INSTHI
\INSTHA
\INSTID
\INSTIF
\INSTCC
\INSTCD
\INSTEJ
\INSTFC
\INSTFI
\INSTHB
\INSTCE
\INSTDF
\INSTFJ
\INSTGJ
\INSTCF
\INSTGG
\INSTIC
\INSTGC
\INSTFA
\INSTE
\INSTGD
\INSTHC
\INSTBC
\INSTFB
\INSTDI
\INSTIA
\INSTBB
\INSTEH
\INSTCH
\INSTBJ
\INSTCG
\INSTHF
\INSTGI
\INSTHG
\INSTF
\INSTB
\INSTG
\INSTDJ
\INSTDH
\INSTFD
\INSTGH
\INSTEA
\INSTHE
\INSTH

\author[\noINSTBJ]{K.\,Abe,}
\author[\noINSTFA]{N.\,Akhlaq,}
\author[\noINSTCG]{R.\,Akutsu,}
\author[\noINSTCD]{A.\,Ali,}
\author[\noINSTEF]{C.\,Alt,}
\author[\noINSTEH,\noINSTFC]{C.\,Andreopoulos,}
\author[\noINSTEI]{L.\,Anthony,}
\author[\noINSTEC]{M.\,Antonova,}
\author[\noINSTCC]{S.\,Aoki,}
\author[\noINSTEE]{A.\,Ariga,}
% ...

\bigbreak
\NNINFN
\NNJPARC
\NNKAVLI
\NNNRNU
\NNVIETNAM
\NNDUBNA
\NNYOICHIRO
\NNMBCC
\NNDECEASED

\collaboration{The T2K Collaboration}

\emailAdd{georgios.christodoulou@cern.ch}

\abstract{The electron (anti-)neutrino component of the T2K neutrino beam constitutes the largest background in the measurement of electron (anti-)neutrino appearance at the far detector. The electron neutrino scattering is measured directly with the T2K off-axis near detector, ND280. The selection of the electron (anti-)neutrino events in the plastic scintillator target from both neutrino and anti-neutrino mode beams is discussed in this paper. The flux integrated single differential charged-current inclusive electron (anti-)neutrino cross-sections, $d\sigma/dp$ and $d\sigma/d\cos(\theta)$, and the total cross-sections in a limited phase-space in momentum and scattering angle ($p > 300$~MeV/c and $\theta \leq 45^{\circ}$) are measured using a binned maximum likelihood fit and compared to the neutrino Monte Carlo generator predictions, resulting in good agreement.}

\arxivnumber{2002.11986}

\begin{document}

\maketitle
\flushbottom

\section{Introduction}
\label{sec:introduction}

The measurement of the $\nu_{\mu}\rightarrow\nu_{e}$ (and $\bar\nu_{\mu}\rightarrow\bar\nu_{e}$) oscillations - which is the main goal of the T2K experiment~\cite{T2KExperiment} - is affected by two main background sources. The first is the intrinsic $\nu_{e}$ and $\bar\nu_{e}$ beam contaminations and the second is the neutral current (NC) $\pi^{0}$ production, where the $\pi^{0}$ can mimic an electron from a charged-current (CC) $\nu_{e}$ or $\bar\nu_{e}$ interaction at the far detector, Super-Kamiokande. In addition, the $\nu_{\mu}\rightarrow\nu_{e}$ ($\bar\nu_{\mu}\rightarrow\bar\nu_{e}$) appearance signal is predicted by using a predominantly $\nu_{\mu}$ ($\bar\nu_{\mu}$) sample, which relies on the knowledge of the $\nu_{e}$ ($\bar\nu_{e}$) cross-section relative to the $\nu_{\mu}$ ($\bar\nu_{\mu}$). The modelling of signal and backgrounds is strongly depending on the neutrino cross-sections and the near detector is crucial for measuring them.

The electron (anti-)neutrino flux arises from the decay of kaons, muons and pions produced when the proton beam impinges upon a graphite target. Kaons can decay to electron (anti-)neutrinos through the decay channels, $K^{\pm}\rightarrow \pi^{0}+e^{\pm}+\nu_{e}(\bar\nu_{e})$ and~$K^{0}_{e3}\rightarrow \pi^{\pm} + e^{\mp} + \bar{\nu}_{e} (\nu_{e})$. Muons, mainly produced from pion decay, can also decay to electron (anti-)neutrinos through~$\mu^{\pm}\rightarrow e^{\pm}+\bar{\nu}_{\mu}(\nu_{\mu})+\nu_{e}(\bar\nu_{e})$. The direct contribution of the pion decays to the electron (anti-)neutrino flux is tiny. Together these combinations provide the $\nu_{e}$ and $\bar\nu_{e}$ flux at the near detector. In general, the electron (anti-)neutrinos from kaon decays are more energetic than those from muon decays and populate the high energy tail of the neutrino energy spectrum.

The CC electron (anti-)neutrino selection at the near detector is challenging for two reasons. Firstly, there is a small number of electrons (positrons) produced from CC~$\nu_{e}$~($\bar\nu_{e}$) interactions, compared to the much larger number of muons, pions and protons produced in the final states of CC and NC $\nu_{\mu}$ and $\bar\nu_{\mu}$ interactions. The particle identification (PID) must work extremely well to obtain a pure electron selection. The second reason is the large number of background electrons from sources such as $\pi^{0}$, which can be produced either inside or outside the target detectors. Rejection of background electrons (positrons) is vital for the measurement of the CC $\nu_{e}$ ($\bar\nu_{e}$) interactions.

Electron (anti-)neutrino cross-section measurements in the GeV region are rare since the (anti-)neutrino beams primarily produce muon (anti-)neutrinos. The first CC-$\nu_{e}$ inclusive cross-section measurement and the only CC-$\bar\nu_{e}$ inclusive cross-section measurement so far were made by the Gargamelle bubble chamber experiment in 1978~\cite{Gargamelle}. Thirty-six years later, in 2014, T2K measured the CC-$\nu_{e}$ inclusive cross-section~\cite{ND280NueXs} and in 2016 MINERvA performed the first CC-$\nu_{e}$ cross-section measurement without pions in the final state~\cite{MinervaNue}. Measurements of the electron (anti-)neutrino cross-sections will have a pivotal role for the precision measurements of neutrino oscillations for the current and next generation of long-baseline neutrino oscillation experiments~\cite{DUNETDR, HYPERKTDR}.

Compared to the 2014 results, the work in this paper follows a different approach to measure the CC-$\nu_{e}$ and CC-$\bar\nu_{e}$ cross-sections. Following the developments in the T2K muon neutrino cross-sections measurements~\cite{ND280cc0pi, ND280numucc4pi, ND280nucleff}, the differential cross-sections are measured in a model independent way as a function of electron and positron kinematics (momentum and scattering angle), the quantities which are measured in the near detector. Although cross-section results were calculated in $Q^{2}$ in the 2014 work, such measurements could introduce model dependencies and are not included in this work. Each $Q^{2}$ bin contains contributions from events with different electron kinematics leading to model dependencies when correcting for the efficiencies since our acceptance for backward and high angle events is very poor. Similarly, cross-section measurements in momentum, scattering angle and neutrino energy which are extrapolated to regions with no or very little acceptance are also model dependent since they depend on the underlying model for the efficiency corrections. Such results are not produced in this paper. For the differential cross-section extraction, following the experience from T2K muon neutrino cross-section measurements~\cite{ND280cc0pi, ND280numucc4pi, ND280nucleff}, this work uses a binned likelihood fit with control samples to tune the backgrounds instead of an iterative matrix inversion method~\cite{dagostini}. The likelihood fit method is preferred as the correction of detector smearing effects is independent of the signal model used in the simulation and it allows in-depth validation of the background tuning and of the extracted results. Finally, events with momentum below 200~MeV/c were not considered in the 2014 results. This background enriched region can be used for fit validation studies and it is used in the current work. 

Since the CC-$\nu_{e}$ inclusive cross-section measurement in 2014, T2K has doubled the neutrino data and collected a significant amount of anti-neutrino data. With these new datasets, T2K performs new measurements of the CC-$\nu_{e}$ inclusive cross-sections in neutrino and anti-neutrino modes. In addition, the first CC-$\bar\nu_{e}$ inclusive cross-section in anti-neutrino mode, since Gargamelle, is measured.

\section{Experimental Setup}
\label{sec:experimentalsetup}

\subsection{T2K beam}
\label{subsec:t2kbeam}

The T2K neutrino beam is produced at the Japan Proton Accelerator Research Complex (J-PARC) by colliding 30~GeV protons with a graphite target. The pions and kaons produced in the target are focused by three magnetic horns and decay in flight to produce neutrinos. T2K can run with either forward horn current (FHC) or with reverse horn current (RHC) producing beams in neutrino or anti-neutrino enhanced mode, respectively. 

\begin{figure}[htbp]
	\centering
	\includegraphics[width=0.495\textwidth]{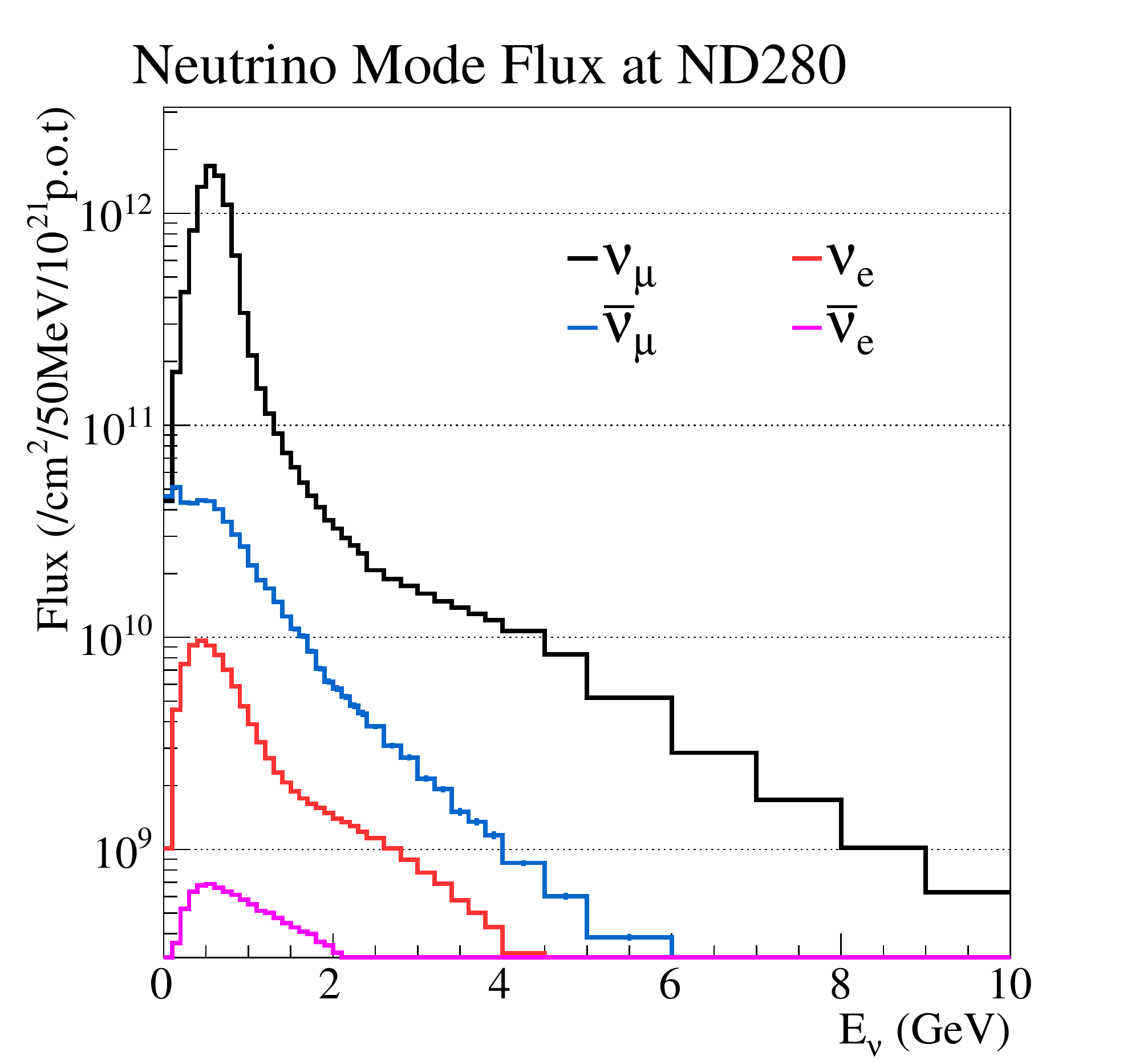}
	\includegraphics[width=0.495\textwidth]{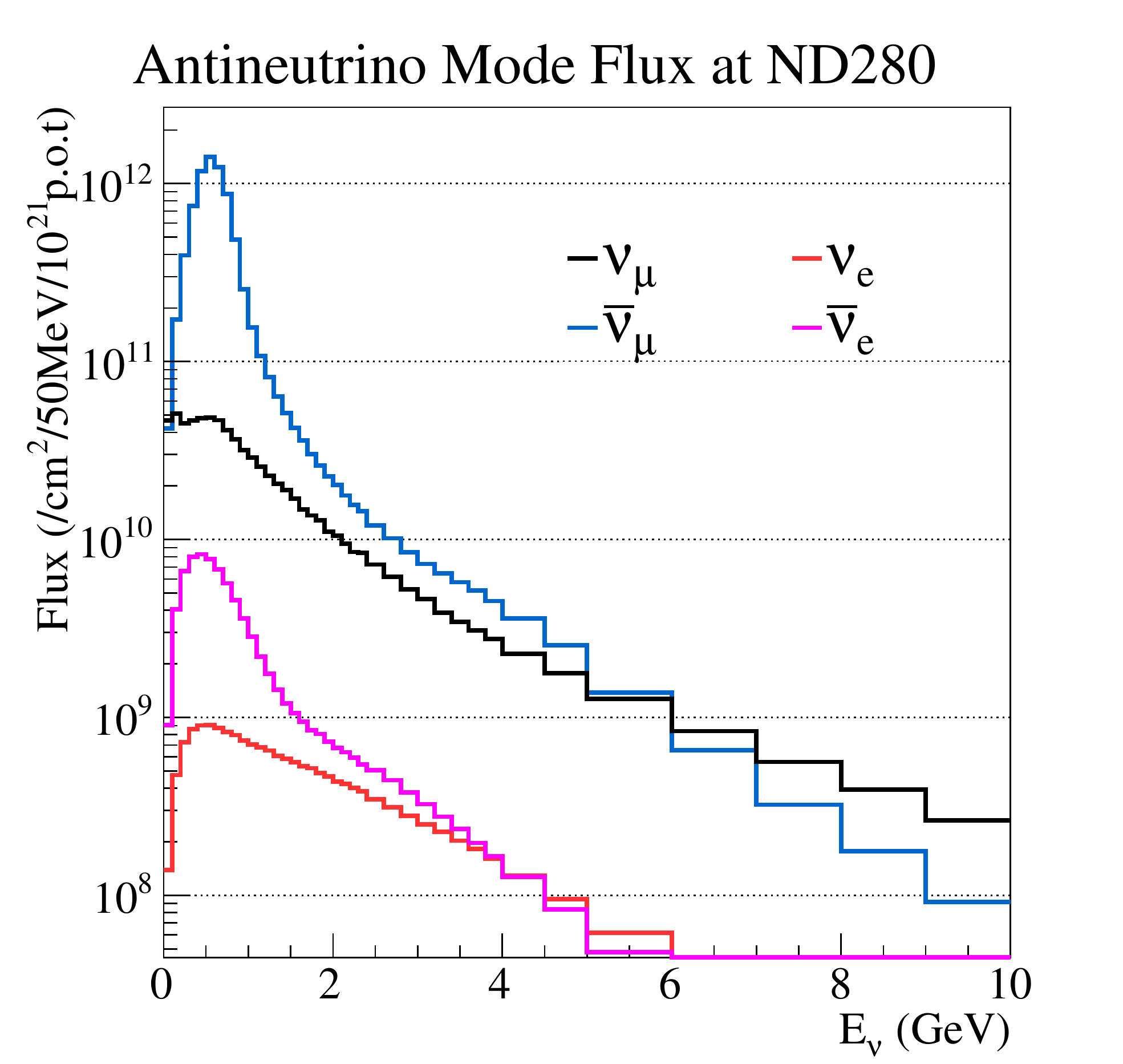}
	\caption{The neutrino and anti-neutrino fluxes at ND280 in neutrino (FHC) mode (left) and in anti-neutrino (RHC) mode (right).}
	\label{fig:neutrinoflux}
\end{figure}

The T2K beamline~\cite{T2Kbeamline} is simulated using FLUKA2011~\cite{fluka1,fluka2}, GEANT3~\cite{geant3} and GCALOR~\cite{gcalor}. The simulated yields of hadronic particles are tuned using the NA61/SHINE~\cite{na61shine1,na61shine2,na61shine3} thin target measurements. The neutrino fluxes at the off-axis near detector ND280 in FHC and RHC are shown in Figure~\ref{fig:neutrinoflux}. The off-axis position of the near detector, from the neutrino beam direction, results in a narrow-band $\nu_{\mu}$ or $\bar\nu_{\mu}$ beam, however, the same does not occur with $\nu_{e}$ and $\bar\nu_{e}$ fluxes due to their production via three-body decays, resulting in broader $\nu_{e}$ and $\bar\nu_{e}$ spectra. The mean of the $\nu_{e}$ energy spectrum at ND280 is 1.28~GeV in FHC and 1.98~GeV in RHC. The mean of the $\bar\nu_{e}$ energy spectrum in RHC is 0.99~GeV.
The total integrated $\nu_{e}$ flux at ND280 in FHC is $\Phi_{\nu_{e}}^{FHC} = \left(2.67\pm0.24\right)\times10^{11}$~$\rm neutrinos/cm^{2}$ and in RHC is $\Phi_{\nu_{e}}^{RHC} = \left(2.65\pm0.21\right)\times10^{10}$~$\rm neutrinos/cm^{2}$. The total integrated $\bar\nu_{e}$ flux at ND280 in RHC is $\Phi_{\bar\nu_{e}}^{RHC} = \left(1.00\pm0.10\right)\times10^{11}$~anti-neutrinos$\rm/cm^{2}$.

\subsection{T2K off-axis near detector ND280}
\label{subsec:ND280}

\begin{figure}[htbp]
	\centering
	\includegraphics[width=0.7\textwidth]{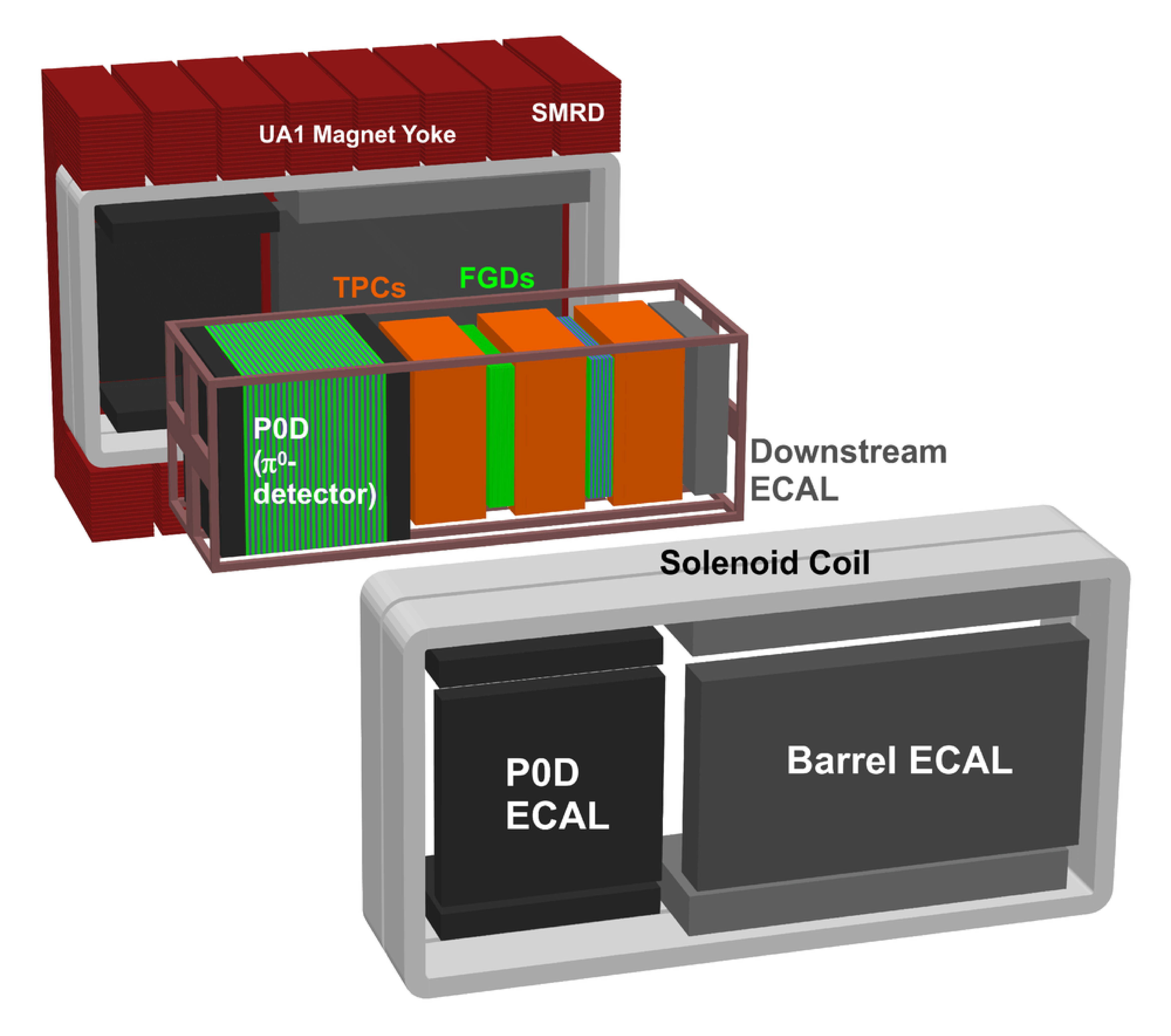}
	\caption{An exploded view of the T2K near detector, ND280. The neutrino beam enters ND280 from the left.}
	\label{fig:nd280detector}
\end{figure}

The $2.5^{\circ}$ off-axis near detector, ND280, is located 280 metres from the proton target. The main goal of ND280 is to constrain the neutrino flux and the interaction cross-sections. It is composed of several sub-detectors located inside a 0.2~T magnet, as depicted in Figure~\ref{fig:nd280detector}. The front part is the $\pi^{0}$ detector (P0D)~\cite{ND280P0D} and is optimised to measure neutrino interactions with $\pi^{0}$ production. The rear part is the tracker and it is optimised to measure charged particles produced in neutrino interactions. It consists of two Fine-Grained Detectors~\cite{ND280FGD}, the first of which is composed of layers of plastic scintillator (FGD1) and the second has alternating layers of plastic scintillators and water (FGD2). 

\begin{figure}[htbp]
 \centering
	\includegraphics[width=0.495\textwidth]{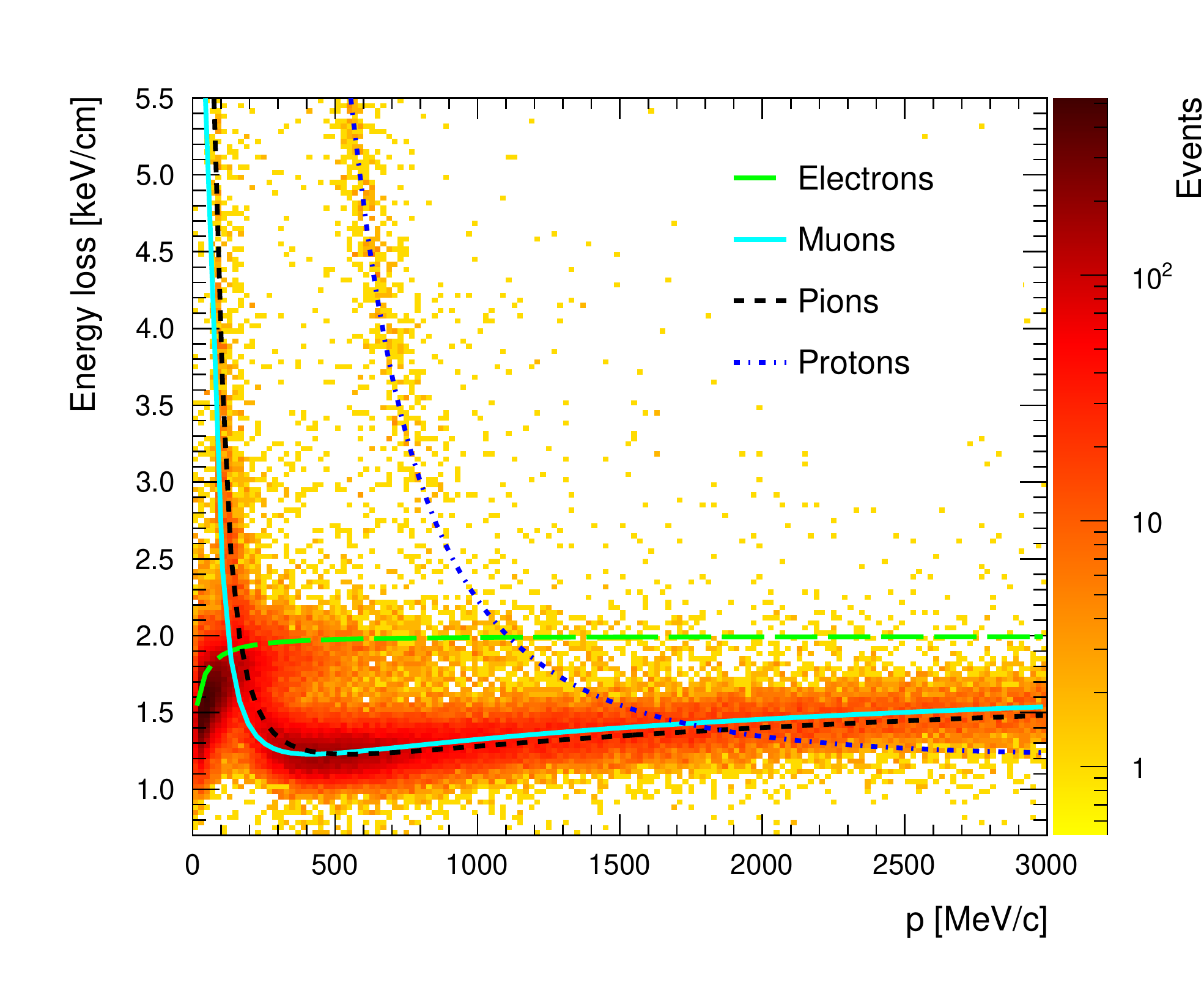}
	\includegraphics[width=0.495\textwidth]{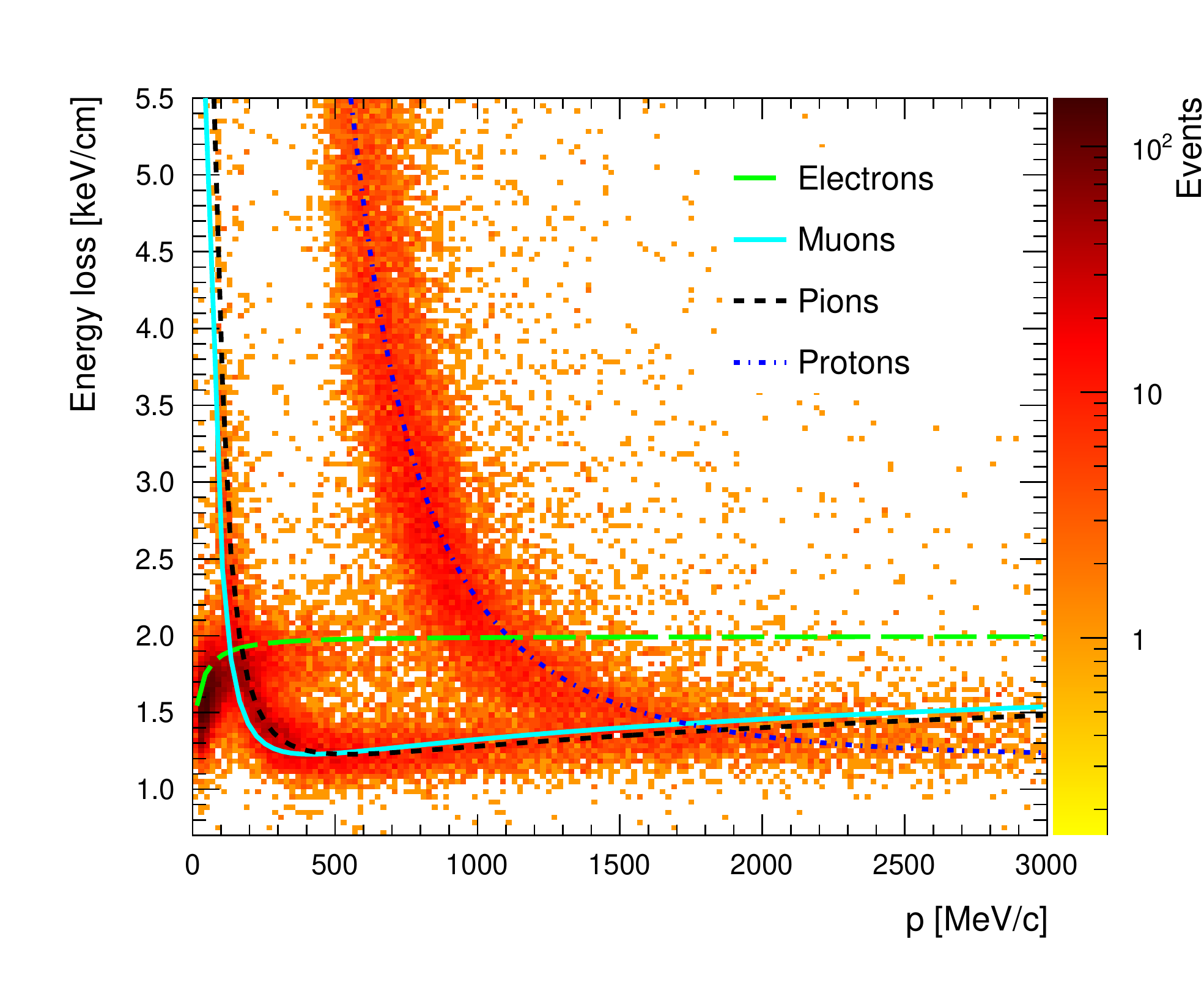}
	\caption{TPC energy loss for tracks in data originating in FGD1. Left: negatively charged tracks. Right: positively charged tracks. The expected energy loss curves for electrons, muons, pions and protons are also shown.}
	\label{fig:tpcdedx}
\end{figure}

The P0D, FGD1 and FGD2 provide the target mass for neutrino interactions and each is followed by a Time Projection Chamber (TPC1, TPC2 and TPC3)~\cite{ND280TPC}. The TPCs are filled with a gas mixture based on argon and provide excellent track reconstruction with a momentum resolution of roughly 8\% for 1~GeV/c tracks. This can be combined with energy loss ($dE/dx$) measurements in order to perform PID of tracks crossing the TPCs. The measured and the expected $dE/dx$ are used to define the "pull" (the difference between the measured mean ionization and the expected one divided by the resolution) of each particle species. The TPC energy loss for negatively and positively charged tracks originating in FGD1 is shown in Figure~\ref{fig:tpcdedx}. Notice the region below 200~MeV/c where the electron $dE/dx$ curve crosses with the muon and pion dE/dx curves, and the region around 1~GeV/c where the proton $dE/dx$ curve crosses with the electron $dE/dx$ curve. 

The P0D and the tracker are surrounded by the lead-scintillator Electromagnetic Calorimeter (ECal)~\cite{ND280ECal} and a Side Muon Range Detector (SMRD)~\cite{ND280SMRD}. The ECal measures the energy of photon and electrons (EM energy) and provides additional PID for minimum ionizing particles (MIP), electromagnetic showers (EM) and highly ionizing stopping particles (HIP) like protons. 

The ECal EM energy is reconstructed under the hypothesis that the energy deposit is due to an electromagnetic shower. Comparing the TPC momentum with the ECal EM energy, electrons can be separated from muons and protons. The ratio of the TPC momentum over the ECal EM energy peaks at unity for electrons and at lower values for muons and protons. The ECal EM energy resolution is approximately 10\% at 1~GeV.

The ECal PID is based on the longitudinal and lateral profile of ECal clusters to generate probability density functions (PDFs). These are combined for each particle type and PID variable to form a likelihood from the products of the PDFs, see \cite{ND280NueSel} for details.

\begin{figure}[htbp]
 \centering
	\includegraphics[width=0.495\textwidth]{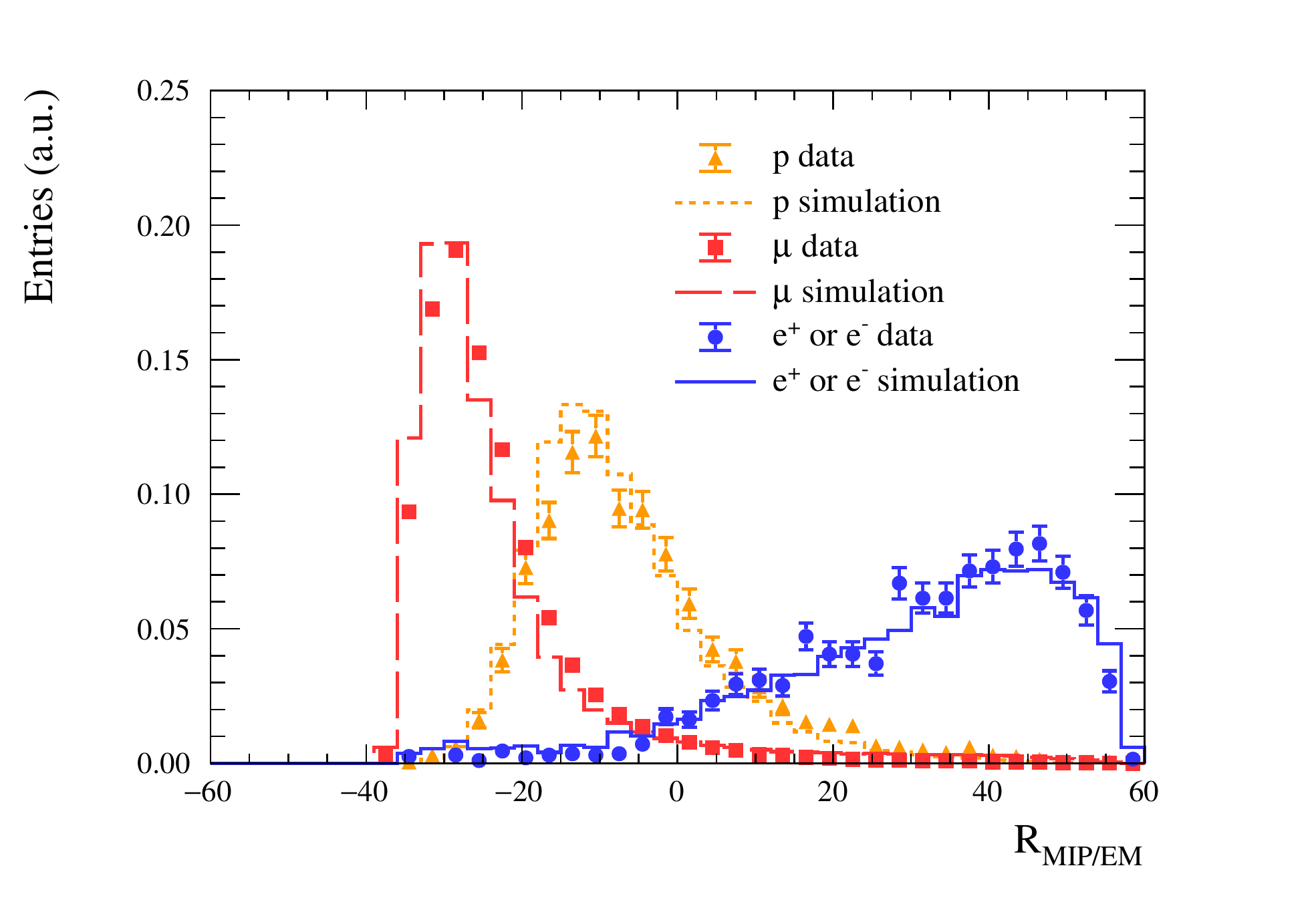}
	\includegraphics[width=0.495\textwidth]{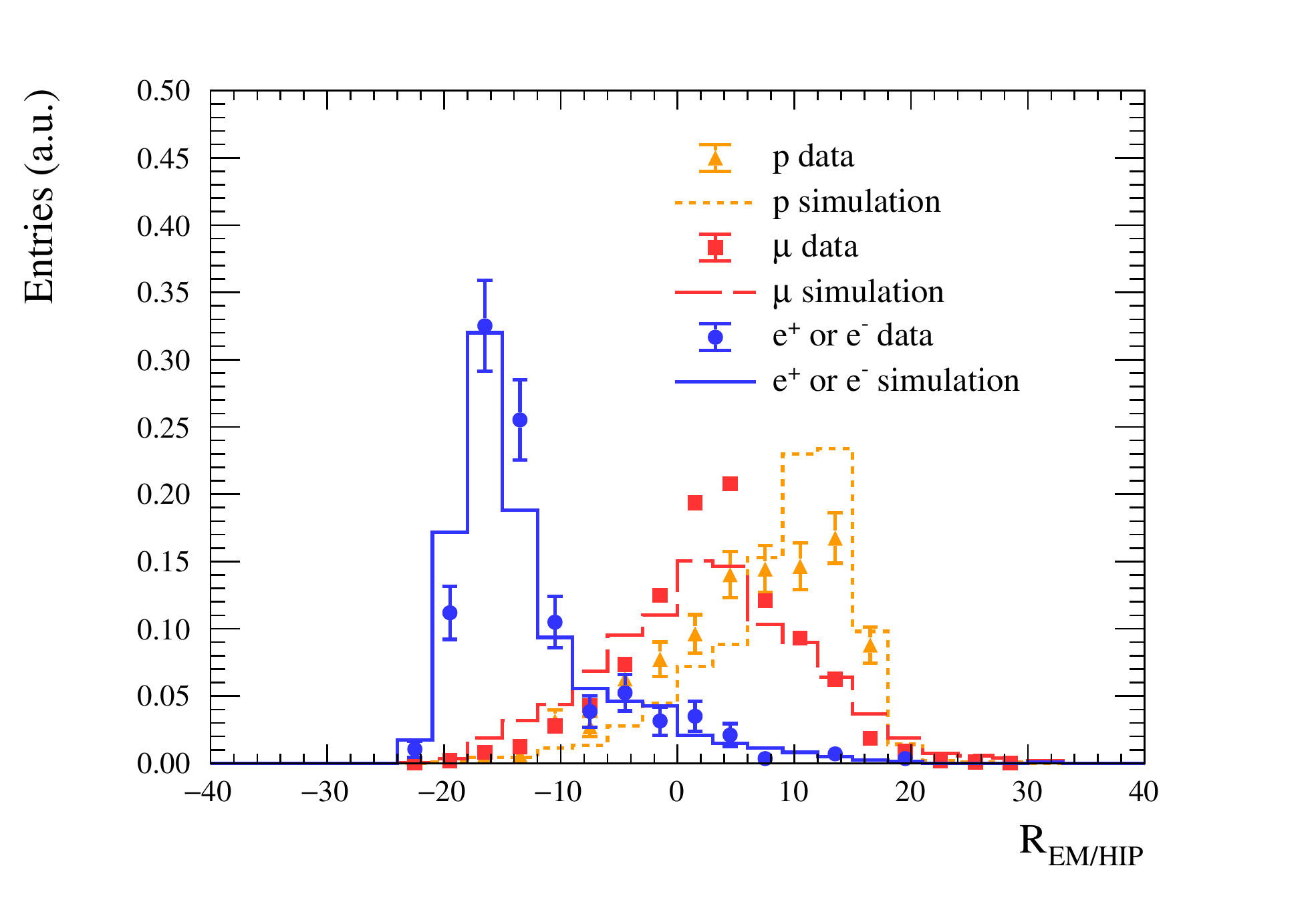}
	\caption{Performance of the ECal PID using high purity control samples of cosmic and through-going muons, electrons and positrons from gamma conversions and protons from neutrino interactions. Left: Log-likelihood ratio of the ECal track-shower ($R_{MIP/EM}$) PID. Right: Log-likelihood ratio of the ECal electron-proton ($R_{EM/HIP}$) PID for showers with $R_{MIP/EM} > 0$ and $p > 600$~MeV/c. Plots are normalised to unity.}
	\label{fig:ecalpid}
\end{figure}

$R_{MIP/EM}$ is the log-likelihood ratio of the MIP and electron hypothesis and $R_{EM/HIP}$ is the log-likelihood ratio of the electron and proton hypothesis. The $R_{MIP/EM}$ for high purity control samples (90\% or better) is shown in Figure~\ref{fig:ecalpid}, where the muon sample comprises cosmic muons and muons produced by neutrino interactions outside ND280 that cross the detector (through-going muons), the electron sample is formed from electron-positron pairs from photon conversions and the protons are from neutrino interactions. The ECal can provide supplementary PID to the TPC, especially in the region around 1~GeV/c where the TPC energy loss curves of electrons and protons cross. Figure~\ref{fig:ecalpid} also shows the $R_{EM/HIP}$ for showers (classified by $R_{MIP/EM} > 0$) with $p > 600$~MeV/c only. Although there are some shape differences in data and simulation for $R_{MIP/EM}$ and $R_{EM/HIP}$, the data and simulation efficiencies to select electron (or positrons) and reject muons and protons are similar. The PID efficiencies in the simulation are corrected using the data control samples. 

\section{Data samples and MC simulation}
\label{sec:analysissamples}

For FHC, $11.92\times10^{20}$ protons-on-target (POT) are analysed corresponding to data collected in the periods 2010-2013 and 2016-2017. For RHC, $6.29\times10^{20}$ POT are analysed corresponding to data collected from 2014 to 2016.

The ND280 flux is simulated as described in section~\ref{subsec:t2kbeam}. The (anti-)neutrino interactions with the ND280 detector materials, including nuclear effects, are simulated using NEUT 5.3.2~\cite{neutmc} and GENIE 2.8.0~\cite{geniemc} Monte Carlo (MC) generators. The neutrino generators account for differences in the lepton mass for the muon and electron neutrino cross-section computations. However, other effects like radiative corrections, modifications of the pseudoscalar form factors and the effect of form factors to second class vector and axial currents are not considered~\cite{NueNumuCCQE}. 

NEUT 5.3.2 uses the Llewellyn-Smith formalism~\cite{LlewellynSmith} to describe the CC quasi-elastic neutrino-nucleon cross sections. The spectral function is used as the nuclear model~\cite{SpectralFunction}. The axial mass used for the CC quasi-elastic process is set to 1.21~$\rm GeV/c^{2}$. The simulation of multi-nucleon interactions, where the neutrino interacts with a correlated pair of nucleons, is described using the Nieves et al. model~\cite{Nieves2p2h}. The resonant pion production process with an invariant mass $W \leq 2$~$\rm GeV/c^{2}$ is described by the Rein-Sehgal model~\cite{ReinSehgal}. The resonant axial mass set to 0.95~$\rm GeV/c^{2}$. The deep inelastic scattering (DIS) is calculated for $W > 1.3$~$\rm GeV/c^{2}$ and is modeled using the GRV98 parton distribution function~\cite{GRV98} including the Bodek and Yang corrections~\cite{BodekYangNeut}. Single pion production with $W \leq 2$~$\rm GeV/c^{2}$ is suppressed to avoid double counting with resonant production. Final state interactions describe the transportation of the hadrons produced from neutrino interaction through the nucleus and are simulated using a semi-classical intra-nuclear cascade model.

GENIE 2.8.0 uses a different value for the axial mass for quasi-elastic process of 0.99~$\rm GeV/c^{2}$. It relies on a different nuclear model using a relativistic Fermi gas with Bodek and Ritchie modifications~\cite{BodekRitchie}. Resonant production is based on Rein-Sehgal model, same as NEUT. In GENIE the resonant model is not restricted to the single pion decay channel. To avoid double counting with the DIS model, the resonant model is switched off when $W > 1.7$~$\rm GeV/c^{2}$. The resonant axial mass is set to 1.12~$\rm GeV/c^{2}$. DIS is simulated similar to NEUT but using slightly different Bodek-Yang corrections~\cite{BodekYangGenie}. A parametrized model of final state interactions (GENIE "hA" model) is used.

Detail description of the NEUT and GENIE models can be found in previous T2K publications~\cite{ND280numucc4pi, T2KOscLong}.

GEANT 4.9.4~\cite{geant4} is used to transport the final state particles through the ND280 detector. Nominal MC is produced by simulating approximately 10 times the data POT for both NEUT and GENIE. 

Data-driven reconstruction efficiency corrections are applied to the nominal MC. These corrections are estimated using high-purity ($>$~90\%) control samples of cosmic and through-going muons, electrons and positrons from photon conversions and protons from neutrino interactions.

The nominal ND280 MC simulates only the neutrino interactions that occur within the ND280 detector.  In reality, neutrino interactions also occur in the surrounding material (sand interactions) and these produce particles that enter ND280.  These particles can then affect the event selection by triggering one of the three veto cuts \footnote{Section \ref{subsec:selcriteria} describes the event selection and the veto cuts in the TPC, ECal and P0D sub-detectors.} during a beam bunch time window (i.e. an ND280 event) and hence causing an ND280 event to fail the selection cut.  This is the sand pile-up effect, which is inherently present in the data, but not in the nominal ND280 MC. To simulate the effect, a second MC simulation is generated (sand MC) to estimate the rate at which sand interactions trigger these veto cuts in coincidence with an ND280 event.  To propagate this effect to the nominal ND280 MC there is a pile-up correction, which is a weight that is applied to all ND280 events, for each of the veto cuts.  If sand interactions are estimated to trigger a given veto for X\% of ND280 events, then a weight of (1 - X/100) is applied to all ND280 events.  Since the pile-up rate depends on the beam intensity and on the beam mode (FHC or RHC), the corrections are computed separately for each data period. For the high intensity neutrino beam in 2017, the total pile-up correction is approximately 5\%.

\section{Selection of electron (anti-)neutrino interactions at ND280}
\label{sec:nueselection}

The selection of electron (anti-)neutrinos in FGD1 closely follows the steps described in the 2014 FHC CC-$\nu_{e}$ analysis~\cite{ND280NueSel} and is summarised below. There are several reconstruction improvements since the 2014 analysis and additional selection criteria are applied to improve purities. The RHC CC-$\nu_{e}$ selection is identical to the FHC selection, but for the CC-$\bar\nu_{e}$ additional selection criteria are applied to remove the proton background. Details are described in section~\ref{subsec:selcriteria}.

\subsection{Signal and background definitions}
\label{subsec:sigbkgdef}

A MC event is defined as signal if the selected primary track is an electron (positron) from a CC-$\nu_{e}$ (CC-$\bar\nu_{e}$) interaction with the vertex inside the FGD1 fiducial volume, which has a total mass of 919.5~kg, corresponding to $\left(5.54\pm0.04\right)\times10^{29}$ nucleons. Backgrounds are separated into four categories: photon, muon, proton and other backgrounds. The photon background category considers events where the selected primary track is an electron or positron from a photon conversion and the true conversion point is inside the FGD1 fiducial volume. Events where the selected primary track is a muon (proton), but misidentified as electron enter the muon (proton) background category. Any other backgrounds including misidentified pions, electrons from photons converting outside of the fiducial volume but reconstructed inside the fiducial volume, electrons from $\pi^{0}$ Dalitz decay and Michel electrons go into the other background category. 

\subsection{Event selection}
\label{subsec:selcriteria}

The event selection for CC-$\nu_{e}$ and CC-$\bar\nu_{e}$ events is described in the following:

\begin{enumerate}[(i)]

\item Only events during periods of good beam and detector quality are used. The event time has to be reconstructed within one of the eight distinct beam bunches.
\item The highest momentum negatively charged (leading negatively charged) FGD1-TPC track, for the CC-$\nu_{e}$ selection, or the highest momentum positively charged (leading positively charged) FGD1-TPC track, for CC-$\bar\nu_{e}$ selection, with a vertex in the FGD1 fiducial volume is selected. The leading positively charged track in the CC-$\bar\nu_{e}$ selection must also be the highest momentum track (from all negatively and positively charged tracks).
\item To ensure reliable PID and momentum measurements, the selected leading track is required to have at least 18 TPC hits if it enters the ECal or 36 TPC hits if it does not enter the ECal. The momentum spectra of the selected leading negatively charged and leading positively charged tracks with the minimum number of TPC hits are shown in Figure~\ref{fig:CCNueBeforePIDMom}. Notice the large number of protons selected as the leading positively charged track in the RHC CC-$\bar\nu_{e}$ selection. Some data-MC discrepancies are visible in the low momentum region which contains the poorly modelled photon and other backgrounds.
\begin{figure}[htbp]
 \centering
	\includegraphics[width=0.495\textwidth]{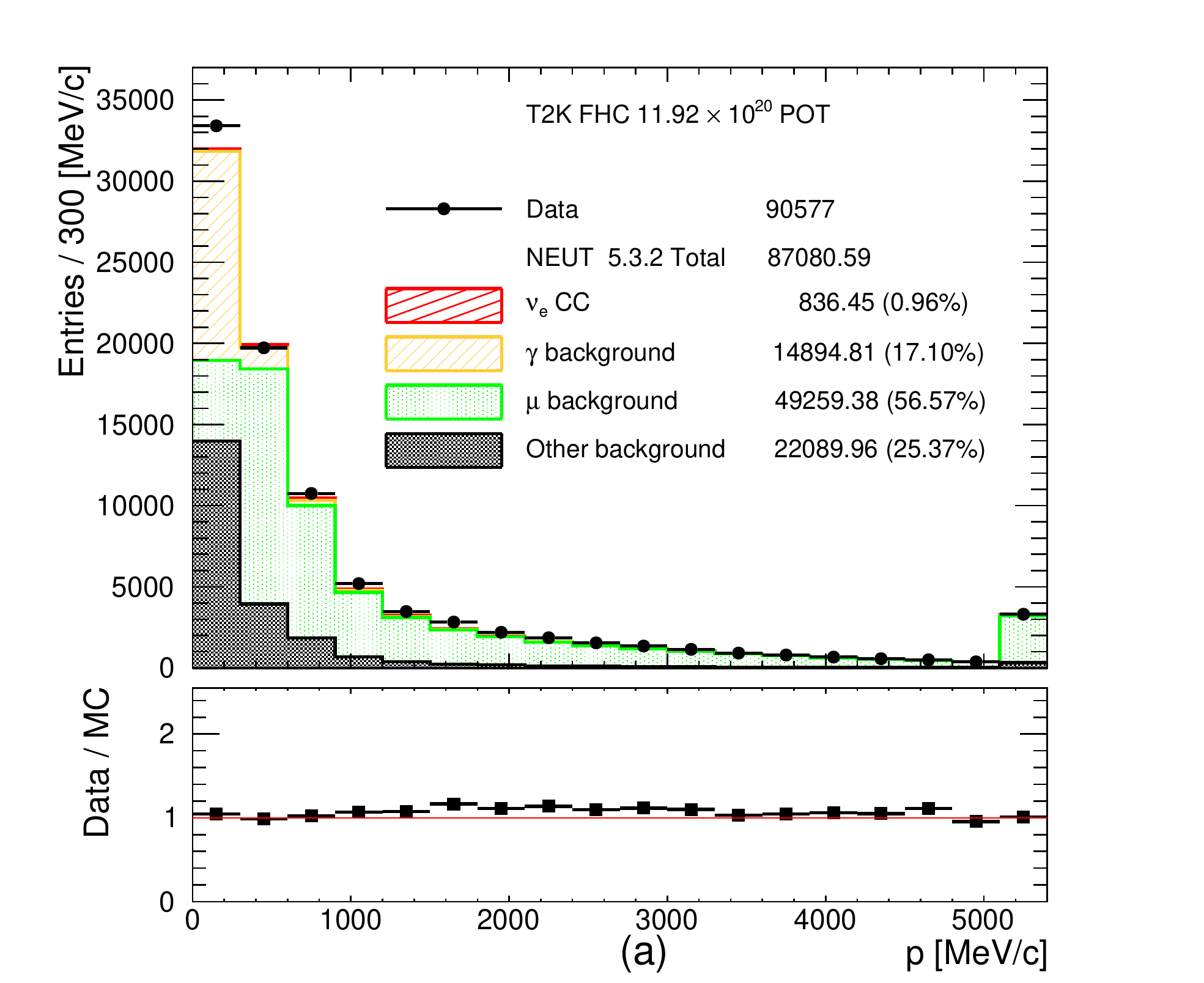}
	\includegraphics[width=0.495\textwidth]{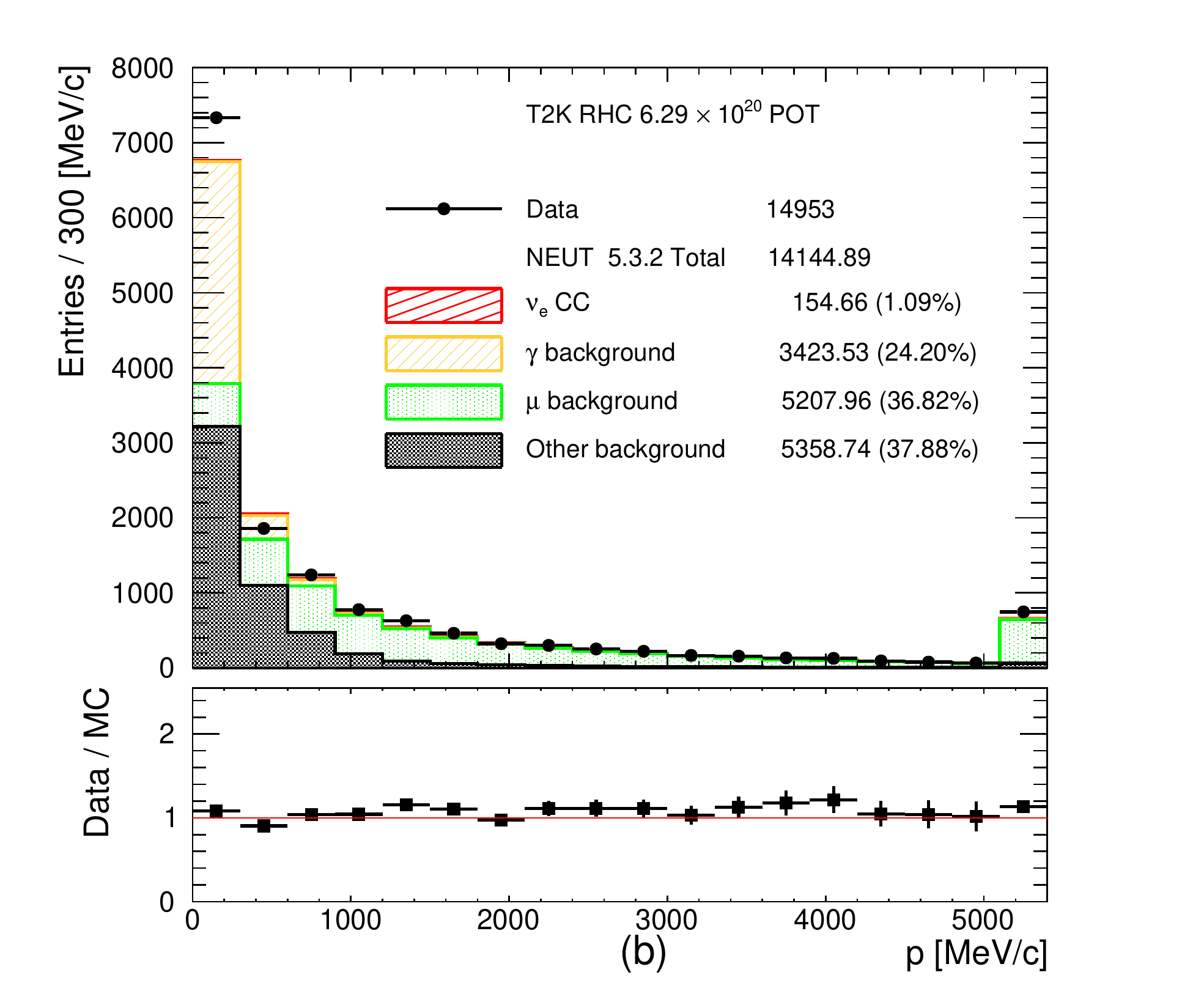}
	\includegraphics[width=0.495\textwidth]{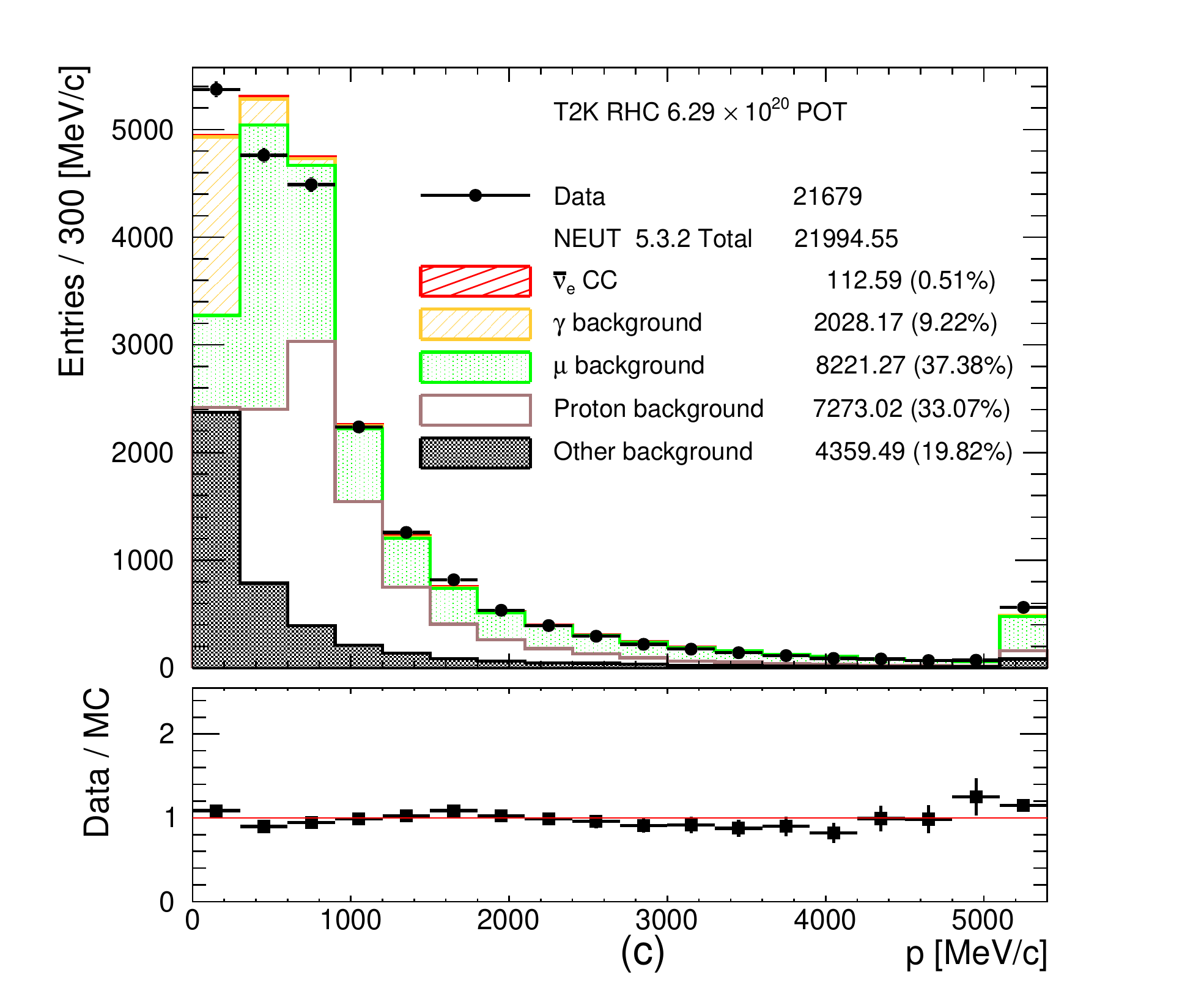}
	\caption{Momentum distribution of the selected leading negatively charged track with a vertex in the FGD1 fiducial volume for (a) FHC CC-$\nu_{e}$, (b) RHC CC-$\nu_{e}$ and (c) leading positively charged track for RHC CC-$\bar\nu_{e}$. The number of MC events is normalized to the data POT. The last bin is the overflow bin.}
	\label{fig:CCNueBeforePIDMom}
\end{figure}
\item TPC PID is applied to select electrons and remove minimum-ionizing tracks. Using the electron TPC pull, the leading track must agree with the electron TPC $dE/dx$ hypothesis. If the leading track does not enter the ECal, then additional cuts on the TPC PID are applied using the muon and pion TPC pulls. The event is rejected if the leading track agrees with the muon or pion TPC hypothesis; events around 150 MeV/c, including electrons, are rejected where the only information (TPC) is unable to distinguish them.
\item Additional PID is applied using either the ECal EM energy or the ECal PID depending on the momentum of the leading track as it enters the ECal. To maximize the efficiency, if the leading track has $p > 1$~GeV/c and is fully contained in the ECal, the reconstructed ECal EM energy is used to separate EM showers from MIPs and it is required to be larger than 1~GeV. Otherwise the ECal MIP/EM shower PID discriminator $R_{MIP/EM}$ has to agree with the EM shower PID hypothesis. Events that pass the TPC and ECal PID are shown in Figure~\ref{fig:CCNueAfterPIDMom}. For the CC-$\bar\nu_{e}$ selection a complication arises since the TPC energy loss curves for positrons and protons cross around 1~GeV/c (see Figure~\ref{fig:tpcdedx}) leaving a significant amount of proton background.
\begin{figure}[htbp]
 \centering
	\includegraphics[width=0.495\textwidth]{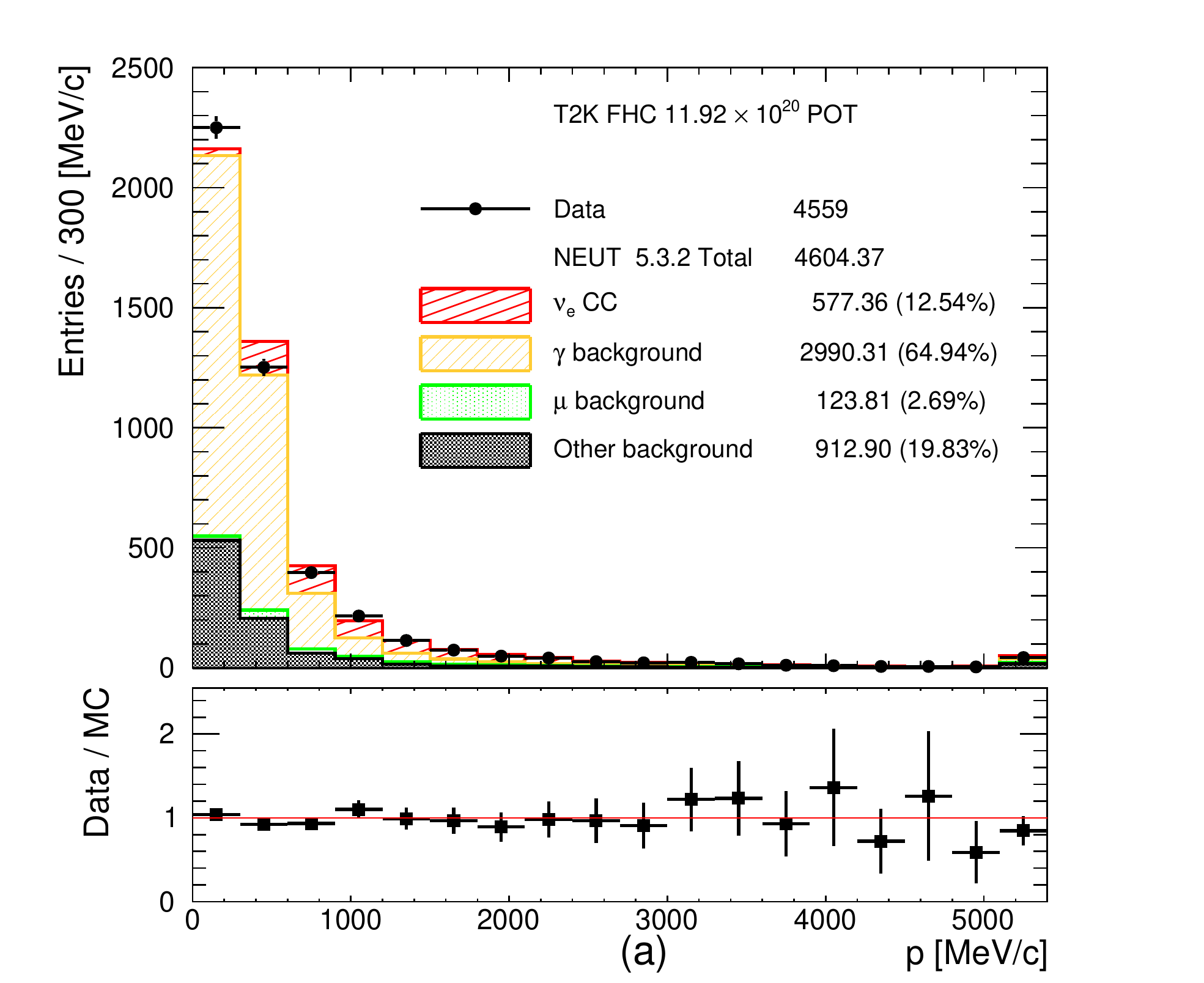}
	\includegraphics[width=0.495\textwidth]{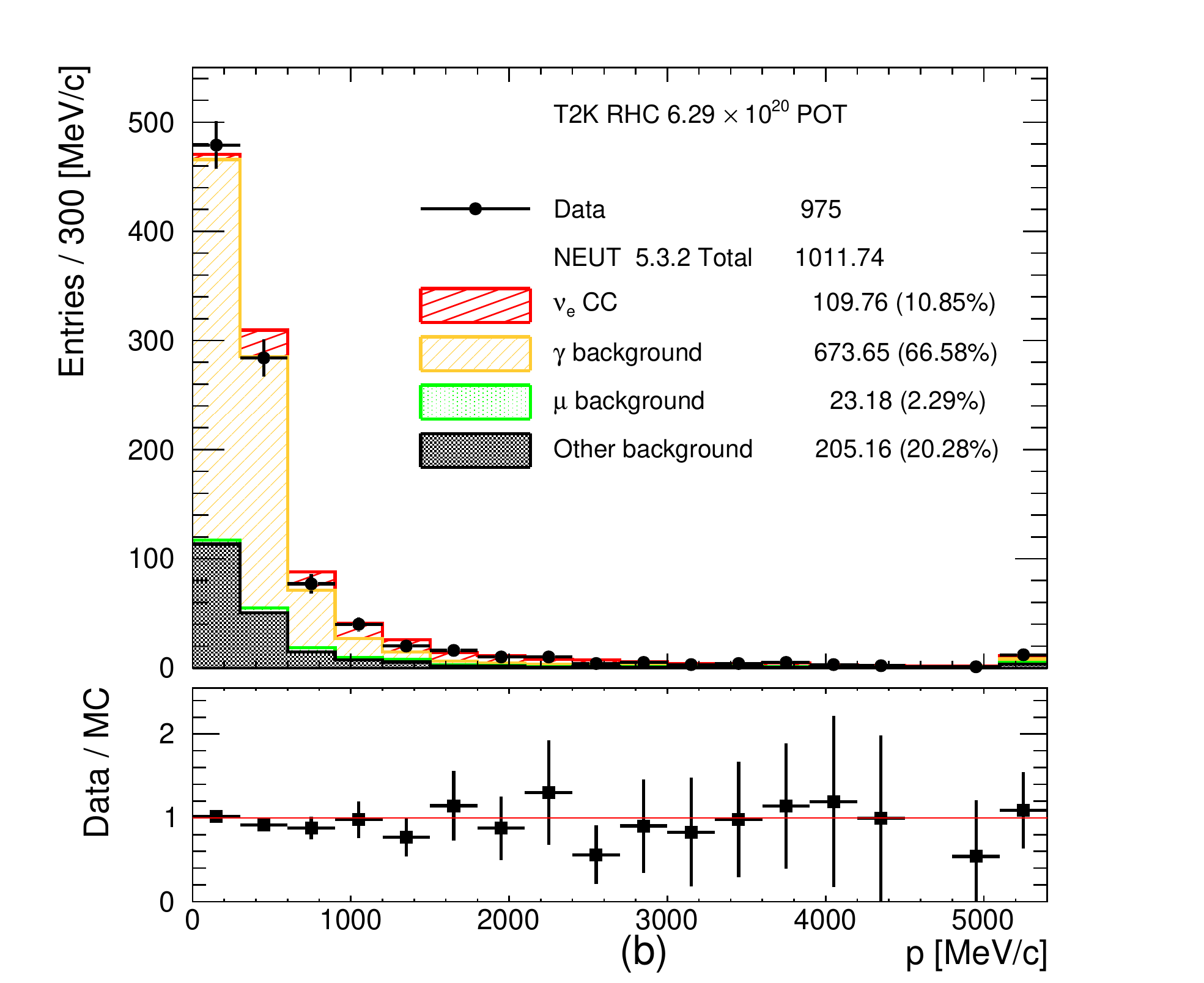}
	\includegraphics[width=0.495\textwidth]{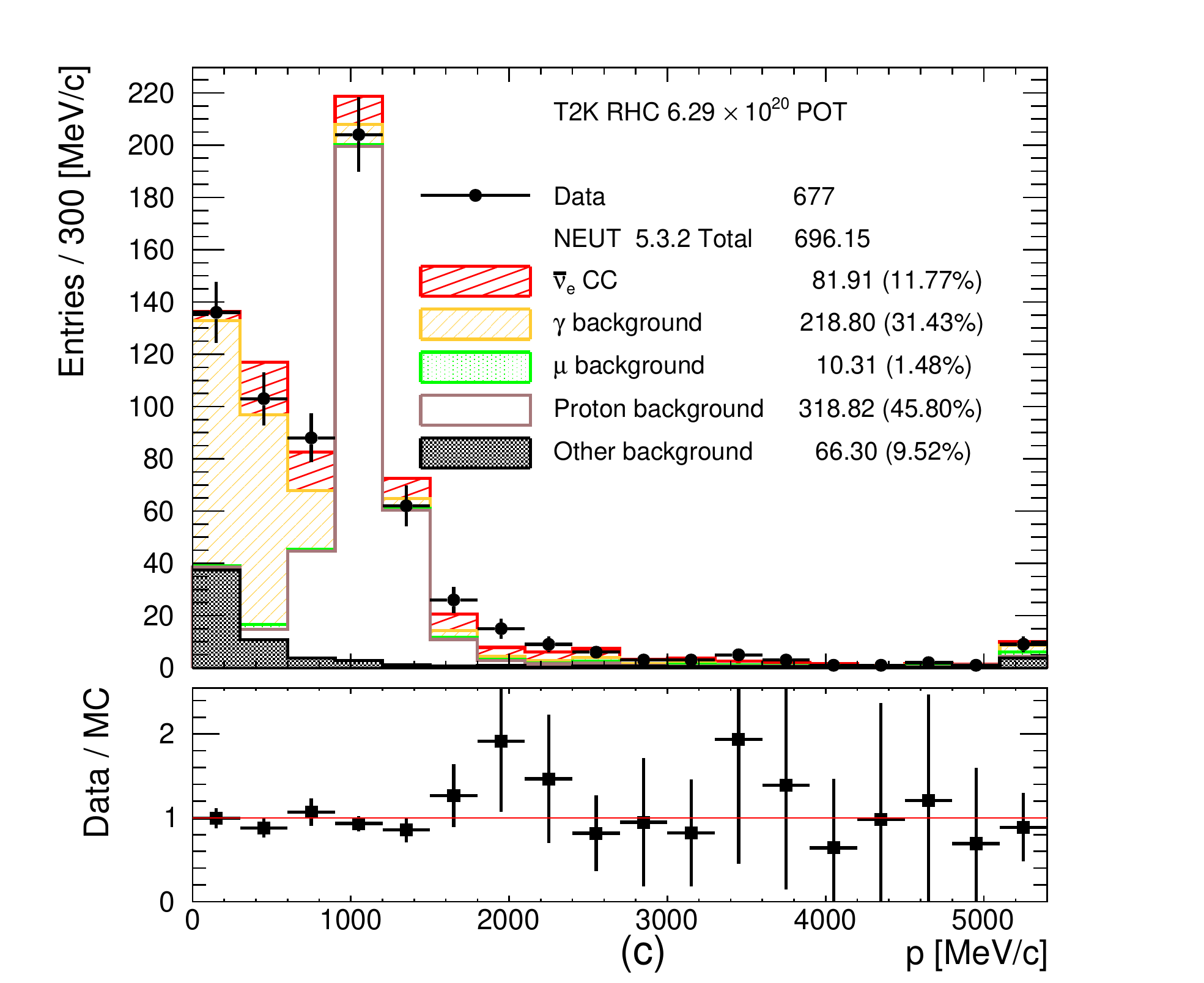}
	\caption{Momentum distribution after the TPC and ECal PID cuts for (a) FHC CC-$\nu_{e}$, (b) RHC CC-$\nu_{e}$ and (c) RHC CC-$\bar\nu_{e}$ candidates. The number of MC events is normalized to the data POT. Notice the significant proton background around 1~GeV/c in the CC-$\bar\nu_{e}$ selection due to the weakness of the TPC PID to separate positrons from protons, see the text and Figure~\ref{fig:tpcdedx} for details. Additional PID is applied to remove this proton background, see the text for more details. The last bin is the overflow bin.}
	\label{fig:CCNueAfterPIDMom}
\end{figure}
\item Search for the paired FGD1-TPC electron or positron track from a potential photon conversion. The paired track must have opposite charge than the leading track, start within 5~cm from the leading track and agree with the electron TPC dE/dx hypothesis. If several paired tracks are found, the pair with the lowest invariant mass is considered since it is more likely to come from a photon conversion. Pairs with invariant mass less than 110~$\rm MeV/c^{2}$ are removed.
\item Veto P0D, TPC and ECal activity upstream of the vertex and remove events with additional vertices in FGD1. Events with multiple vertices more likely to come from a $\nu_{\mu}$ interaction with one or more $\pi^{0}$ in the final state.
\item For the CC-$\bar\nu_{e}$ selection, additional selection criteria are applied if the leading positively charged track has $p > 600$~MeV/c, the region which is contaminated by the proton background. If the leading positively charged track produce shower activity in FGD2 then it is selected. If the leading positively charged track enters the ECal, the proton background can be removed by comparing the ECal EM energy ($E$) and the TPC momentum ($p$) using a cut~$E/p > 0.65$. In addition, the $R_{EM/HIP}$ shower PID discriminator has to agree with the EM shower hypothesis. 
\item For the CC-$\bar\nu_{e}$ selection, if the leading positively charged track stops in FGD2, the FGD2 energy loss must not agree with the proton hypothesis. 
\item Remove external background by comparing the time stamps of the leading track between FGD1 and ECal. This cut aims to remove tracks originating in the ECal and stop in FGD1 but are mis-reconstructed with the wrong direction.
\item Check if the leading track is broken inside FGD1. A track is broken if it originates in FGD1 and is not reconstructed as a single track, but is broken into two or more components. In such pathological cases the leading track could originate outside the fiducial volume but mis-reconstructed within it. If the leading track follows an isolated FGD1 track then the event is removed. 
    
\end{enumerate}

\begin{figure}[htbp]
 \centering
	\includegraphics[width=0.99\textwidth]{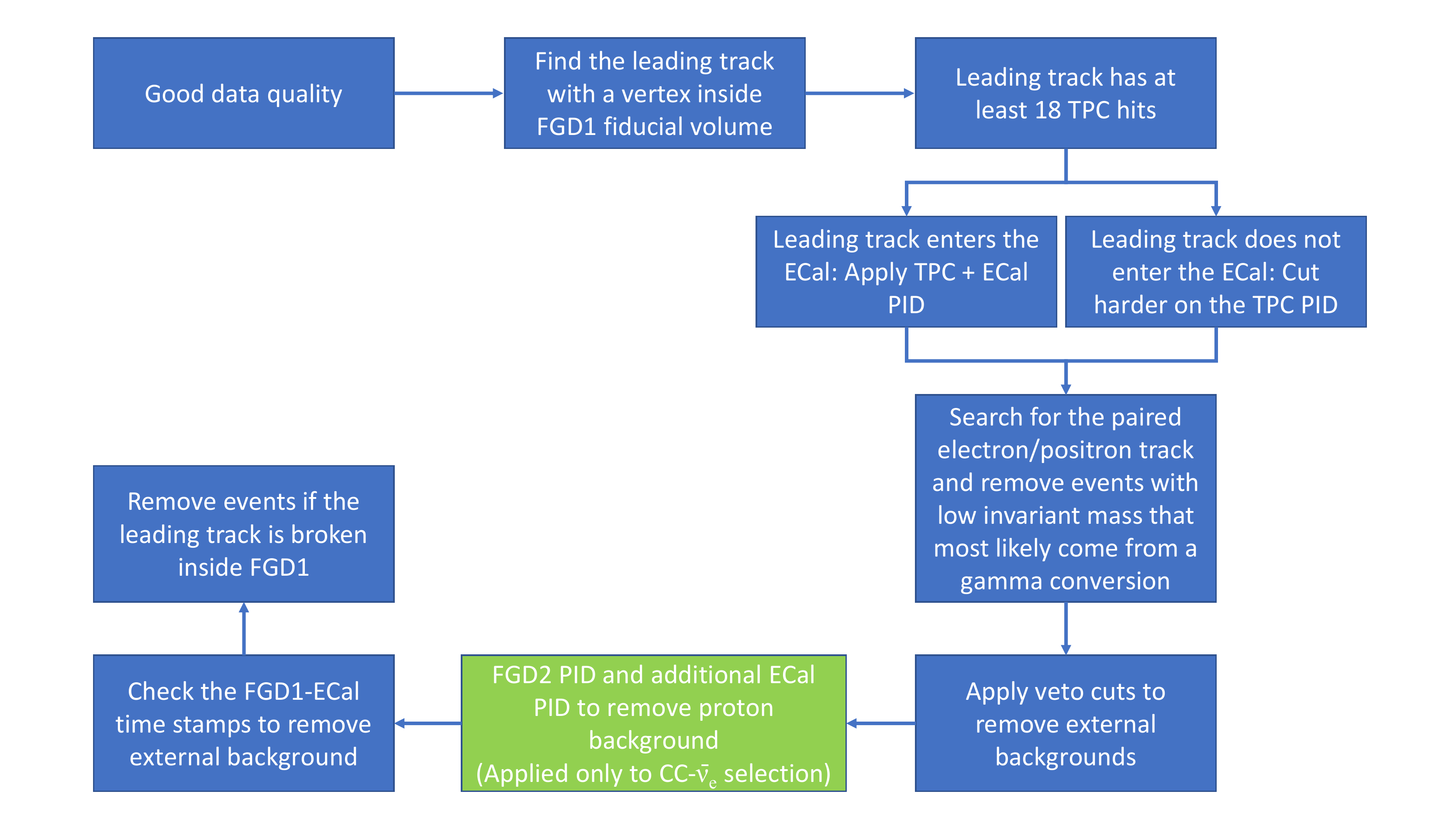}
	\caption{Summary of the CC-$\nu_{e}$ and CC-$\bar\nu_{e}$ selections in FGD1. See the text for the details of each cut.}
	\label{fig:nuecutflow}
\end{figure}

Figure~\ref{fig:nuecutflow} summarises the CC-$\nu_{e}$ and CC-$\bar\nu_{e}$ selections. 

\subsection{Final selection}
\label{subsec:finalsel}

\begin{figure}[htbp]
 \centering
	\includegraphics[width=0.495\textwidth]{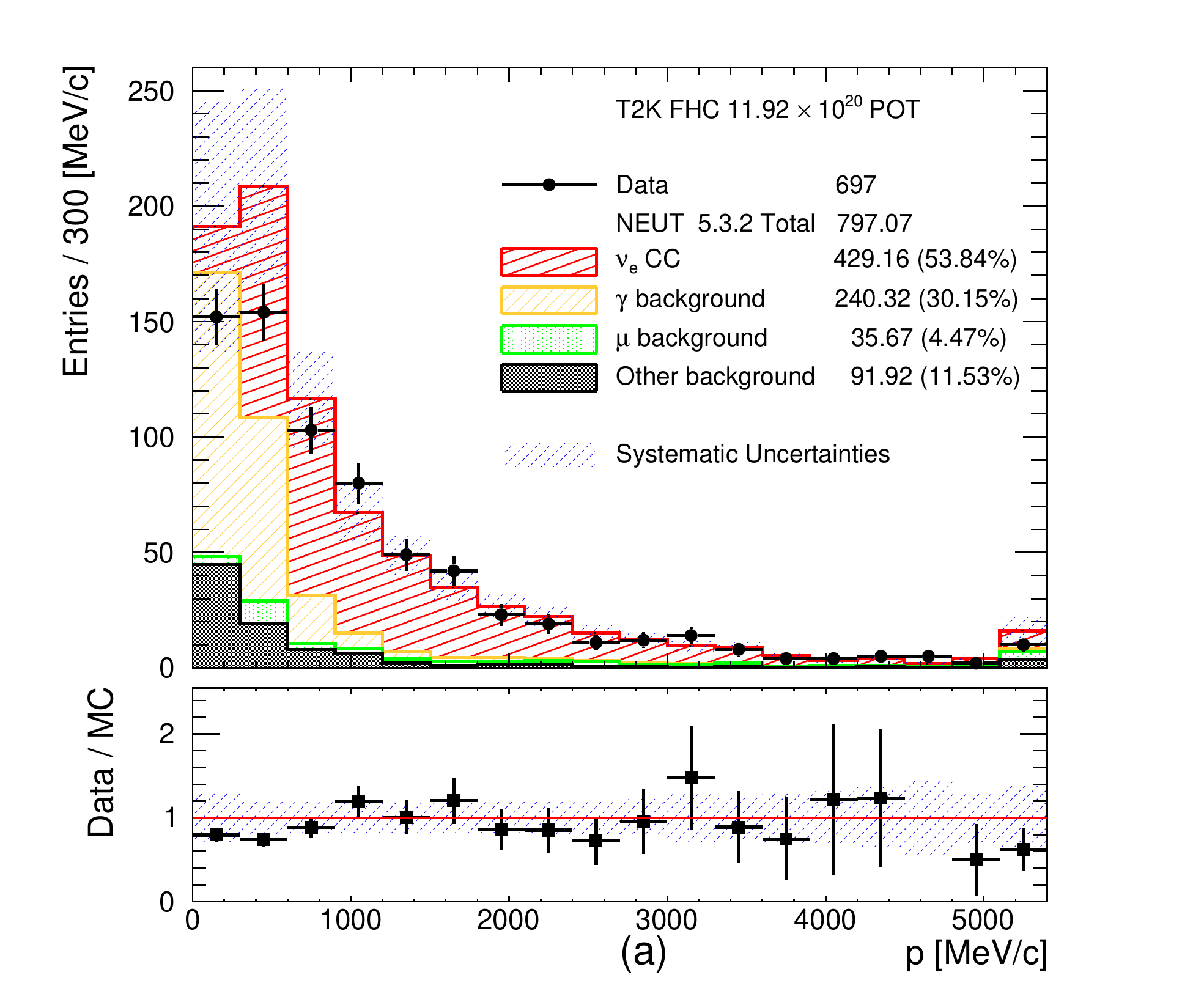}
	\includegraphics[width=0.495\textwidth]{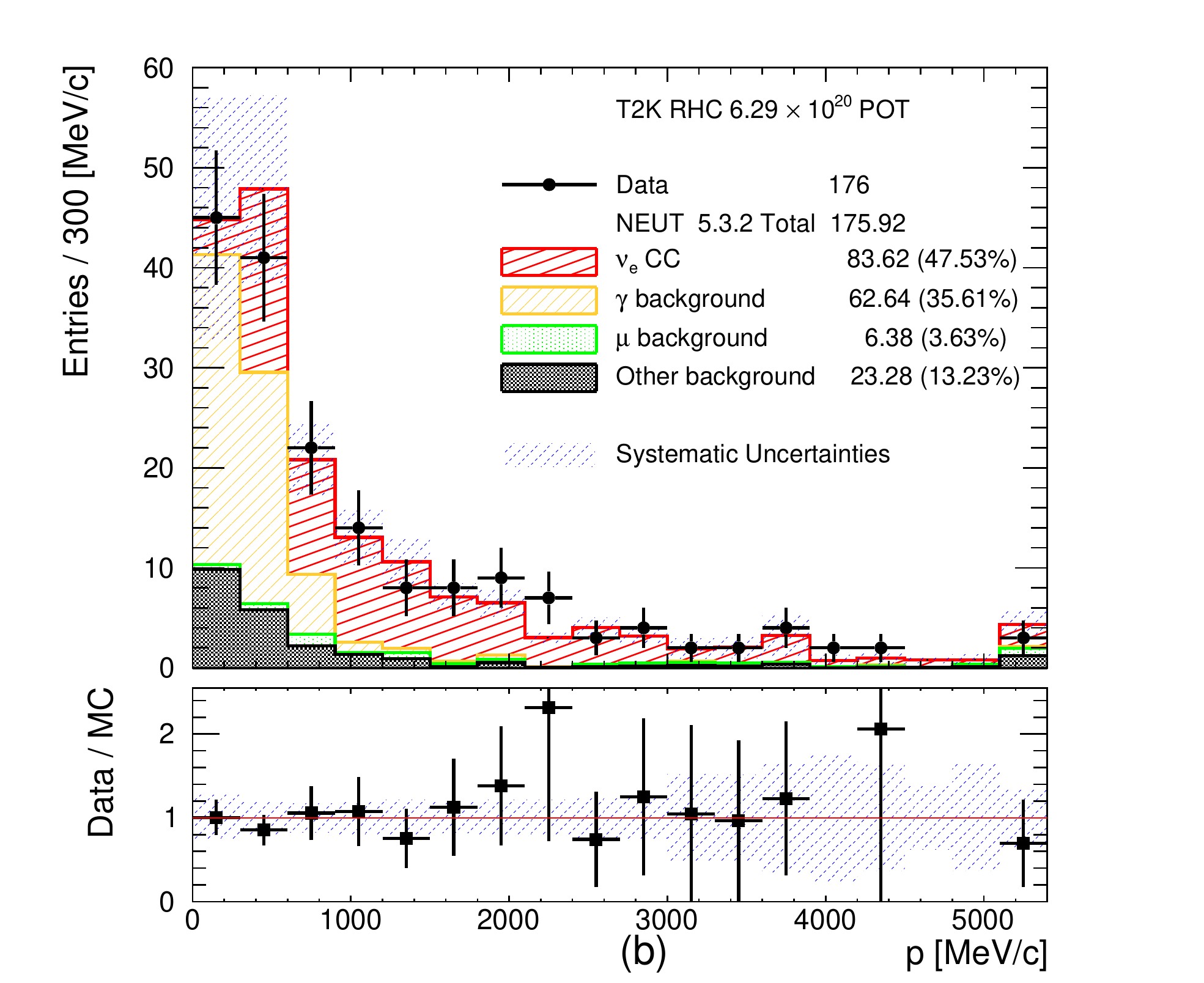}
	\includegraphics[width=0.495\textwidth]{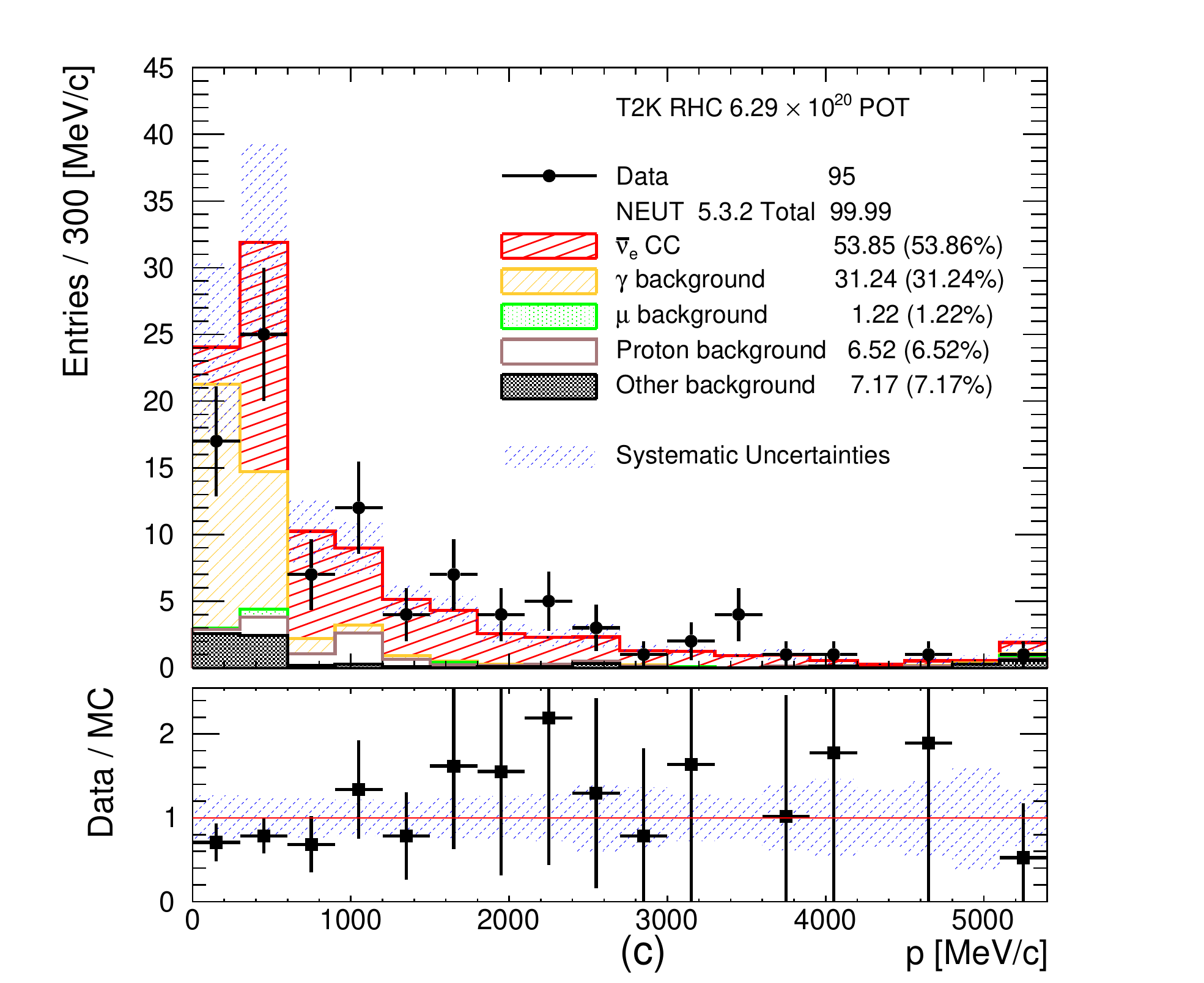}
	\caption{Momentum distribution of the selected electron and positron candidates for (a) FHC CC-$\nu_{e}$, (b) RHC CC-$\nu_{e}$ and (c) RHC CC-$\bar\nu_{e}$. The number of MC events is normalized to the data POT. The effect of the total systematic uncertainty on the MC event yields (see section~\ref{sec:systuncertainties} for details) is also shown on these plots. The last bin is the overflow bin.}
	\label{fig:CCNueMom}
\end{figure}

\begin{figure}[htbp]
 \centering
	\includegraphics[width=0.495\textwidth]{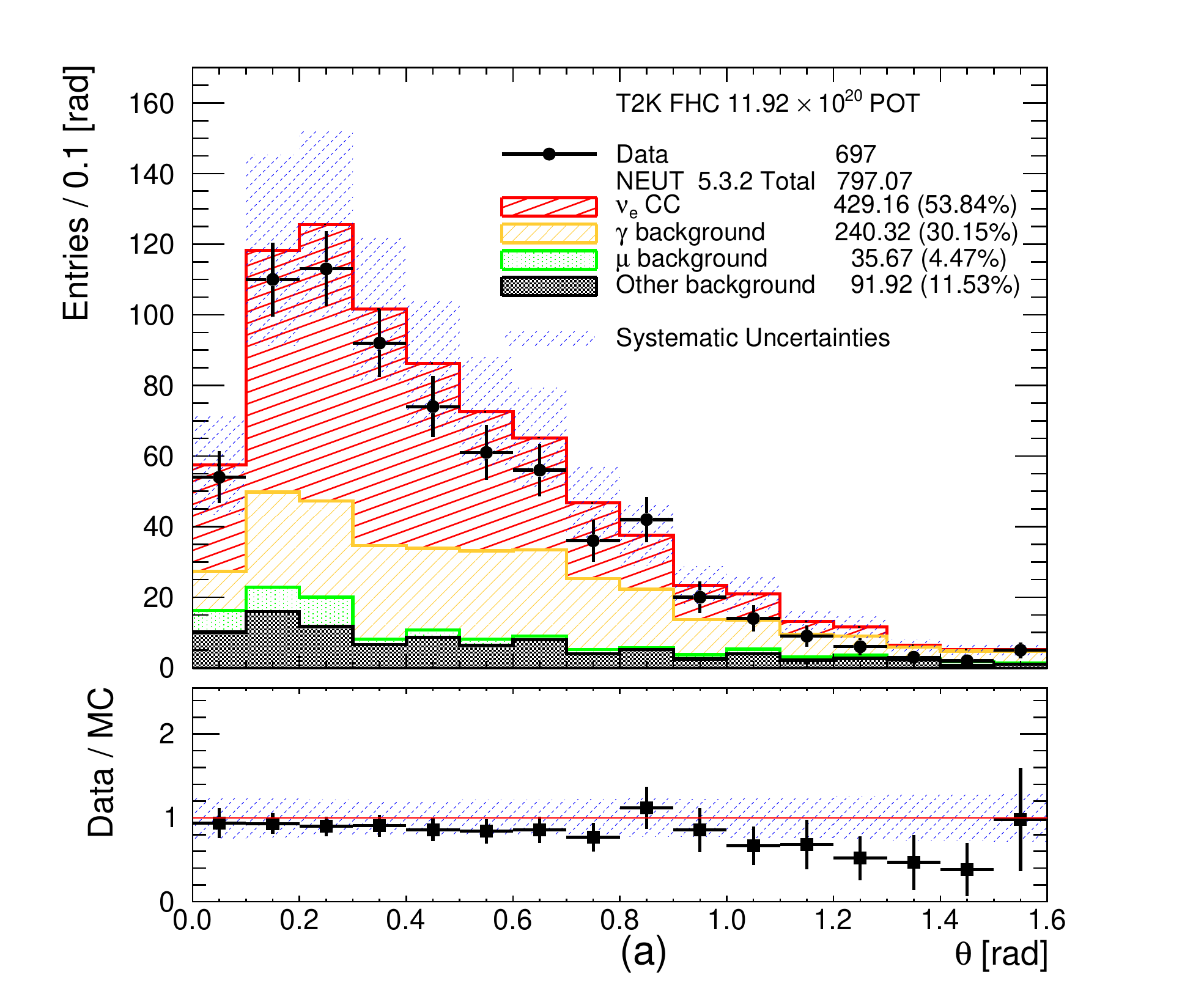}
	\includegraphics[width=0.495\textwidth]{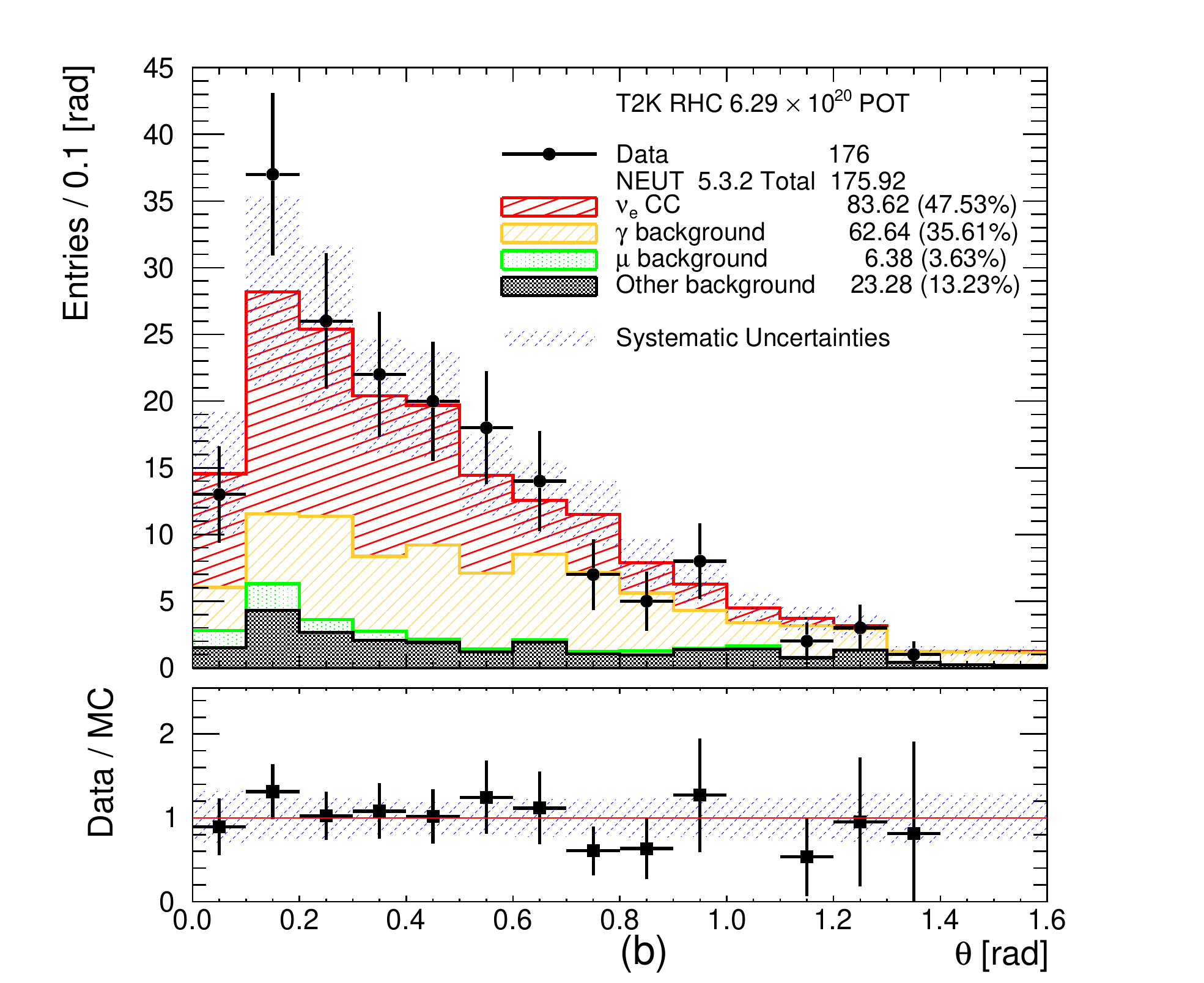}
	\includegraphics[width=0.495\textwidth]{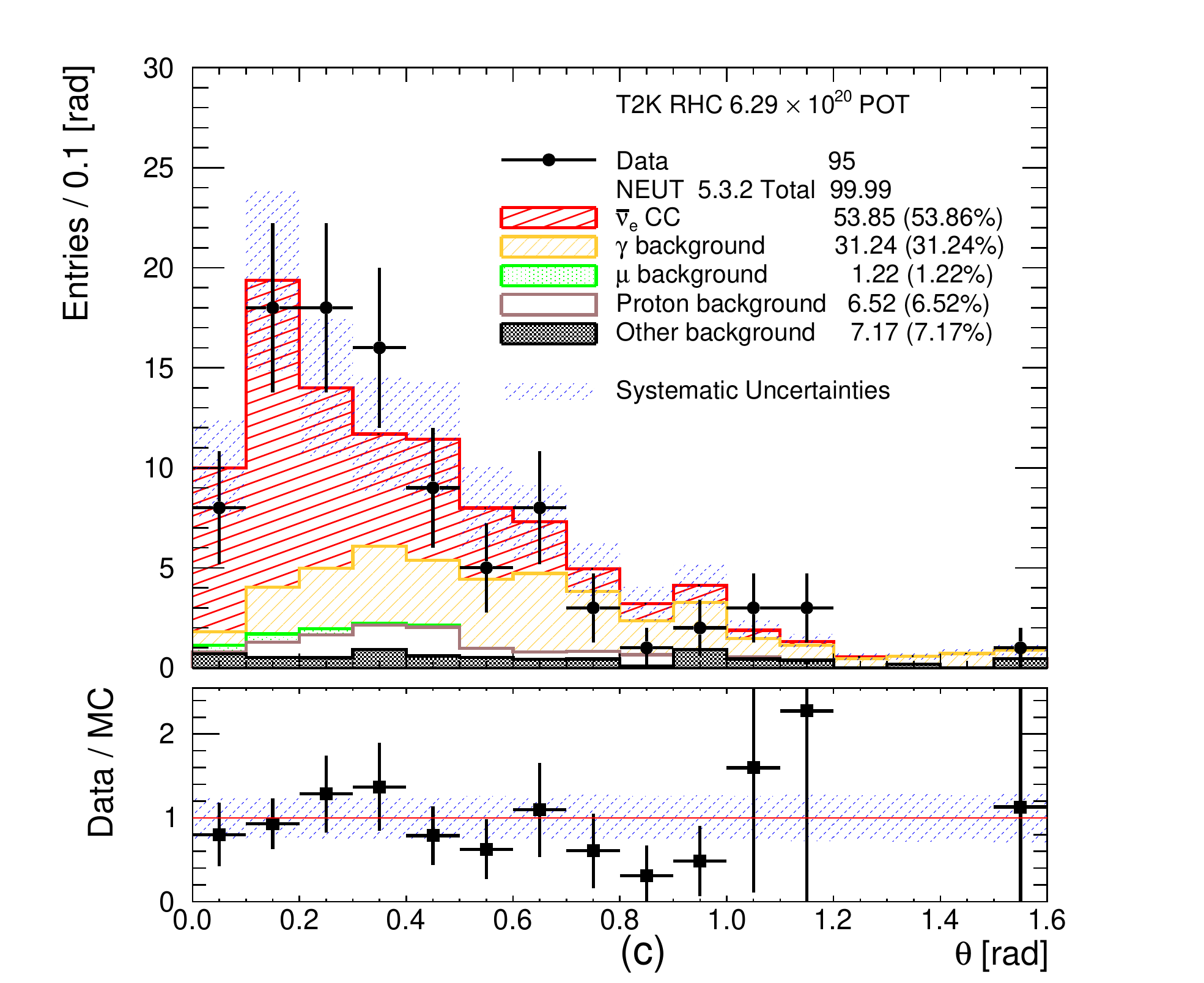}
	\caption{Angular distribution of the selected electron and positron candidates for (a) FHC CC-$\nu_{e}$, (b) RHC CC-$\nu_{e}$ and (c) RHC CC-$\bar\nu_{e}$. The number of MC events is normalized to the data POT. The effect of the total systematic uncertainty on the MC event yields (see section~\ref{sec:systuncertainties} for details) is also shown on these plots. The last bin includes all backward-going candidates.}
	\label{fig:CCNueAng}
\end{figure}

The momentum and angular (with respect to the neutrino direction) distributions of all selected CC-$\nu_{e}$ and CC-$\bar\nu_{e}$ candidates are shown in Figures~\ref{fig:CCNueMom} and~\ref{fig:CCNueAng}, respectively. These plots also show the total systematic uncertainty on the MC event yields, which is discussed in section~\ref{sec:systuncertainties}. A significant data deficit is observed at low momentum ($p < 600$~MeV/c) in the FHC CC-$\nu_{e}$ channel. In this region the photon background is dominant which has significant systematic uncertainties associated with the $\pi^{0}$ production. Roughly a third of the photon background comes from neutrino interactions outside of FGD1. These external backgrounds could originate from neutrino interactions on heavy targets, like iron, copper or lead with significant final state interaction systematic uncertainties. In addition, roughly another third of the photon background come from NC interactions which are poorly measured. The final third of the photon background come from CC-$\nu_{\mu}$ and CC-$\bar\nu_{\mu}$ interactions, usually when the muon is emitted in high angles and it is lost. In such occasions the most energetic of the other tracks is selected as the leading track. A similar data deficit is also observed in the statistically poorer RHC CC-$\bar\nu_{e}$ channel. In addition, an excess of events has been observed in the RHC channels at high momenta (more visible in the RHC CC-$\bar\nu_{e}$ channel). For the photon background produced from $\nu_{\mu}$ and $\bar\nu_{\mu}$ interactions in FGD1, roughly 10\% is coming from NC DIS interactions in all three selections. The relevant fraction of FGD1 CC DIS events entering the photon background is approximately 4\% in CC-$\bar\nu_{e}$ selection, 16\% in RHC CC-$\nu_{e}$ selection and 20\% in FHC CC-$\nu_{e}$ selection. The differences are due to the additional selection criteria applied to CC-$\bar\nu_{e}$ and the presence of protons which can be selected as the leading track instead of the primary muon or background positron.

\begin{figure}[htbp]
 \centering
	\includegraphics[width=0.495\textwidth]{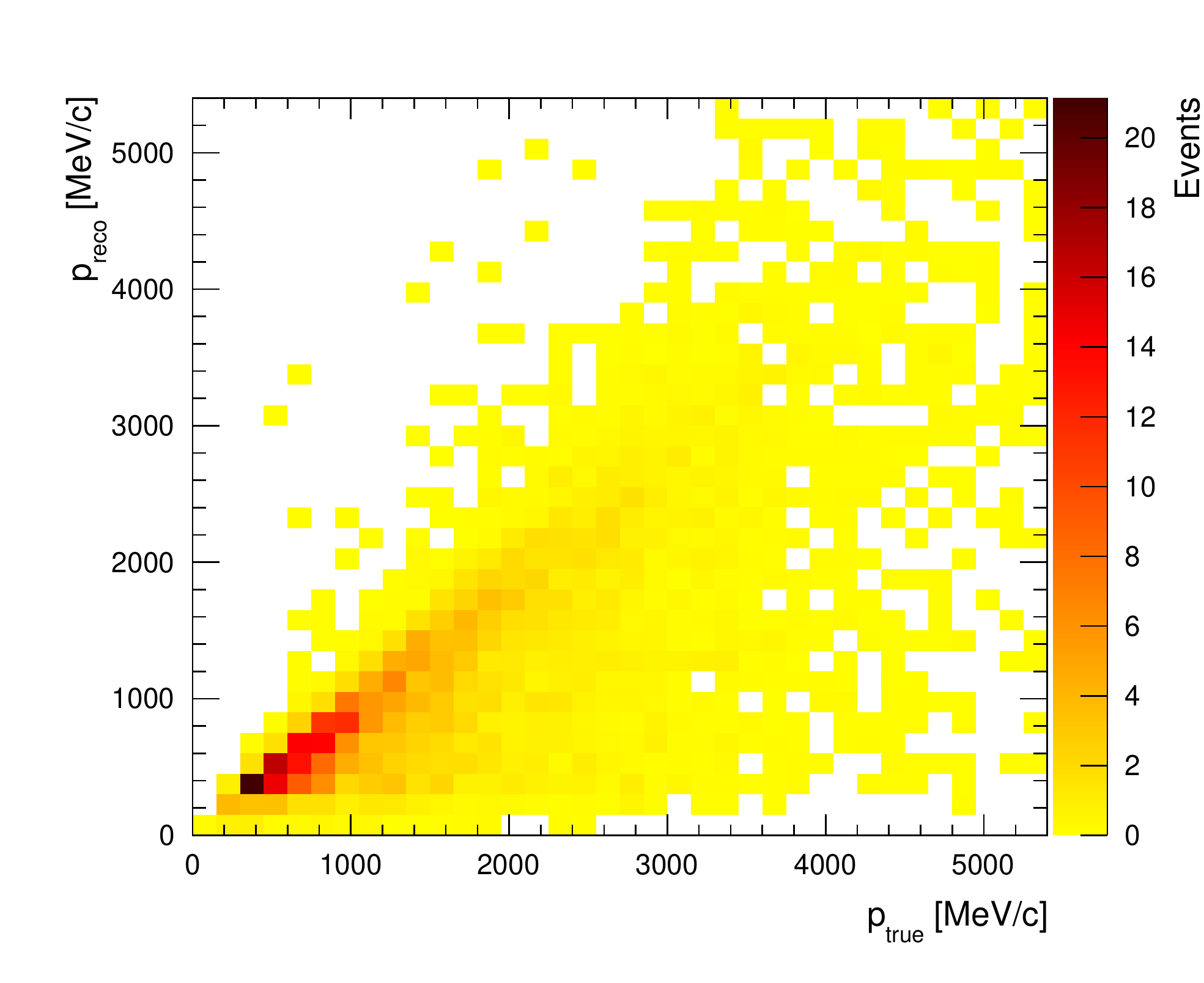}
	\includegraphics[width=0.495\textwidth]{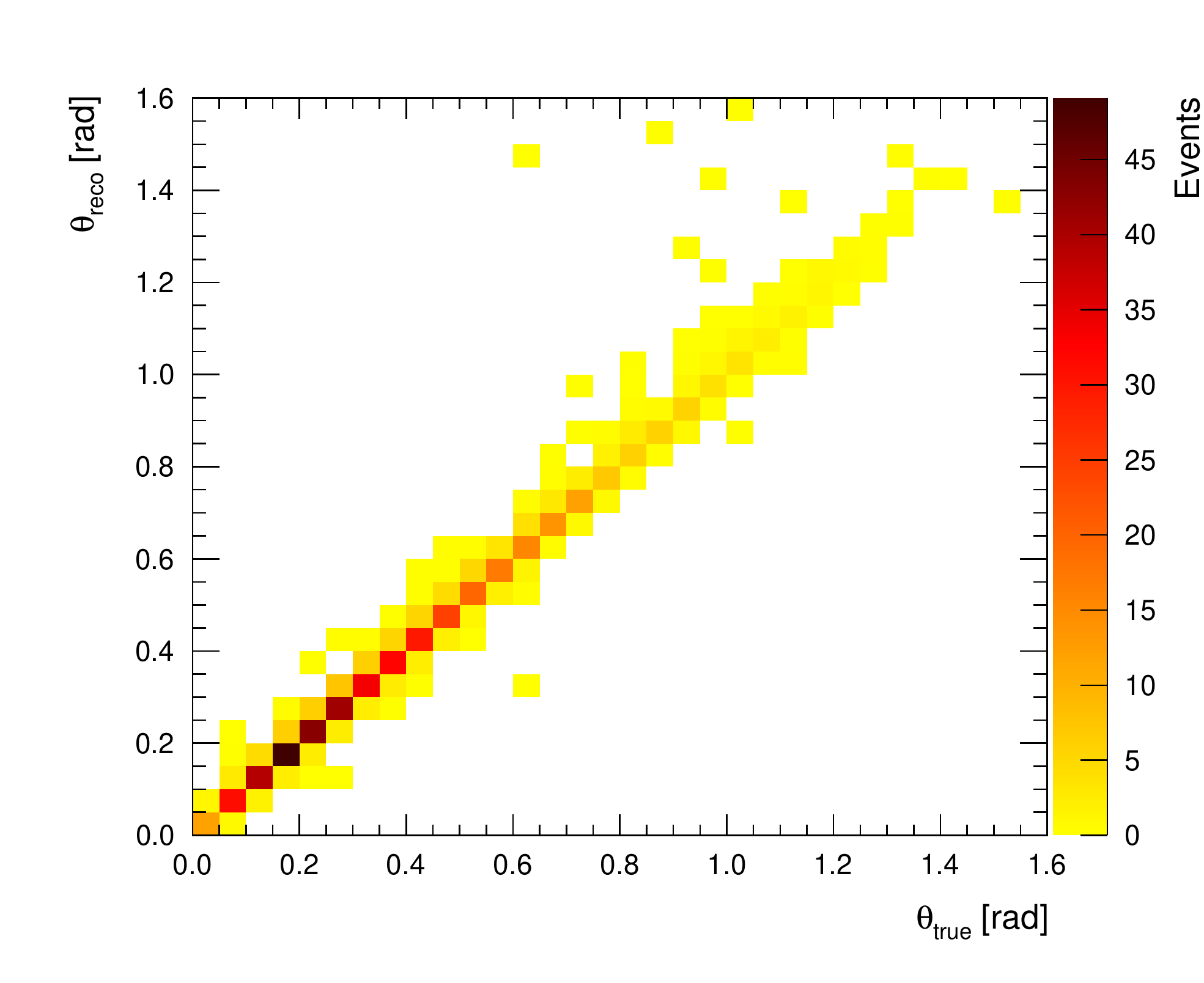}
	\caption{Distribution of the true vs reconstructed values of momentum (left) and angle (right) for signal electrons and positrons that passed all cuts in the MC. The effect of bremsstrahlung is visible on the left plot, see the text for details.}
	\label{fig:CCNueMomAngRes}
\end{figure}

Most of the efficiency loss is observed at low momentum since the electron and muon/pion $dE/dx$ energy loss curves cross around 150~MeV/c (see Figure~\ref{fig:tpcdedx}). In addition, high angle tracks that do not enter the TPC are not selected and the events are lost. Another important source of efficiency loss is due to electron shower or bremsstrahlung in FGD1. As a result the primary electron track does not enter the TPC or another track is selected as the leading track. As estimated from the MC, 35-45\% of the signal electrons or positrons are lost because the primary electron track does not enter the TPC. The efficiency loss is larger in the FHC CC-$\nu_{e}$ channel since the electron momentum spectrum is softer and at higher angles. The true vs reconstructed momentum and angular distributions in the MC for the selected signal electrons and positrons are shown in Figure~\ref{fig:CCNueMomAngRes}. The effect of bremsstrahlung is visible as the reconstructed momentum spectrum is biased towards lower momenta. A summary of efficiencies and purities is shown in Table~\ref{tab:neutgeniesumtable}.

The muon mis-identification probability (probability of a muon to be mis-identified as an electron after applying the PID) was studied in previous T2K publications~\cite{ND280NueSel} with very good agreement between data and MC. Similarly, the proton mis-identification probability is important for the CC-$\bar\nu_{e}$ selection. A high-purity independent sample of protons has the same PID criteria as the CC-$\bar\nu_{e}$ selection applied and the number of protons that survive is checked. An independent control sample that can be used is the FHC CC-$\bar\nu_{e}$ selection. This channel has a tiny signal contribution and a much larger proton background and it is not used in the cross-section measurements. Before applying the proton rejection cuts (viii) and (ix), approximately 94\% of the leading tracks selected with $p > 600$~MeV/c and not entering the ECal are protons. The measured proton mis-identification probability is the fraction of protons that survive from these independent proton enriched samples and is $\left(4.6\pm0.8\right)$\% for the data compared to $\left(5.0\pm0.3\right)$\% in the MC. The errors are statistical only. The proton purity is lower in the case where the leading track enters the ECal and is approximately 70\% with the rest to be mostly positrons. Due to the relatively low proton purity of this sample, only an approximate proton mis-identification probability can be measured in this case, $\left(9.4\pm0.1\right)$\% in the data compared to $\left(11.9\pm0.05\right)$\% in the MC.    


\subsection{Event selection using alternative MC}
\label{subsec:neutgeniecomp}

The CC-$\nu_{e}$ and CC-$\bar\nu_{e}$ selections in the MC are repeated using GENIE (2.8.0) instead of NEUT (5.3.2) MC. There are some differences between these two neutrino generators, see section~\ref{sec:analysissamples} and for more details the description in~\cite{ND280numucc4pi}. One of the most important is that the neutrino multi-nucleon interaction simulations are turned-off in this version of GENIE. Efficiencies and purities for NEUT and GENIE agree quite well. Compared to the selected events, both NEUT and GENIE predictions disagree with data at low momenta with the FHC CC-$\nu_{e}$. The prediction of the photon background in particular is similar in both neutrino generators. Tables~\ref{tab:neutgeniesumtable} and~\ref{tab:genieneutEventRate} summarize the event selections using NEUT and GENIE. 

\begin{table}[!ht]
 \caption{Summary of efficiency, purity and number of MC events normalised to the $11.92\times10^{20}$~POT in the FHC beam and $6.29\times10^{20}$~POT in the RHC beam for the CC-$\nu_{e}$ and CC-$\bar\nu_{e}$ channels using NEUT (5.3.2) and GENIE (2.8.0) MC, in addition to the number of data events that survive all cuts in each channel.}
 \begin{center}
 \begin{tabular}{l|c|c|c|c}
 \hline
 \hline
 Channel                  & Efficiency & Purity & MC Events & Data events \\
 \hline
 NEUT  FHC CC-$\nu_{e}$     & 0.26       & 0.54   & 797.07    & 697         \\
 GENIE FHC CC-$\nu_{e}$     & 0.27       & 0.53   & 769.17    & 697         \\
 \hline
 NEUT  RHC CC-$\nu_{e}$     & 0.33       & 0.48   & 175.92    & 176         \\
 GENIE RHC CC-$\nu_{e}$     & 0.33       & 0.44   & 168.10    & 176         \\
 \hline
 NEUT  RHC CC-$\bar\nu_{e}$ & 0.31       & 0.54   & 99.99     & 95          \\
 GENIE RHC CC-$\bar\nu_{e}$ & 0.30       & 0.51   & 99.21     & 95          \\	
 \hline
 \hline
 
\end{tabular}
\end{center}
       
\label{tab:neutgeniesumtable}
            
\end{table}
\begin{table}[!ht]
 \caption{Breakdown of the number of CC-$\nu_{e}$ and CC-$\bar\nu_{e}$ events selected in FGD1 according to their category for NEUT (5.3.2) and GENIE (2.8.0) MC. The number of events is normalized to data POT. The photon background is separated to events with a true vertex in FGD1 (In-FGD) and to events with a true vertex out of FGD1 (OO-FGD).}
 \begin{center}
 \resizebox{\textwidth}{!}{
 \begin{tabular}{l|c|c|c|c|c|c}
 \hline
 \hline
 Channel               & Signal (\%)     & In-FGD $\gamma$ (\%) & OO-FGD $\gamma$ (\%) & $\mu^{\pm}$ (\%)  & Proton (\%) & Other (\%) \\
 \hline
 NEUT  FHC  CC-$\nu_{e}$ & 429.16 (53.9)   & 162.23 (20.4)        & 78.09 (9.8)          & 35.67 (4.5) &  -    & 91.92 (11.4)   \\
 GENIE FHC  CC$-\nu_{e}$ & 409.23 (53.5)   & 152.56 (20.0)        & 78.00 (10.2)         & 33.29 (4.4) &  -    & 96.10 (12.0)   \\
 \hline
 NEUT  RHC  CC-$\nu_{e}$ & 83.62 (47.5)    & 42.41 (24.1)         & 20.23 (11.5)         & 6.38 (3.6)  &  -    & 23.28 (13.2)   \\
 GENIE RHC  CC-$\nu_{e}$ & 73.28 (43.6)    & 43.46 (25.9)         & 21.67 (12.9)         & 6.33 (3.8)  &  -    & 23.35 (13.9)   \\
 \hline
 NEUT  RHC  CC-$\bar\nu_{e}$ & 53.85 (53.9) & 18.76 (18.8)        & 12.47 (12.5)         & 1.22 (1.2) &  6.52 (6.5)   & 7.17 (7.2)   \\
 GENIE RHC  CC-$\bar\nu_{e}$ & 50.49 (51.2) & 21.28 (21.5)        & 11.43 (11.5)         & 1.74 (1.7) &  7.20 (7.3)   & 7.07 (7.1)   \\
\hline
\hline
 
\end{tabular}}
\end{center}
       
\label{tab:genieneutEventRate}
            
\end{table}

\section{Photon background control samples}
\label{sec:gammaselection}

Since the photon background is the most important in the electron (anti-)neutrino selections, a dedicated photon control sample of electrons and positrons from photon conversions is selected to constrain this background. Photon candidates are selected from two nearby electron-like FGD1-TPC tracks of opposite charge with low invariant mass that start in the FGD1 fiducial volume. 

\subsection{Selection of photon candidates}
\label{subsec:gammaselection}

The steps to select photon candidates are:

\begin{enumerate}[(i)]

\item Only events during periods of good beam and detector quality are used. The event time has to be reconstructed within one of the eight distinct beam bunches.
\item The highest momentum negatively charged or highest momentum positively charged FGD1-TPC track (leading track) with a vertex in the FGD1 fiducial volume is selected.
\item The leading track must be compatible with the electron TPC $dE/dx$ hypothesis. If the leading track enters the ECal, and has momentum $p > 1$~GeV/c and ECal energy E, then $E/p > 0.5$ is required in order to clean up the high momentum tail.  
\item Require a second track with opposite charge to the leading track, also compatible with the electron TPC $dE/dx$ hypothesis and with a starting position within 5~cm from the primary track.
\item The invariant mass calculated from the leading and paired tracks must be less than 55~$\rm MeV/c^{2}$. The distributions of the invariant mass of the selected $e^{-}e^{+}$ pairs are shown in Figure~\ref{fig:gammaIM}. The invariant mass cut is very effective to remove backgrounds from misidentified muons, protons and electrons from CC-$\nu_{e}$ interactions.

\begin{figure}[htbp]
	\centering
	\includegraphics[width=0.495\textwidth]{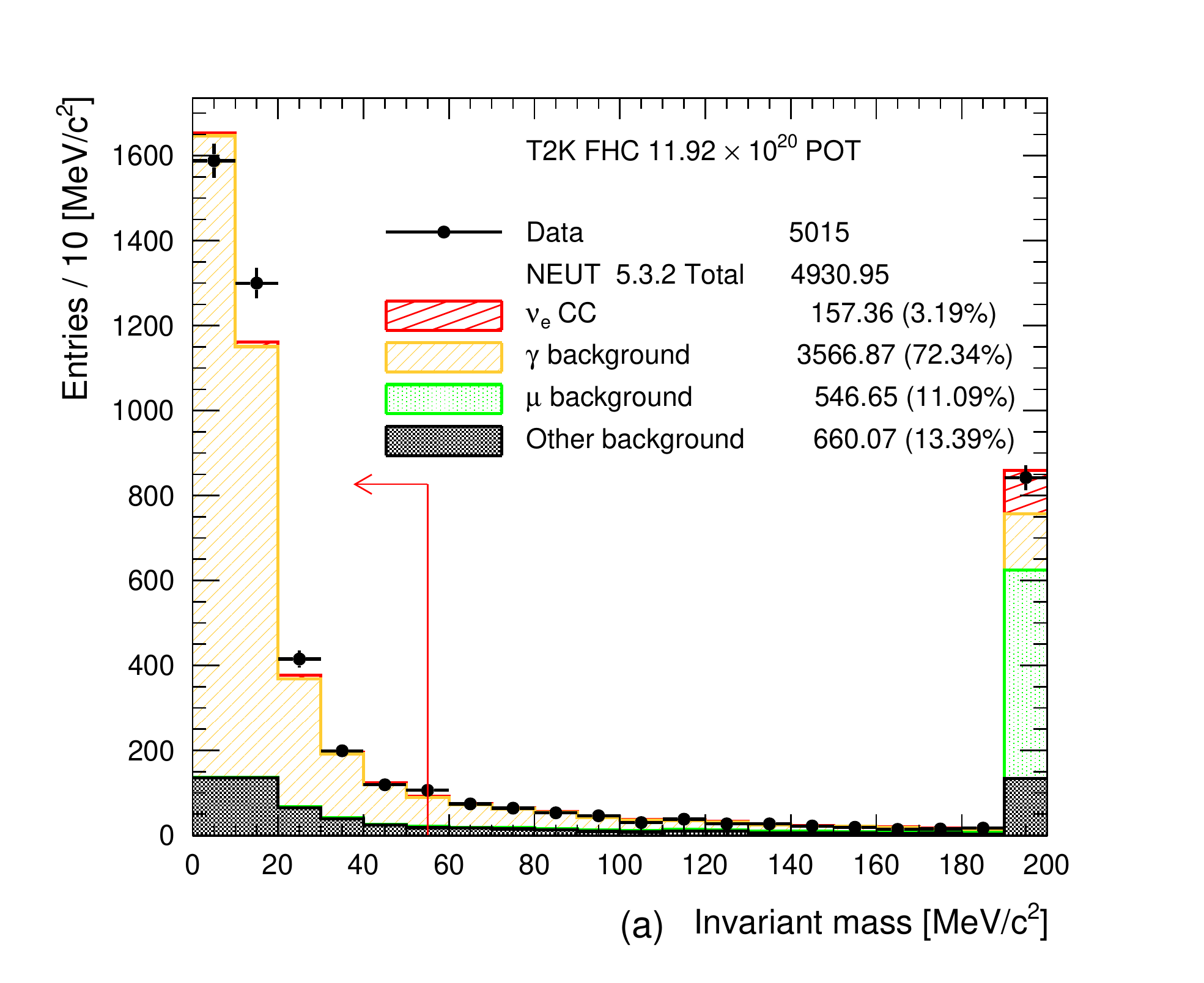}
	\includegraphics[width=0.495\textwidth]{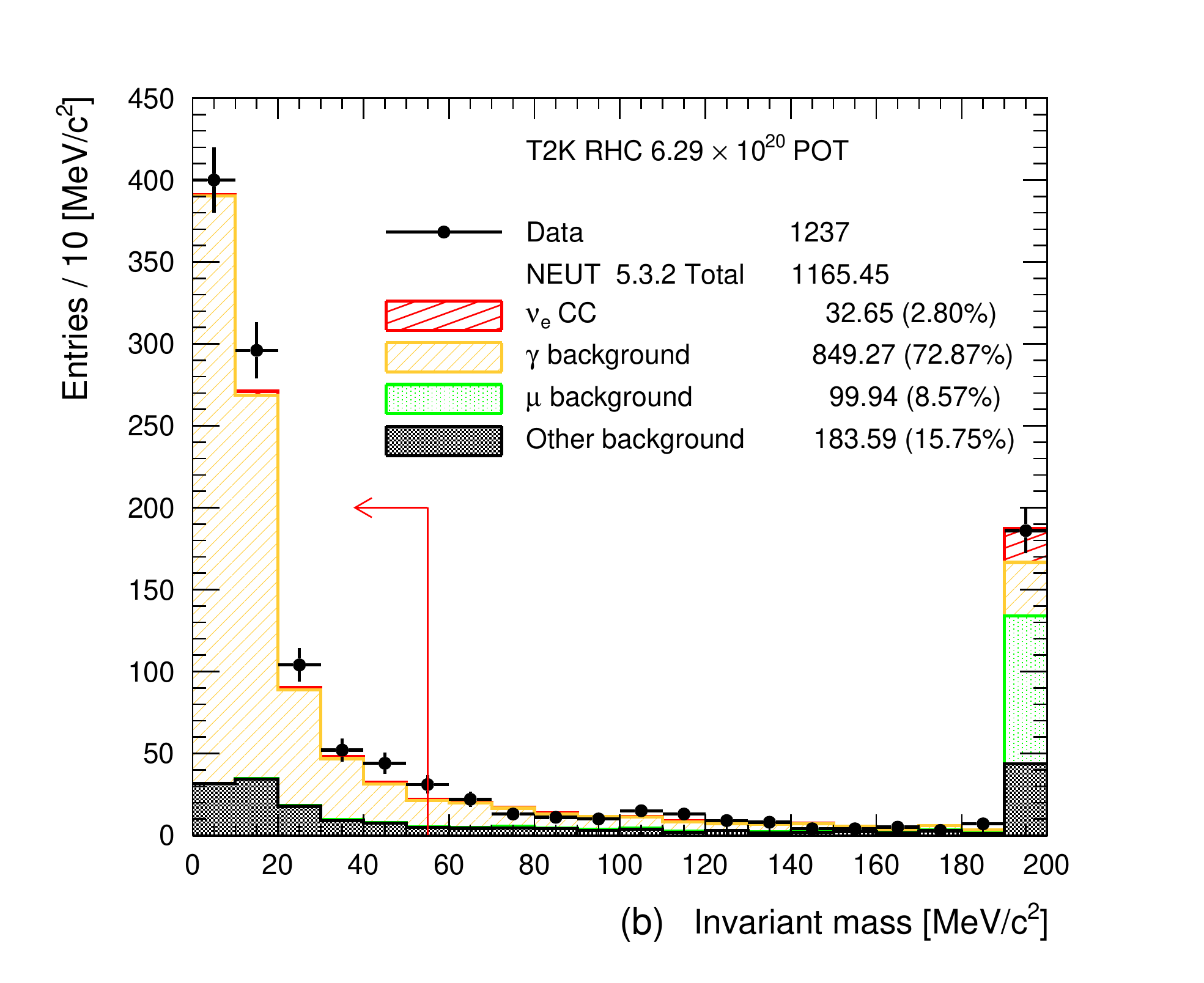}
	\includegraphics[width=0.495\textwidth]{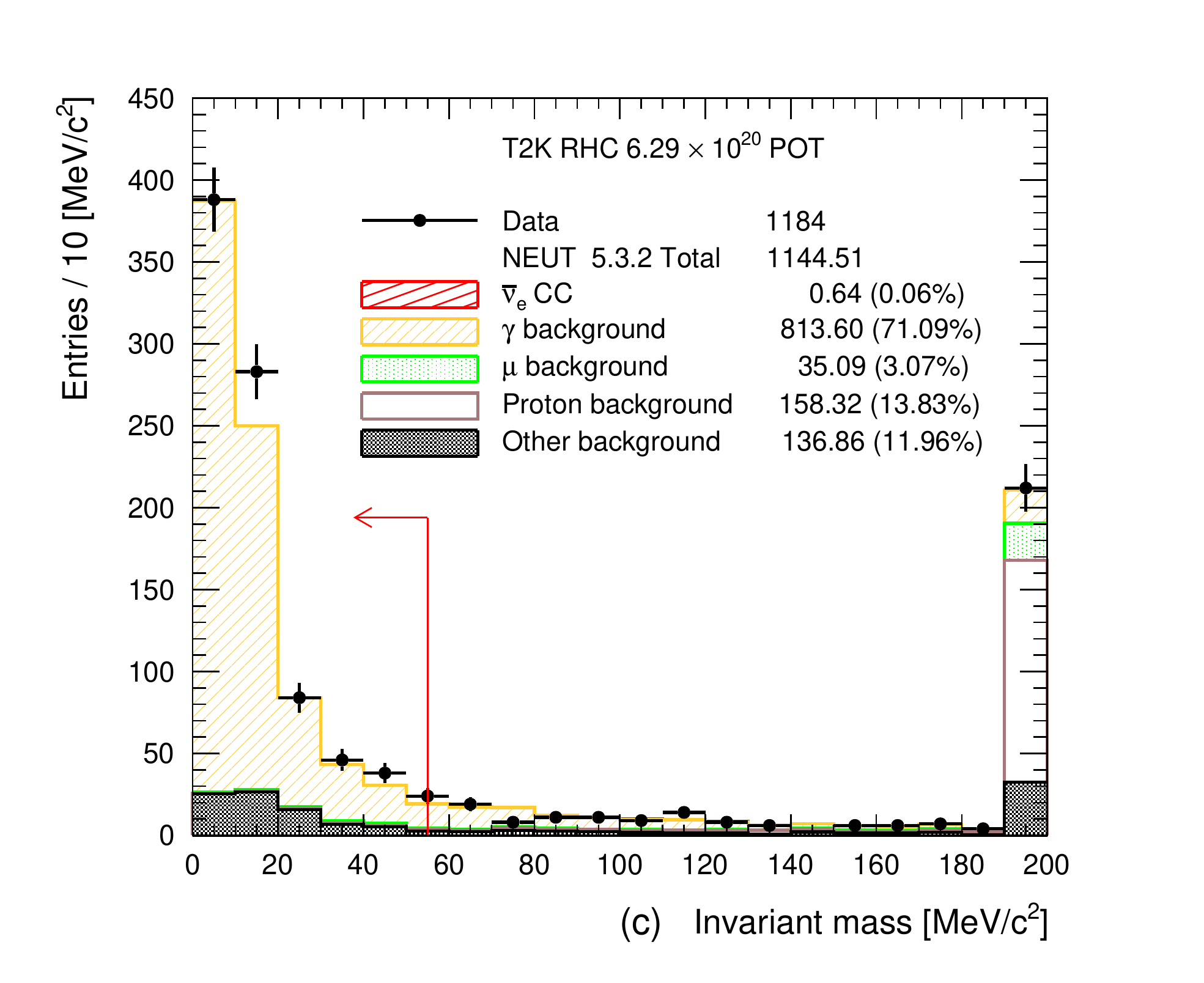}
	\caption{Invariant mass of electron-like FGD1-TPC pairs with opposite charge for (a) FHC selecting electron as the leading track, (b) RHC selecting electron as the leading track and (c) RHC selecting positron as the leading track. The number of MC events is normalized to the data POT. Last bin is the overflow bin. The arrow at 55~MeV/$c^2$ indicates the final photon to $e^{-}e^{+}$ conversion cut.}
	\label{fig:gammaIM}
\end{figure}

\item Although the photon selection at this stage is very pure, it is contaminated by external photons (photons from neutrino interactions outside FGD1). To remove external photons the same veto cuts used in the CC-$\nu_{e}$ and CC-$\bar\nu_{e}$ selections are applied.

\end{enumerate}

\begin{figure}[htbp]
	\centering
	\includegraphics[width=0.495\textwidth]{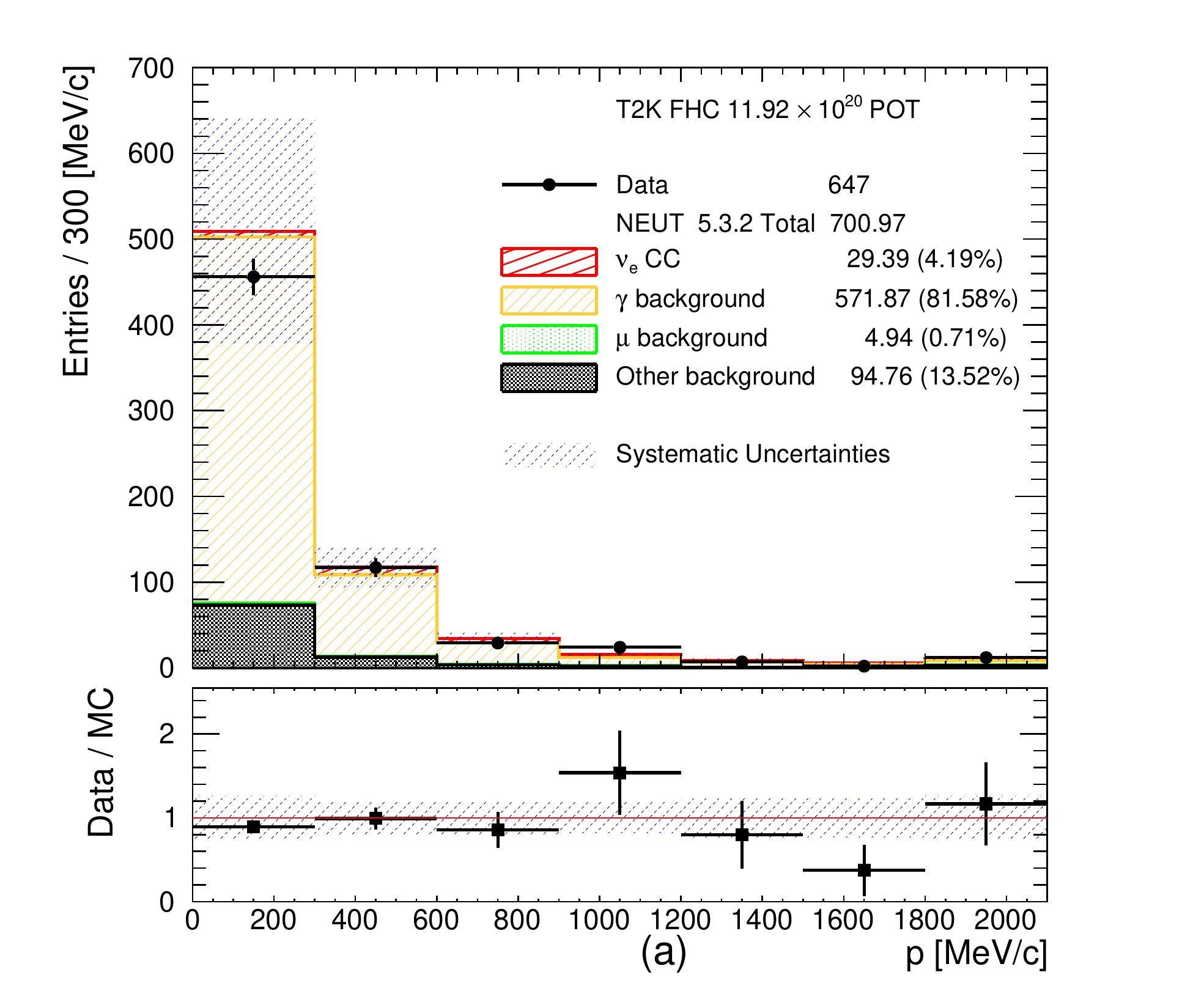}
	\includegraphics[width=0.495\textwidth]{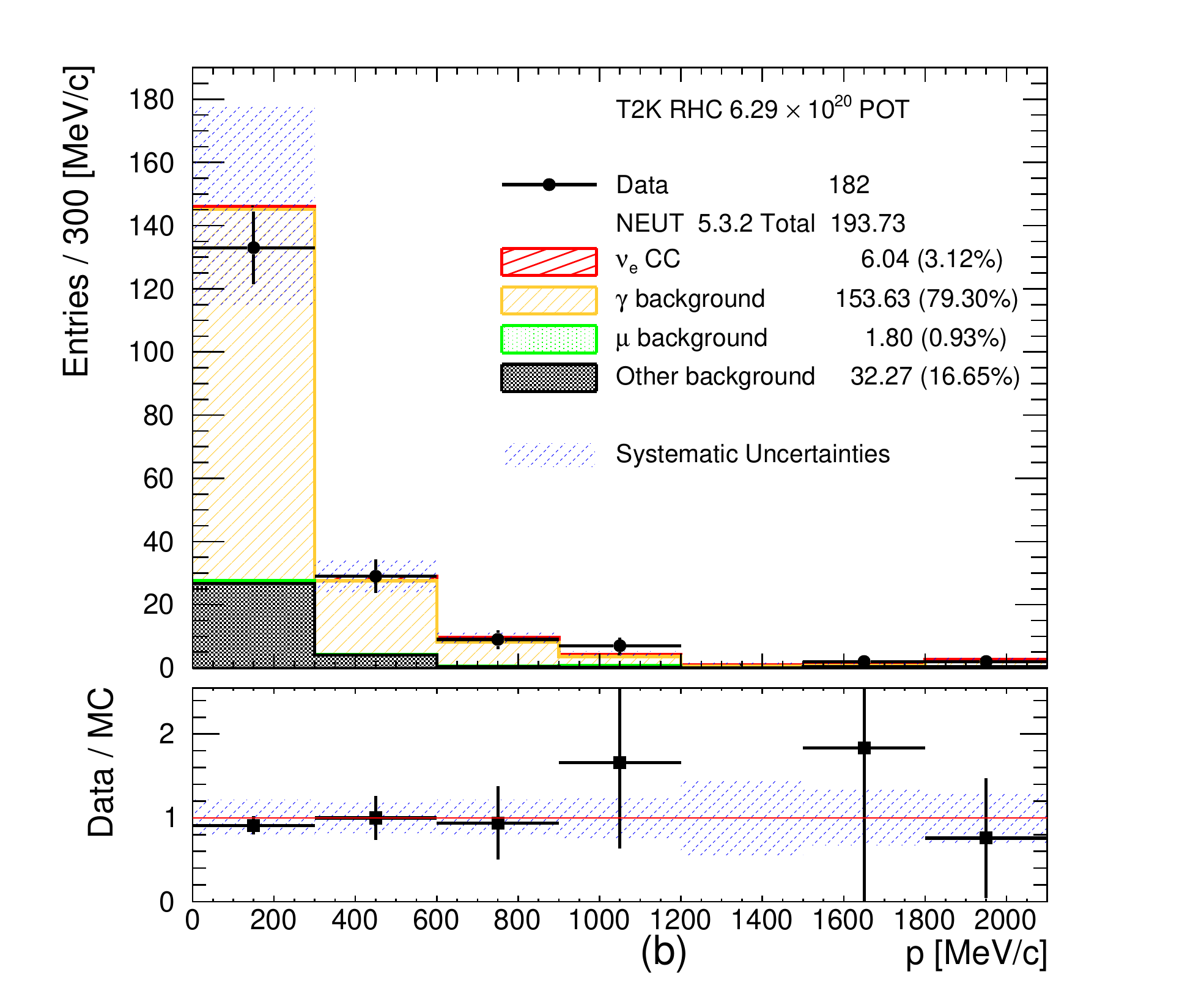}
	\includegraphics[width=0.495\textwidth]{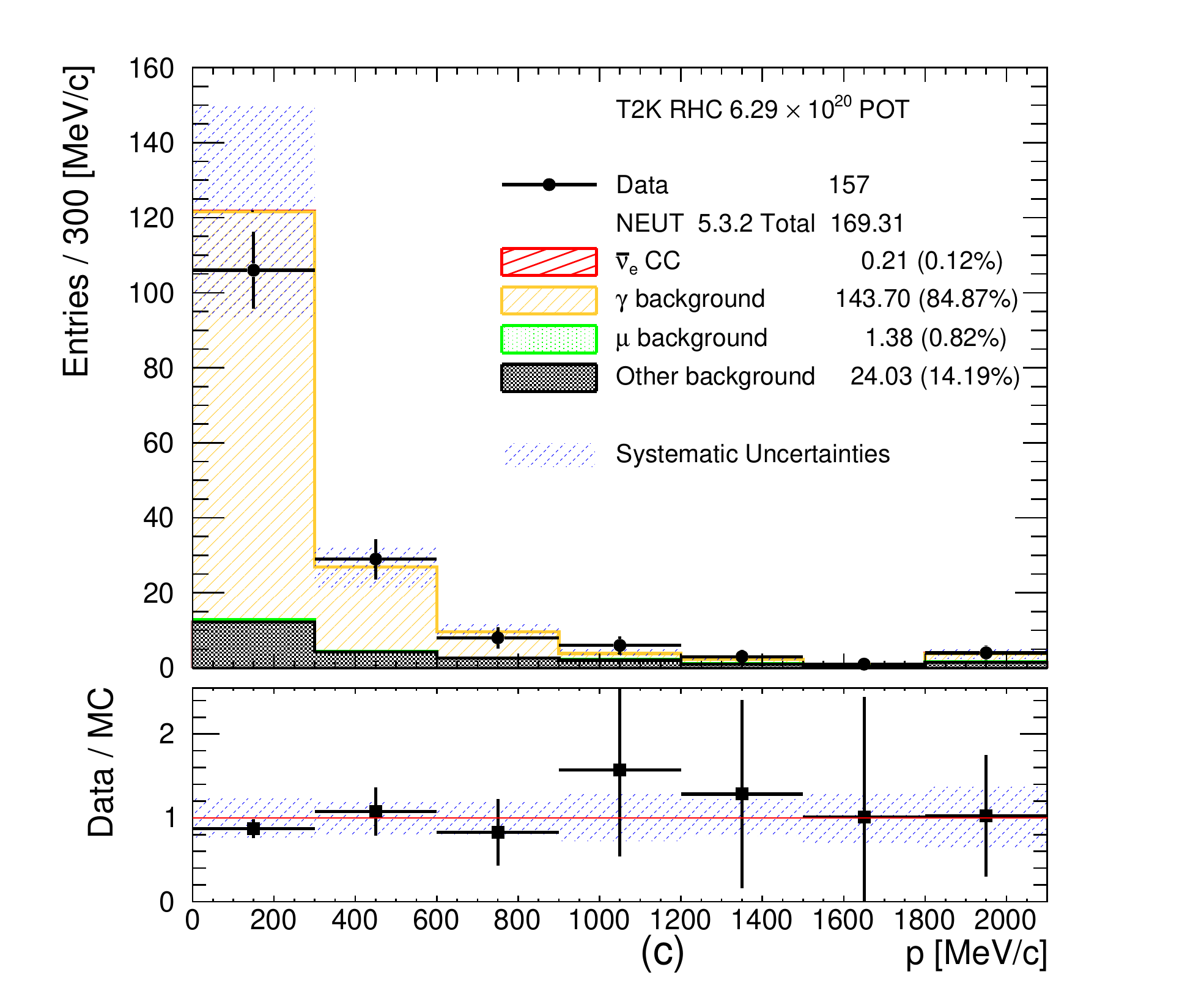}
	\caption{Momentum distribution of the selected photon candidates for (a) FHC selecting electron as the leading track, (b) RHC selecting electron as the leading track and (c) RHC selecting positron as the leading track. The number of MC events is normalized to the data POT. The effect of the total systematic uncertainty on the MC event yields (see section~\ref{sec:systuncertainties} for details) is also shown on these plots. Last bin is the overflow bin.}
	\label{fig:gammaSidebandMom}
\end{figure}

\begin{figure}[htbp]
	\centering
	\includegraphics[width=0.495\textwidth]{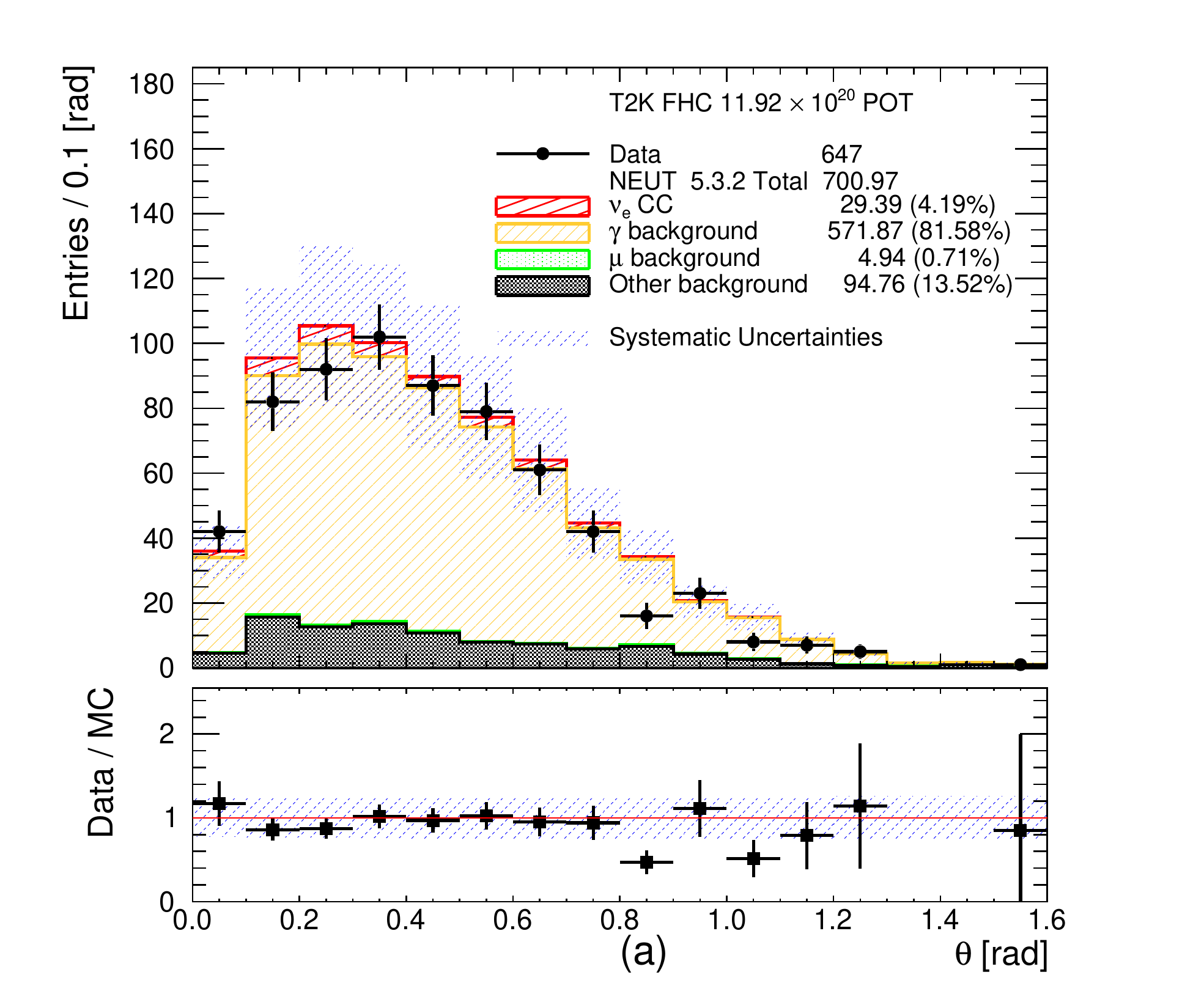}
	\includegraphics[width=0.495\textwidth]{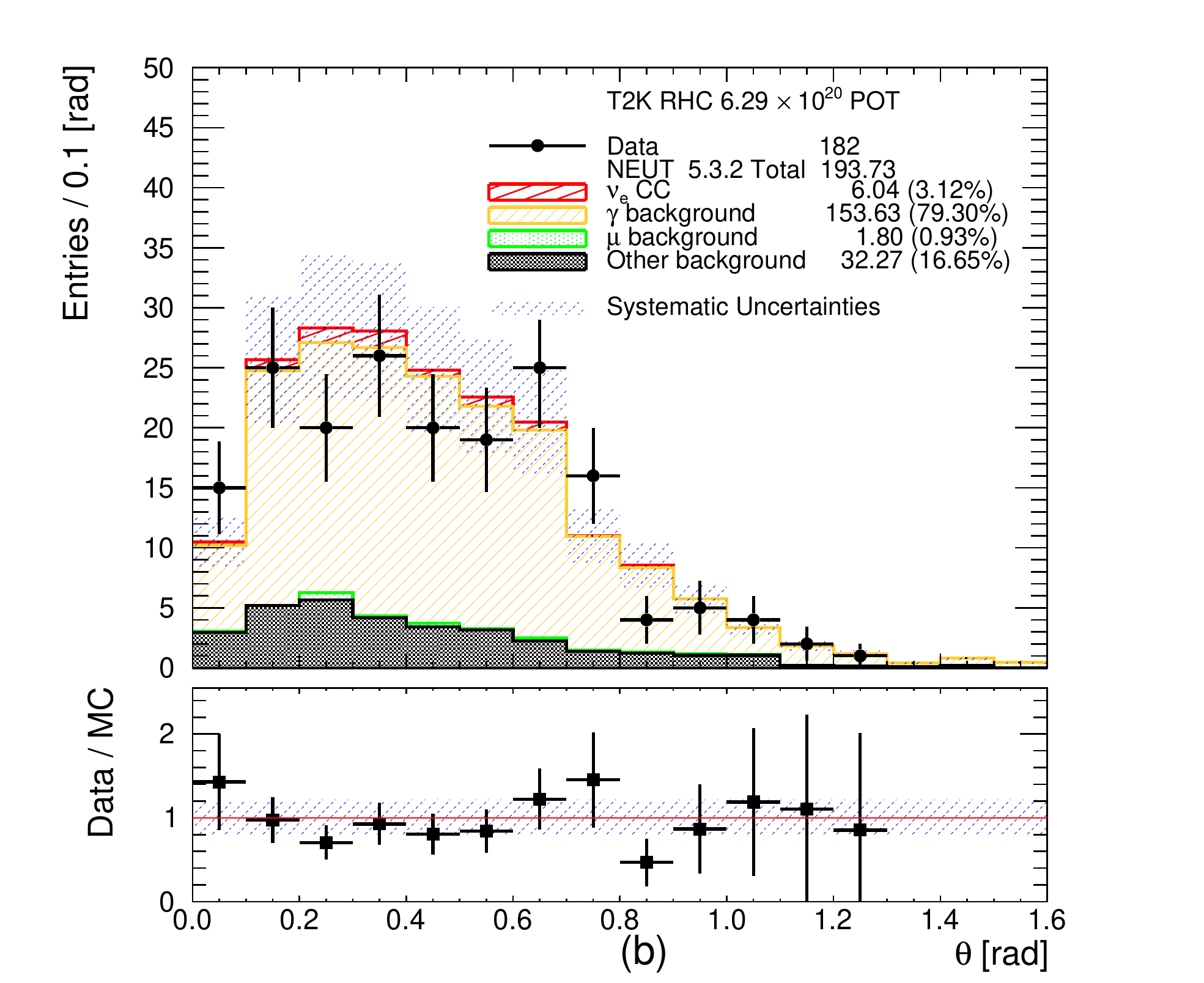}
	\includegraphics[width=0.495\textwidth]{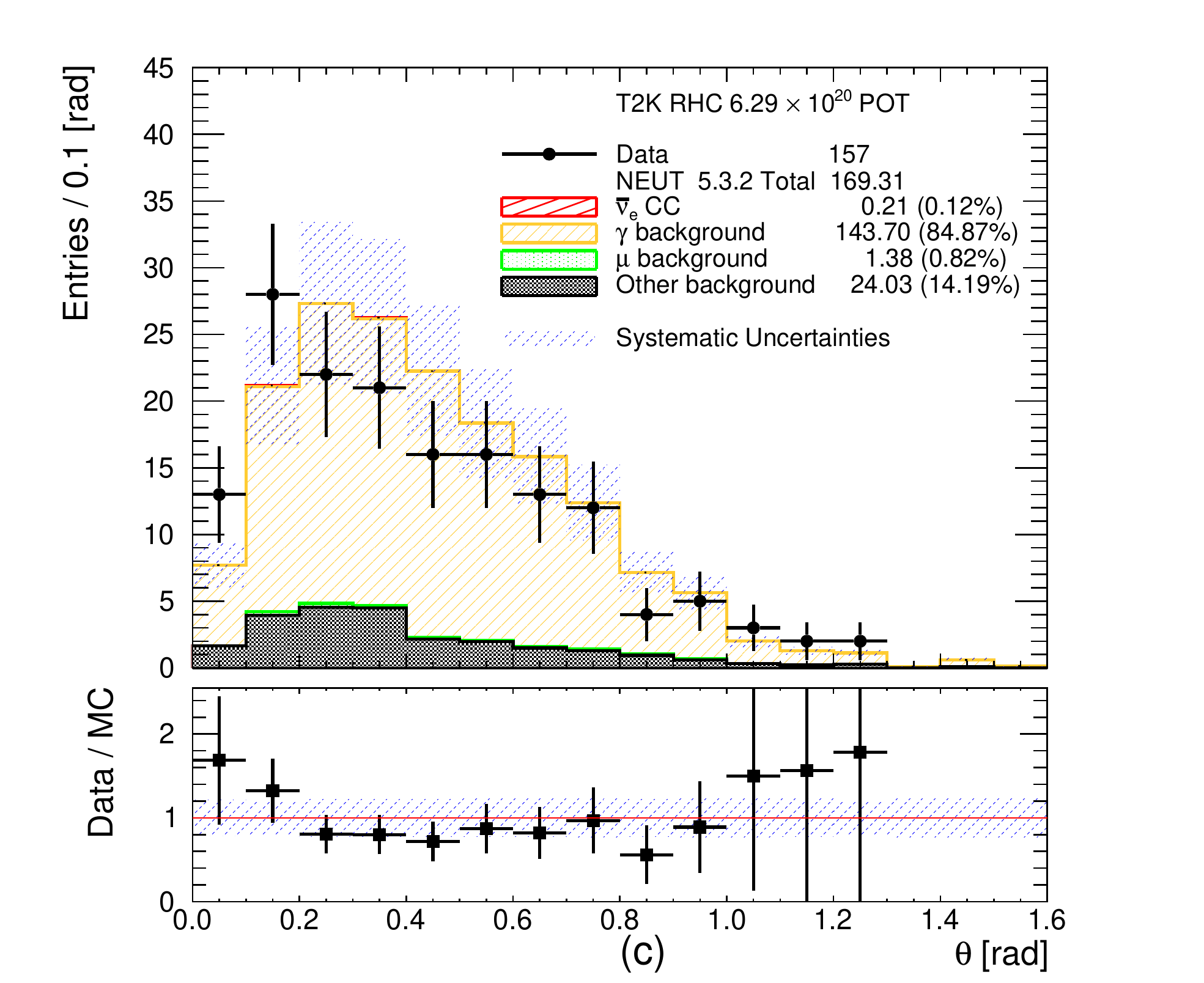}
	\caption{Angular distribution of the selected photon candidates for (a) FHC selecting electron as the leading track, (b) RHC selecting electron as the leading track and (c) RHC selecting positron as the leading track. The number of MC events is normalized to the data POT. The effect of the total systematic uncertainty on the MC event yields (see section~\ref{sec:systuncertainties} for details) is also shown on these plots. The last bin includes all backward-going candidates.}
	\label{fig:gammaSidebandAng}
\end{figure}

The signal and background categories are the same as for the CC-$\nu_{e}$ and CC-$\bar\nu_{e}$ selections. The momentum and angular distributions of the selected photon candidates are shown in Figures~\ref{fig:gammaSidebandMom} and~\ref{fig:gammaSidebandAng}, respectively. The systematic uncertainties on the MC event yields are also shown in these plots, see section~\ref{sec:systuncertainties} for details. A MC excess below 300~MeV/c is visible. In the angular distributions a significant MC excess is observed at high angles in the FHC CC-$\nu_{e}$ selection but not in the photon control selection (Figures~\ref{fig:CCNueAng} and~\ref{fig:gammaSidebandAng}).

The purity of the photon control samples is approximately 80\% when selecting electrons and 85\% when selecting positrons. A significant fraction of the selected photon candidates is classified in the other background category where the photons are coming from a true conversion point outside the FGD1 fiducial volume, but are mis-reconstructed inside of it. Including these events in the photon category definition increases the purity to approximately 90\%. The rest of the other background contributes (5 - 6)\% in the photon control samples and comes from $\pi^{0}$ Dalitz decay, general mis-reconstructions like broken tracks and accidental matching when at least one of the two tracks selected in the pair is not electron or positron. The signal leakage (CC-$\nu_{e}$ or CC-$\bar\nu_{e}$) in the photon control samples is around (3 - 4)\% when the selected leading track is an electron. The leakage is otherwise negligible when the selected leading track is a positron. The muon background entering the photon control samples is less than 1\% in all of the cases.

When selecting electrons as the leading track in the photon control selections, approximately 40\% of the photon candidates come from external photons, approximately 30\% come from NC interactions in FGD1 and the other 30\% come from CC interactions in FGD1. When selecting positrons as the leading track in the photon control selections the contributions are slightly different. Approximately 45\% of the photon candidates come from external photons, approximately 35\% come from NC interactions in FGD1 and 20\% come from CC interactions in FGD1.  
Often the event is rejected if the selected highest positively charged momentum track is a proton. However, since the protons are invisible when selecting negatively charged tracks the same event could be selected when searching for the highest momentum negatively charged track. This explains the difference in the number of photon candidates in RHC when the leading track selected is the electron or the positron.


\subsection{Comparisons with the photon background in the standard selections}
\label{subsec:gammacomparison}

Although the photon control samples are of high purity they have some differences compared to the photon background entering the CC-$\nu_{e}$ and CC-$\bar\nu_{e}$ selections. The main reason is that the photon control selection requires both the electron and positron to be reconstructed in the TPC, while the photon background is mostly related to events where either the electron or positron is lost, usually when it is not very energetic or emitted at high angles. As a result, the photon background consists mostly of highly asymmetric events where most of the energy of the photon goes into one of two electrons. For high angle events it is predominantly due to one of the two electrons being lost, resulting in more high angle photon background in the CC-$\nu_{e}$ and CC-$\bar\nu_{e}$ selections. 

This angular dependence will introduce different external photons to the photon background and the photon control selection. Most of the external photons entering the photon control samples come from neutrino interactions in the P0D or in the aluminium frame of TPC1. For the photon background, however, a significant population of external photons are also from neutrino interactions in the ECals. The photons mostly come from $\pi^{0}$ decays and Table~\ref{tab:gammanuecomp} shows the different contributions to the photon background and the photon control selections from CC and NC interactions and from external photons. Despite the differences discussed, the origin of the photon background entering the CC-$\nu_{e}$ and CC-$\bar\nu_{e}$ selections and the photon control selections is similar. This provides confidence that the photon control samples can be used to constrain the photon background in the signal channels. Additional simulation studies are also performed to check for shape variations in momentum and angle in the photon selections and in the photon background in the signal selections. These studies include the variation of the relevant fraction of CC/NC photon events by a factor of 2, weighting the nominal MC by varying the Delta resonance width by $\pm 1\sigma$ and varying the external photon background between (40 - 75)\% based on the target material the neutrino interaction occurred. In all the cases the effect on the momentum and angular shapes is found to be very small. 

\begin{table}[!ht]
 \caption{Comparison of the photon background entering the CC electron (anti-)neutrino selections and the photon control selections split down to different $\pi^{0}$ contributions from CC and NC interactions in FGD1 and to external photons. Out of fiducial volume (OOFV) photons are separated into events where the true neutrino vertex is in FGD1 (In-FGD) and into events where the true neutrino vertex is out of FGD1 (OO-FGD).}
 \begin{center}
 \resizebox{\textwidth}{!}{
 \begin{tabular}{l|cc|cc|cc}
 \hline
 \hline
 Interaction Type & FHC CC-$\nu_{e}$ (\%) & Photon Selection (\%)   & RHC CC-$\nu_{e}$ & Photon Selection (\%)   & RHC CC-$\bar\nu_{e}$ & Photon Selection (\%)\\
 \hline
 CC $0\pi^{0}$    & 4.5                   & 4.3                  & 4.8              & 6.9   & 1.1  & 5.4 \\
 CC $1\pi^{0}$    & 15.7                  & 14.6                 & 14.7             & 12.8  & 6.7  & 11.8 \\
 CC $>1\pi^{0}$   & 6.1                   & 4.7                  & 5.4              & 4.5   & 1.9  & 3.8 \\
 NC $0\pi^{0}$    & 3.6                   & 3.6                  & 2.6              & 3.0   & 1.9  & 2.3 \\
 NC $1\pi^{0}$    & 24.8                  & 28.5                 & 26.7             & 30.5  & 35.1 & 31.1 \\
 NC $>1\pi^{0}$   & 4.3                   & 5.1                  & 4.7              & 4.2   & 2.8  & 3.6 \\
 OOFV (In-FGD)     & 8.5                   & 7.4                  & 8.8              & 7.8   & 10.7 & 8.9 \\
 OOFV (OO-FGD)     & 32.5                  & 31.8                 & 32.3             & 30.2  & 39.9 & 33.1 \\
 \hline
 \hline
 
\end{tabular}}
\end{center}
       
\label{tab:gammanuecomp}
            
\end{table} 
\section{Systematic uncertainties}
\label{sec:systuncertainties}

Systematic uncertainties affecting the MC prediction on event yields are separated into five main categories: cross-section modelling, final state interactions, detector, external backgrounds and flux.
 
\textbf{Cross-section modelling.} The cross-section interaction modelling in NEUT and GENIE is briefly described in section~\ref{sec:analysissamples} and in detail in previous T2K publications~\cite{ND280numucc4pi, T2KOscLong}. In this section, the systematic uncertainties relevant to cross-section modelling parameters will be briefly discussed. Neutrino cross-section parameters in NEUT relevant to charged-current quasi-elastic interactions are the axial mass ($M_{A}^{QEL}$ = 1.21~$\pm$~0.41~$\rm GeV/c^2$), binding energy ($E_{B}^{C}$ = 25.0~$\pm$~9.0~MeV) and  Fermi momentum ($p_{F}^{C}$ = 223.0~$\pm$~31.0~MeV/c). Binding energy and Fermi momentum are target dependent, for this analysis only those relevant to carbon are considered. For multi-nucleon interactions, a 100\% normalization uncertainty is assumed. The CC resonant production model has three parameters in NEUT: the axial mass ($M_{A}^{RES}$ = 0.95~$\pm$~0.15~$\rm GeV/c^2$), the normalization of the axial form factor for resonant pion production ($CA_{5}^{RES}$ = 1.01~$\pm$~0.12) and the normalisation of the isospin non-resonant component ($I_{\frac{1}{2}}$ = 1.3~$\pm$~0.2). For the CC DIS process an energy dependent normalisation uncertainty (10\% at 4~GeV) is considered. For CC coherent interactions a 100\% normalisation uncertainty is considered. For neutral-current interactions, due to poor constraints from external data, a 30\% normalisation uncertainty is applied. The effect of the cross-section uncertainties on the event yields is evaluated by shifting each cross-section parameter by $\pm1\sigma$ and shifting the nominal MC. 

\textbf{Final State Interactions} The pion final state interaction systematic uncertainties include the effects of absorption, inelastic scattering, charge exchange and quasi-elastic scattering inside the nucleus. A full description can be found in previous T2K publications~\cite{ND280numucc4pi, T2KOscLong}. Similarly with the cross-section uncertainties, the effect of final state interaction systematic uncertainties on the event yields is evaluated by varying simultaneously the final state interaction effects by $\pm1\sigma$ and shifting the nominal MC.

\textbf{Detector.} Detector systematic uncertainties encapsulate the performance of each ND280 sub-detector (FGDs, TPCs and ECals). They are applied to simulated events and are separated in three categories: normalization, selection efficiency and  variation of the observable. Normalization systematics are applied as a single weight to all events. Efficiency systematics are applied as a weight that depends on one or more observables. Variation systematics are treated by varying the observables and redoing the event selections. Detector systematic uncertainties considered and their treatment are summarised in Table~\ref{tab:detsysts}.

Detector systematics are evaluated using high purity ($>$~90\%) control samples from cosmic and through-going muons, electrons and positrons from photon conversions and protons from neutrino interactions. ECal related uncertainties are evaluated using the same methodology described in~\cite{ND280NueSel}. All other detector systematics, except FGD2 shower efficiency, are evaluated in the same way as explained in~\cite{ND280NueSel, ND280numucc4pi, T2KOscLong}. 

The FGD2 shower efficiency describes the probability of electrons and protons originating in FGD1 to shower in FGD2. Since FGD2 is a thin detector and cannot contain showers, a shower is defined when multiple FGD2-TPC3 tracks are produced when the leading track passes through FGD2. Since this systematic is only relevant for the CC-$\bar\nu_{e}$ channel, the uncertainty is evaluated using events with single electron or proton tracks in the neutrino beam originating in FGD1, passing through FGD2 and comparing the FGD2 shower efficiencies for data and MC.

\begin{table}[!ht]
 \caption{List of detector systematic uncertainties and their treatment for simulated events. Normalization systematics are applied as a single weight to all events. Efficiency systematics are applied as a weight that depends on one or more observables. Variation systematics are treated by varying the observables and redoing the event selection.}
 \begin{center}
 \begin{tabular}{l|c|c}
 \hline
 \hline
  Systematic  & Treatment & Comment \\
  \hline
   TPC tracking efficiency & efficiency \\
   TPC charge mis-identification & efficiency \\
   TPC momentum resolution and scale & variation \\
   B-field distortions & variation \\
   TPC PID             & variation \\
   FGD-TPC matching efficiency & efficiency \\
   TPC-ECal matching efficiency & efficiency \\
   FGD2 PID & variation & Only applied to CC-$\bar\nu_{e}$ \\
   FGD2 shower efficiency & efficiency & Only applied to CC-$\bar\nu_{e}$\\
   FGD1 mass & normalisation \\
   TPC, P0D and ECal pile-up & normalisation \\
   ECal $R_{MIP/EM}$ PID & efficiency \\
   ECal $R_{EM/HIP}$ PID & efficiency & Only applied to CC-$\bar\nu_{e}$\\
   ECal EM energy resolution and scale & variation \\
   Pion and proton secondary interactions & efficiency \\
   Sand interactions & efficiency \\
   FGD1-ECal time resolution & variation \\
	\hline
	\hline
	\end{tabular}
 \end{center}
       
\label{tab:detsysts}

\end{table} 

\textbf{External backgrounds.} These are related to the uncertainties associated with photons (or other particles) produced outside of the FGD1, either in other sub-detectors or outside of ND280, that propagate inside FGD1. A large number of these neutrino interactions are on heavier nuclear targets (aluminium, iron and lead) with considerable cross-section modelling uncertainties. A detailed study of the external photon propagation in ND280 was performed in~\cite{ND280SingleGamma} but only in limited angular regions. Outside these angular regions there are large data/MC differences due to the poor simulation of inactive material. Since the method developed in~\cite{ND280SingleGamma} is very sensitive to the material density and composition (small changes can cause large variation in systematic uncertainties), conservatively a 100\% systematic uncertainty on the external photon production and propagation is assumed. The effect on the momentum and angular shapes for both the photon background in the signal selections and the photon selections is studied with additional simulations. The external photon events in the simulation are varied between (40 - 75)\% based on the target material the neutrino interaction occurred. The effect on both momentum and angular shapes is found to be negligible.  

\begin{figure}[htbp]
 \centering
	\includegraphics[width=0.495\textwidth]{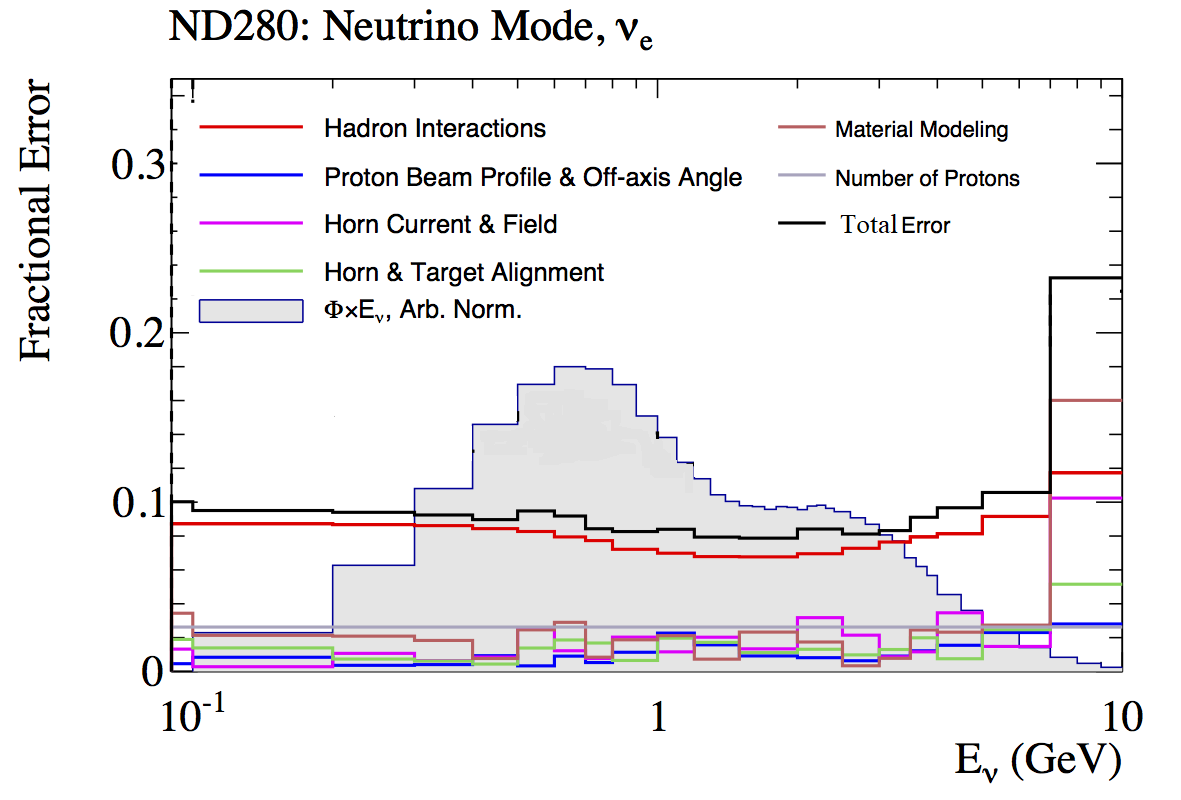}
	\includegraphics[width=0.495\textwidth]{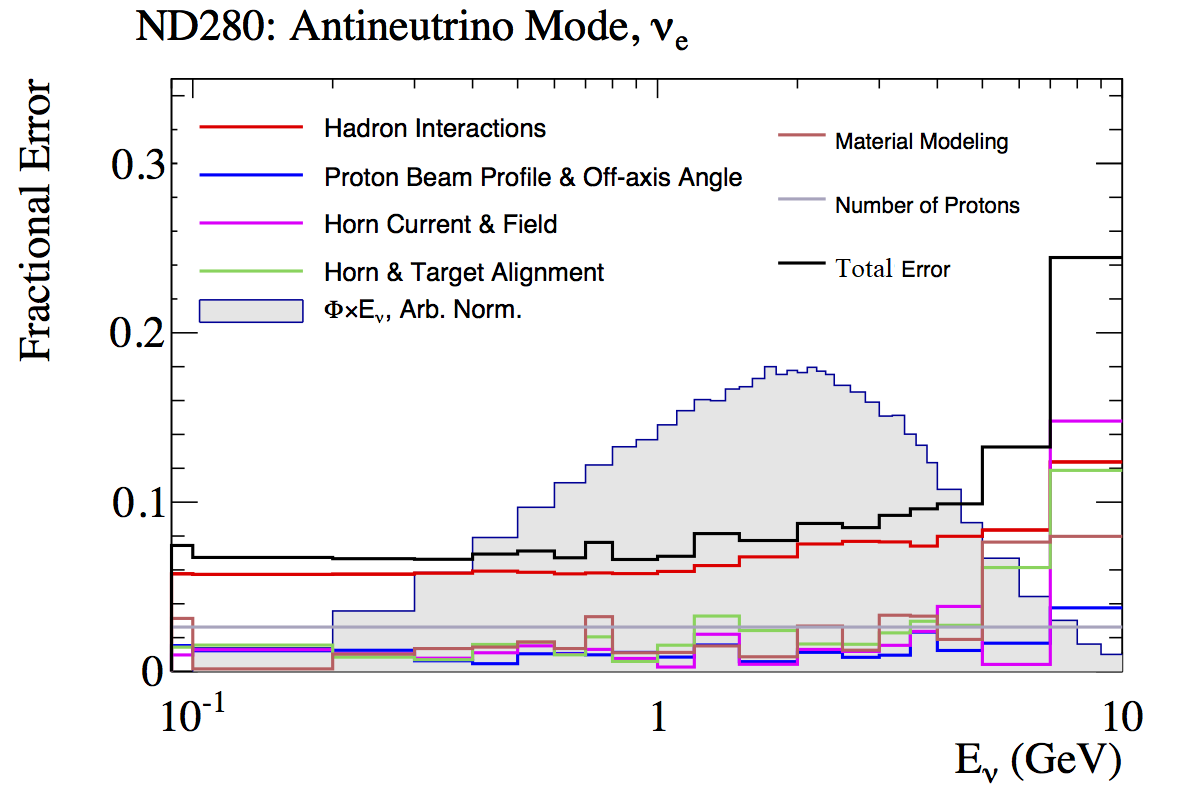}
	\includegraphics[width=0.495\textwidth]{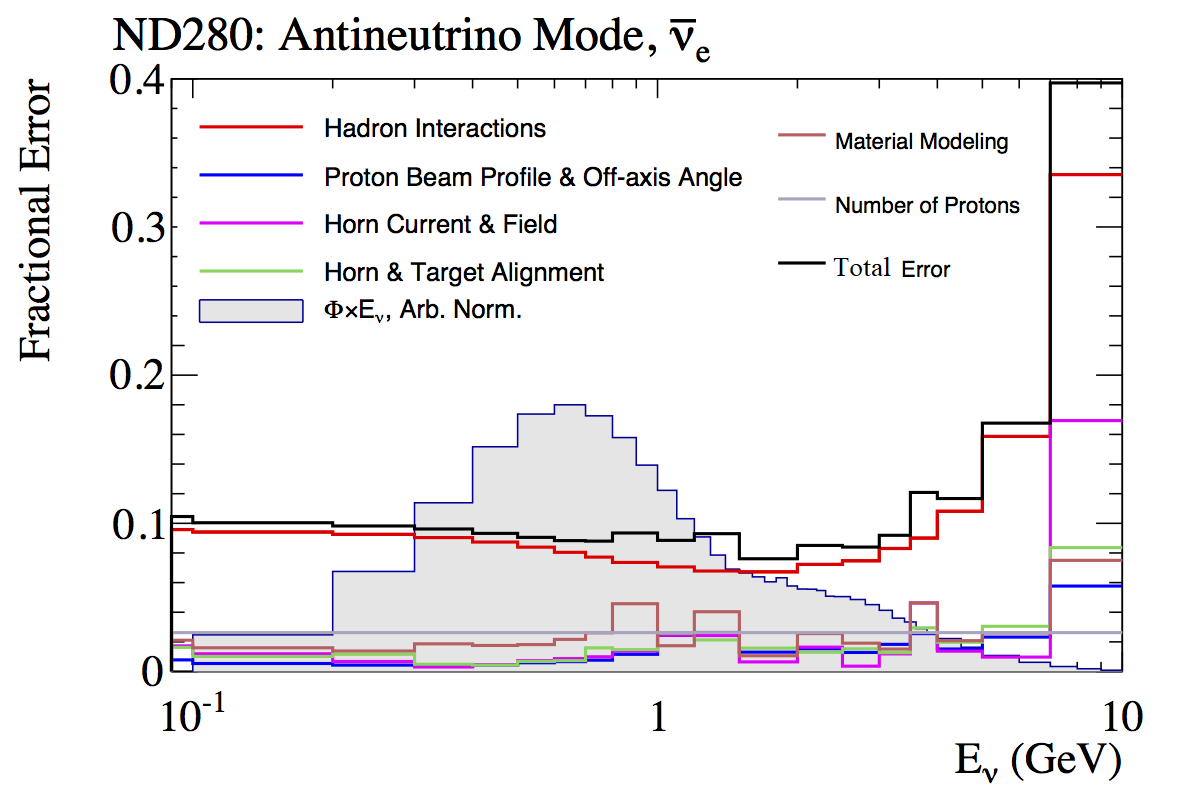}
	\caption{Flux systematic uncertainties for the FHC $\nu_{e}$ flux (top left), RHC $\nu_{e}$ flux (top right) and RHC $\bar\nu_{e}$ flux (bottom).}
	\label{fig:nuefluxsysts}
\end{figure}

\textbf{Flux.} Flux systematic uncertainties are calculated as a function of the neutrino energy and they are correlated between the neutrino flavours and between the neutrino and anti-neutrino beams. Flux systematic uncertainties are larger at the high energy tail of the neutrino spectrum and for the $\nu_{\mu}$ and $\bar\nu_{\mu}$ fluxes are in the range (7.0 - 14)\%. The $\nu_{e}$ and $\bar\nu_{e}$ flux systematic uncertainties are shown in Figure~\ref{fig:nuefluxsysts} and are dominated by the systematic uncertainties on hadron production. The evaluation of the flux systematic uncertainties can be found in previous T2K publications~\cite{T2KOscLong, t2kfluxerror}. 

\subsection{Effect of systematic uncertainties on the event yields}
\label{subsec:systeffect}
 
A summary of systematic uncertainties on signal and background MC event yields for the CC-$\nu_{e}$ and CC-$\bar\nu_{e}$ selections is shown in Table~\ref{tab:allsystsccnue}. The systematic uncertainties on signal yields are dominated by the flux (8 -- 10\%) and cross-section modelling (13 -- 14\%). The larger cross-section systematic uncertainties come from the large uncertainties considered on the quasi-elastic axial mass $M_A^{QEL}$ and multi-nucleon interactions, each contributing (6.5 -- 8.5)\% to the total cross-section systematic uncertainty. Detector systematic uncertainties on signal yields are (2 -- 4)\% with the most important being the TPC PID and TPC-ECal matching efficiencies. For CC-$\bar\nu_{e}$, the ECal PID and FGD2 shower efficiency, which are related to the proton background rejection, are also important. For an inclusive CC selection, final state interaction systematic uncertainties on signal yields are small. They are only considered if a charged pion, after final state interactions, becomes more or less energetic than the primary electron or when there is a $\pi^{0}$ involved as the secondary electrons can be more or less energetic than the primary electron. The total systematic uncertainty on the signal yields is approximately (16 -- 17)\% in all the channels.

The systematic uncertainties on the MC background event yields are separated into photon background and all other backgrounds. The total systematic uncertainties on the MC photon background event yields are approximately (23 -- 26)\% in all channels and are dominated by the cross-section and external systematic uncertainties. Cross-section systematic uncertainties (16 -- 19)\% are dominated by the charged-current and neutral-current resonant and DIS production models. The flux systematic uncertainties are around 8\% and the final state interaction systematic uncertainties are (1.5 -- 3.0)\%. Detector systematic uncertainties are (3 -- 6)\%, with TPC PID, FGD1 and ECal time resolutions, TPC-ECal matching efficiency and pion secondary interactions being the most important. Approximately a third of the photon background comes from neutrino interactions outside FGD1, either in other sub-detectors or outside the ND280 and the majority of these events populate the low momentum and/or high angle regions.

The systematic uncertainties on the other backgrounds MC event yields vary from (19 -- 33)\% since different sources of backgrounds contribute to each channel. The biggest difference comes from the external background which dominates the systematic uncertainties on the other background event yields and is different in each channel since the neutrino flux is different. Flux systematic uncertainties are around 8\% and the cross-section systematic uncertainties are around (11 -- 12)\%. Detector systematic uncertainties are (4.0 -- 6.5)\%, which are larger than the corresponding detector systematic uncertainties for signal and photon background event yields.

\begin{table}[!ht]
 \caption{Summary of systematic uncertainties on MC signal and background event yields. The total systematic uncertainty is the quadratic sum of all the systematic sources. Possible correlations between the different systematic sources are ignored.}
 \begin{center}
 \begin{tabular}{c|l|c|c|c}
 \hline
 \hline
  & Source of uncertainty & Signal (\%) & $\gamma$ background (\%) & Other backgrounds (\%) \\
  \hline
	\multirow{6}{*}{\rotatebox[origin=c]{90}{FHC CC-$\nu_{e}$}}
	& Detector                 &  2.96         &  3.02                    &  3.91                  \\
	& External background      &  0.00         & 17.25                    & 29.07                  \\
	& Flux                     &  8.92         &  7.61                    &  7.60                  \\
	& Final State Interactions &  0.52         &  2.78                    &  3.72                  \\
	& Cross-section            & 13.60	       & 16.54                    & 11.18                  \\
	& Total                    & 16.54         & 25.41                    & 32.51                  \\
	\hline
	\multirow{6}{*}{\rotatebox[origin=c]{90}{RHC CC-$\nu_{e}$}}
	& Detector                 &  2.12         &  3.09                    &  5.12                  \\
	& External background      &  0.00         & 12.71                    & 17.56                  \\
	& Flux                     &  8.11         &  8.28                    &  8.23                  \\
	& Final State Interactions &  0.98         &  1.48                    &  4.97                  \\
	& Cross-section            & 13.45	       & 17.71                    & 10.67                  \\
	& Total                    & 15.88         & 23.57                    & 23.26                  \\
	\hline
	\multirow{6}{*}{\rotatebox[origin=c]{90}{RHC CC-$\bar\nu_{e}$}} 
	& Detector                 &  3.46         &  5.68                    &  6.46                  \\
	& External background      &  0.00         & 14.90                    &  7.20                  \\
	& Flux                     &  9.95         &  8.33                    &  8.01                  \\
	& Final State Interactions &  0.39         &  1.95                    &  7.94                  \\
	& Cross-section            & 12.98	       & 18.88                    & 12.01                  \\
	& Total                    & 16.72         & 26.15                    & 19.11                  \\
	\hline
	\hline
 \end{tabular}
 \end{center}
       
 \label{tab:allsystsccnue}

\end{table}	

\subsection{Effect of systematic uncertainties on the photon control samples}
\label{subsec:systgamma}

The systematic uncertainties on the photon control samples are roughly (20 -- 23)\% and are summarised in Table~\ref{tab:allsystsgamma}. The dominant sources are coming from the external background and cross-section modelling.

\begin{table}[!ht]
 \caption{Effect of the systematic uncertainties on the photon control sample MC event yields selecting either electron as the leading track ($\gamma$-Elec.) or positron as the leading track ($\gamma$-Posi.). The total systematic uncertainty is the quadratic sum of all the systematic sources. Possible correlations between the different systematic sources are ignored.}
	\begin{center}
	\begin{tabular}{l|c|c|c}
	\hline
	\hline
	Systematic uncertainty   & FHC $\gamma$-Elec. (\%)   & RHC $\gamma$-Elec. (\%)   & RHC $\gamma$-Posi. (\%) \\
	\hline
	Detector                 &  2.35          & 1.81           &  1.72        \\
	External background      & 14.24          & 9.57           & 11.10        \\
	Flux                     &  7.62          & 8.29           &  8.26        \\
	Final State Interactions &  2.62          & 1.49           &  1.93        \\
	Cross-section            & 16.49          & 15.28          & 15.67        \\
	Total                    & 23.35          & 19.98          & 21.06        \\ 
  \hline
	\hline	
	\end{tabular}
	\end{center}
       
	\label{tab:allsystsgamma}

	\end{table}
\section{Fit model}
\label{sec:fitmodel}

The flux integrated single differential cross-section as a function of the electron or positron true momentum $p$ or true scattering angle $\rm \cos(\theta)$ is expressed as  

\begin{equation}
 \frac{d\sigma_{i}}{dk_{i}} = \frac{N_{i}}{\epsilon_{i}} \times \frac{1}{T\Phi\Delta k_{i}},
 \label{eq:xs}
\end{equation}

where $k$ is either $p$ or $\cos(\theta)$, $N_{i}$ is the number of signal events in bin $i$, $\epsilon_{i}$ is the efficiency in bin $i$, $\Phi$ is the neutrino flux, $T$ the number of target nucleons and $\Delta k_{i}$ is the true momentum or true scattering angle bin interval. 

The number of signal events in each bin is calculated using an extended, binned maximum likelihood fit. The PDFs are constructed from histogram templates using ROOT's histogram-based HistFactory fit package~\cite{histfactory}, which is based on the RooStats~\cite{roostats} and RooFit~\cite{roofit} packages. The fit is performed simultaneously on all the signal channels (FHC and RHC CC-$\nu_{e}$ and RHC CC-$\bar\nu_{e}$) and their corresponding photon control channel. Each channel is broken down to angular regions and each angular region is broken down to one dimensional templates in momentum for signal, photon background and other backgrounds. For the photon control channels the small signal contribution is merged in the other backgrounds. 

A likelihood is constructed from all the signal and background templates, nuisance parameters $\vec\theta$ and their constraints $C\left(\theta_{\kappa}^{0},\theta_{\kappa}\right)$ and a set of scaling parameters $c$ and $g$ controlling the signal and photon background respectively, given the observed data $\vec{N}$
 
\begin{equation}
\begin{aligned}
 &L\left(\vec{N}|c,g,\vec\theta\right) = \\
 &\left[ \prod_{i=1}^{N_{region}}\prod_{j=1}^{N_{bin}}\frac{\left[c_{ij}S_{ij}(\vec\theta) + g_{i}B_{ij}^{\gamma}(\vec\theta) + B_{ij}^{other}(\vec\theta)\right]^{n_{ij}}}{n_{ij}!}e^{-\left[c_{ij}S_{ij}(\vec\theta) + g_{i}B_{ij}(\vec\theta) + B_{ij}^{other}(\vec\theta)\right]} \right] \\
 &\times \left[ \prod_{i=1}^{N_{region}}\prod_{l=1}^{N_{bin;PC}}\frac{\left[ g_{i}B_{il;PC}^{\gamma}(\vec\theta) + B_{il;PC}^{other}(\vec\theta)\right]^{m_{il}}}{m_{il}!} e^{-\left[g_{i}B_{il;PC}^{\gamma}(\vec\theta) + B_{il;PC}^{other}(\vec\theta)\right]} \right] \\
 &\times \prod_{k=1}^{N_{syst}}C\left(\theta_{\kappa}^{0},\theta_{\kappa}\right),
 \end{aligned}
 \label{eq:likelihood}
\end{equation}

where $N_{region}$ is the number of angular regions which is the same for signal and photon control channels, $N_{bin}$ ($N_{bin;PC}$) is the number of bins in signal (photon control) region, $S_{ij}$ are the signal templates contributing to reconstructed bin $j$ for region $i$, $B_{ij}^{\gamma}$ ($B_{il;PC}^{\gamma}$) is the number of photon events in reconstructed bin $j$ ($l$) for signal (photon control) region $i$, $B_{ij}^{other}$ ($B_{il;PC}^{other}$) is the number of other background events in reconstructed bin $j$ ($l$) for signal (photon control) region $i$, $n_{ij}$ ($m_{il}$) are the number of entries in each bin in signal (photon control) region and $N_{syst}$ is the number of nuisance parameters.

\subsection{Propagation of systematic uncertainties}
\label{subsec:systspropagation}

Systematic uncertainties are included in the fit as nuisance parameters and are calculated as $\pm 1 \sigma$ variations of the nominal samples in the signal and photon control samples, $S_{ij}(\vec\theta)$, $B_{ij}^{\gamma} (\vec\theta)$, $B_{ij}^{other} (\vec\theta)$, $B_{il;PC}^{\gamma} (\vec\theta)$ and $B_{il;PC}^{other} (\vec\theta)$. These variations can either change the normalisation or produce bin-dependent shape differences or have a correlated effect on shape and normalisation. For each variation (or set of variations) a nuisance parameter is used to interpolate between the $\pm 1\sigma$ uncertainties with a Gaussian constraint. Systematic uncertainties that are common between samples or channels are fully correlated in the fit. A summary of the nuisance parameters included in the fit is shown in Table~\ref{tab:allsystsfit}. 

Variations from the cross-section uncertainties are calculated by varying each cross-section parameter by $\pm 1 \sigma$ and changing the nominal samples. Some of the cross-section uncertainties may produce asymmetric variations and these are considered in the fit. Variations related to the final state interaction systematic uncertainties, including their correlations, are estimated following the methodology described in~\cite{T2KOscLong}. 

Variations from the flux uncertainties are calculated using the full beam covariance taking into account all the correlations between the neutrino beams, neutrino flavours and energy bins. 

Variations of the nominal samples arising from the detector, pile-up and external background systematics are evaluated using MC simulations varying the systematics to change the number of events in each reconstructed bin. Three nuisance parameters are used for the three pile-up systematics in each beam mode (FHC or RHC). Four nuisance parameters in each beam mode are used to describe the external background systematic uncertainties. The external backgrounds are separated based on their origin (ND280 or sand interactions), their background category (photon or other backgrounds) and their beam mode (FHC or RHC). 

MC statistical uncertainties, describing the finite size of the simulated events in each sample, are also included as nuisance parameters in the fit following the Barlow-Beeston~\cite{barlowbeeston} approach considering one nuisance parameter per channel and bin.

\begin{table}[!ht]
 \caption{Summary of nuisance parameters related to systematic uncertainties considered in the fit.}
 \begin{center}
 \resizebox{\textwidth}{!}{
 \begin{tabular}{l|c|c|c}
 \hline
 \hline
 Source of uncertainty   & Number of parameters & Constraint & Variation type \\
 \hline
 MC statistical          & 29                   & Poissonian & One per bin    \\
 Pile-up                 & 6                    & Gaussian   & Normalisation  \\
 External backgrounds     & 8                    & Gaussian   & Shape and normalisation \\
 Detector and flux       & 15                   & Gaussian   & Shape and/or normalisation \\
 Cross-section and final state interactions   & 13                   & Gaussian   & Shape and normalisation \\   
 \hline
 \hline
 \end{tabular}}
 \end{center}
       
 \label{tab:allsystsfit}

\end{table}	


\subsection{Binning choice}
\label{subsec:binning}

The choice of the binning depends on a number of factors, some of the most important are: sufficient number of signal events in each bin, isolation of the backgrounds in specific $p - \theta$ regions, event migration due to bremsstrahlung and flat efficiency corrections.

The first criterion for the binning choice is to not consider high angle events ($\theta > 45^{\circ}$) since the acceptance due to detector effects is almost zero. In addition, the photon background is large and the statistics in the photon control channels is poor. The high angle regions and the low momentum ($p < 300$~MeV/c) bins are background enriched and are kept in the fit as background validation regions. Approximately 75\% of the external photon background is contained in the low momentum and high angle regions. The signal contribution in the low momentum bins ($p < 300$~MeV/c) is tiny and, to help the fit performance, it is kept constant in the fit. Two angular regions are considered to better describe the middle and forward angular cases. The momentum bins are identical in each angular region.

The momentum bins are optimised to minimize the effect of bremsstrahlung. Since bremsstrahlung is not a detector smearing effect, but a physics effect depending on the initial electron kinematics and the material propagated, special requirements are considered to minimize this effect. 
  
The (anti-)correlations between the momentum bins introduced by bremsstrahlung are studied with MC simulations requiring them to be less than 50\%. If the chosen momentum binning fails this requirement, the momentum bins are expanded to reduce the migration of events due to bremsstrahlung and the MC simulations are repeated. Due to the large momentum bins chosen in this analysis, the effect of bremsstrahlung can be efficiently handled in the fit.

\begin{figure}[htbp]
	\centering
	\includegraphics[width=0.99\textwidth]{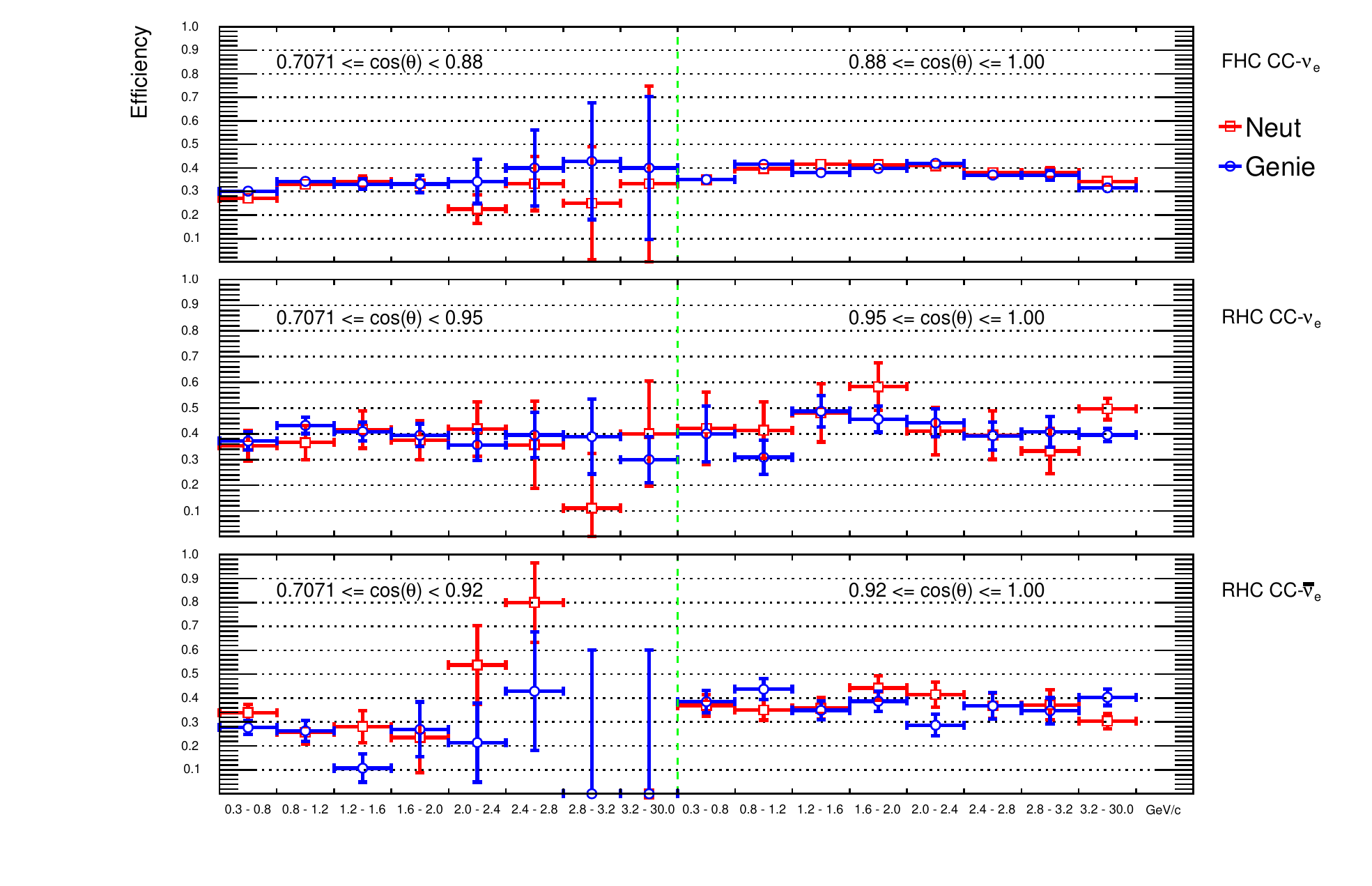}
	\caption{Signal efficiencies for NEUT and GENIE MC using a finer binning than used in the cross-section measurements. Errors are statistical only.}
	\label{fig:signaleff_finebin}
\end{figure}

\begin{figure}[htbp]
	\centering
	\includegraphics[width=0.99\textwidth]{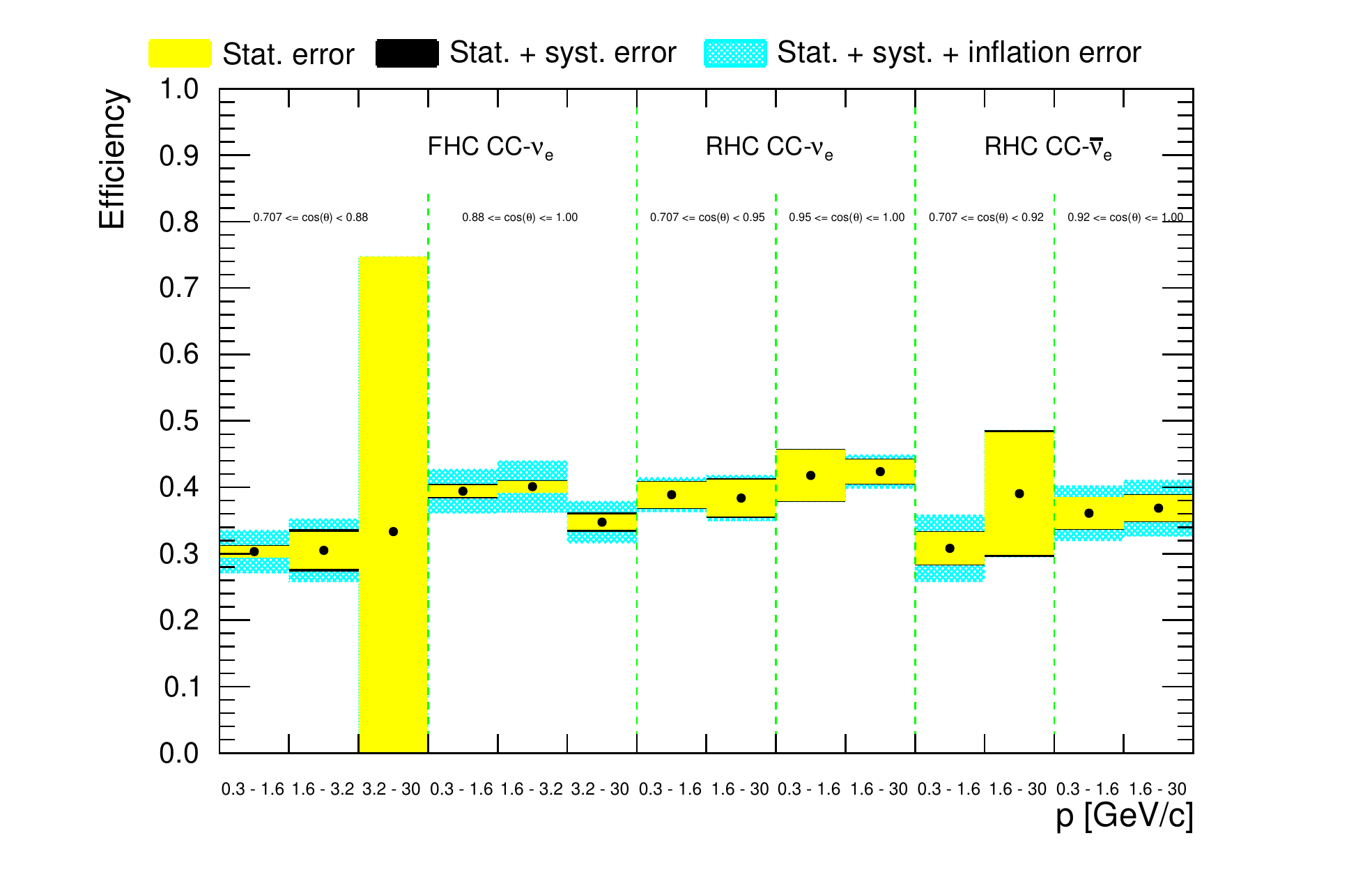}
	\caption{Signal efficiencies in different angular regions for the three samples for NEUT MC with statistical, systematics and inflation uncertainties.}
	\label{fig:signaleffsysts}
\end{figure}

Signal efficiencies are also a significant factor for the binning choice as they should be flat to avoid model dependencies. The efficiencies in the two angular regions in each signal channel are shown in Figure~\ref{fig:signaleff_finebin} and are relatively flat with some small fluctuations observed between NEUT and GENIE and in the low statistics bins. Although the cross-section measurements are calculated in one dimension, fitting in $p-\theta$ is important to check for model dependencies due to the efficiency corrections. After the total statistical and systematic uncertainties are applied on signal efficiencies, the efficiency errors are artificially inflated to cover differences between NEUT and GENIE and variations between momentum bins. The efficiencies with statistical, systematic and inflation uncertainties are shown in Figure~\ref{fig:signaleffsysts}. 

The binning choice for each signal channel is shown in Table~\ref{tab:binsummary}. In total there are 9 free parameters controlling the photon background (one for each angular region) and 17 free parameters controlling the signal (one for each bin in the table, except for the six lowest momentum bins which are kept constant in the fit since the number of signal events is negligible).  

\begin{table}[!ht]
 \caption{Summary of the binning for CC-$\nu_{e}$ and CC-$\bar\nu_{e}$ inclusive channels included in the fit. Validation bins are background enriched and are used as extra fit validation regions. These bins are excluded from the cross-section measurements.}
 \begin{center}
 \begin{tabular}{c|c|c|c}
 \hline
 \hline 
   & Angular region ($\cos(\theta)$)  & Momentum bin (GeV/c) & Comment \\
 \hline
  \multirow{8}{*}{\rotatebox[origin=c]{90}{FHC CC-$\nu_{e}$}}
  & -1.00 - 0.7071                   & 0 - 30     & Validation bin \\
  & 0.7071 - 0.88                    & 0 - 0.3    & Validation bin \\
  & 0.7071 - 0.88                    & 0.3 - 1.6  & \\
  & 0.7071 - 0.88                    & 1.6 - 3.2  & \\
  & 0.7071 - 0.88                    & 3.2 - 30   & \\
  & 0.88 - 1.00                      & 0 - 0.3    & Validation bin \\
  & 0.88 - 1.00                      & 0.3 - 1.6  & \\
  & 0.88 - 1.00                      & 1.6 - 3.2  & \\
  & 0.88 - 1.00                      & 3.2 - 30   & \\
  \hline
  \multirow{6}{*}{\rotatebox[origin=c]{90}{RHC CC-$\nu_{e}$}}
  & -1.00 - 0.7071                   & 0 - 30      & Validation bin \\
  & 0.7071 - 0.95                    & 0 - 0.3     & Validation bin \\
  & 0.7071 - 0.95                    & 0.3 - 1.6   & \\
  & 0.7071 - 0.95                    & 1.6 - 30    & \\
  & 0.95 - 1.00                      & 0 - 0.3     & Validation bin \\
  & 0.95 - 1.00                      & 0.3 - 1.6   & \\
  & 0.95 - 1.00                      & 1.6 - 30    & \\
  \hline
  \multirow{6}{*}{\rotatebox[origin=c]{90}{RHC CC-$\bar\nu_{e}$}}
  & -1.00 - 0.7071                   & 0 - 30     & Validation bin \\
  & 0.7071 - 0.92                    & 0 - 0.3    & Validation bin \\
  & 0.7071 - 0.92                    & 0.3 - 1.6  & \\
  & 0.7071 - 0.92                    & 1.6 - 30   & \\
  & 0.92 - 1.00                      & 0 - 0.3    & Validation bin \\
  & 0.92 - 1.00                      & 0.3 - 1.6  & \\
  & 0.92 - 1.00                      & 1.6 - 30   & \\
 \hline
 \hline
 
\end{tabular}
\end{center}
       
\label{tab:binsummary}
\end{table}


\section{Cross-section results}
\label{sec:dataresults}

The fit is used to measure the number of signal events in all channels including all systematic uncertainties as described in section~\ref{sec:fitmodel}. The best fit results and the fit covariance matrix are used to measure the flux-integrated single differential cross-sections $d\sigma/dp$ and $d\sigma/d\cos(\theta)$ using eq.~(\ref{eq:xs}).

Prior to fitting the data, the signal and background events are varied under different model assumptions
to create pseudo datasets generated from variations of nominal MC (toy experiments). These pseudo datasets are used to check the fit performance, possible biases, over-constraining the nuisance parameters and the impact of nuisance parameters to the signal normalisation parameters and understand the dependencies on signal and background models. In addition, the cross-sections are measured using two generators to test different model assumptions. The results are in good agreement with all the tests providing confidence that our measurements are free from model dependencies.

\begin{figure}[htbp]
 \centering
	\includegraphics[width=0.99\textwidth]{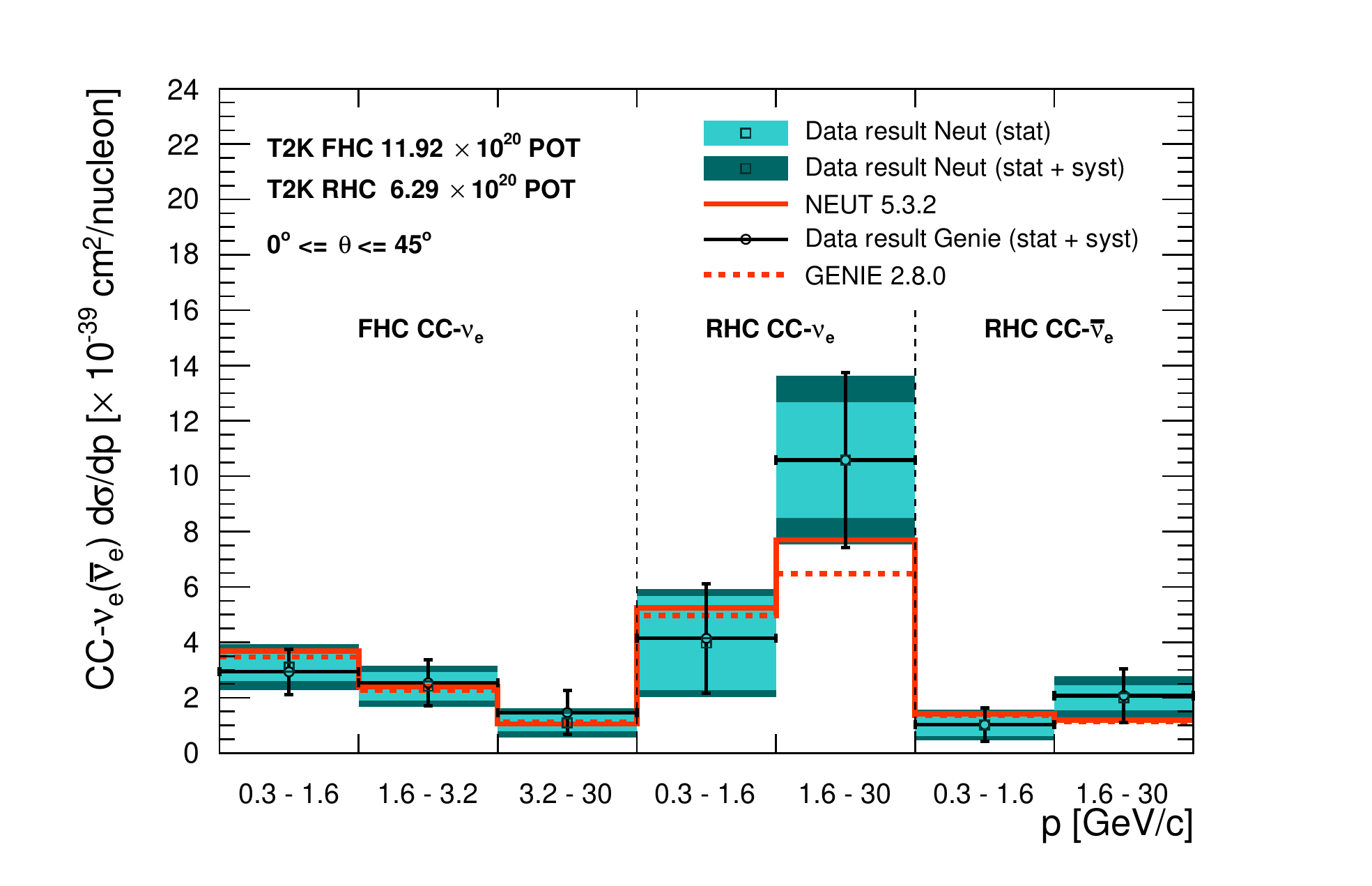}
	\includegraphics[width=0.99\textwidth]{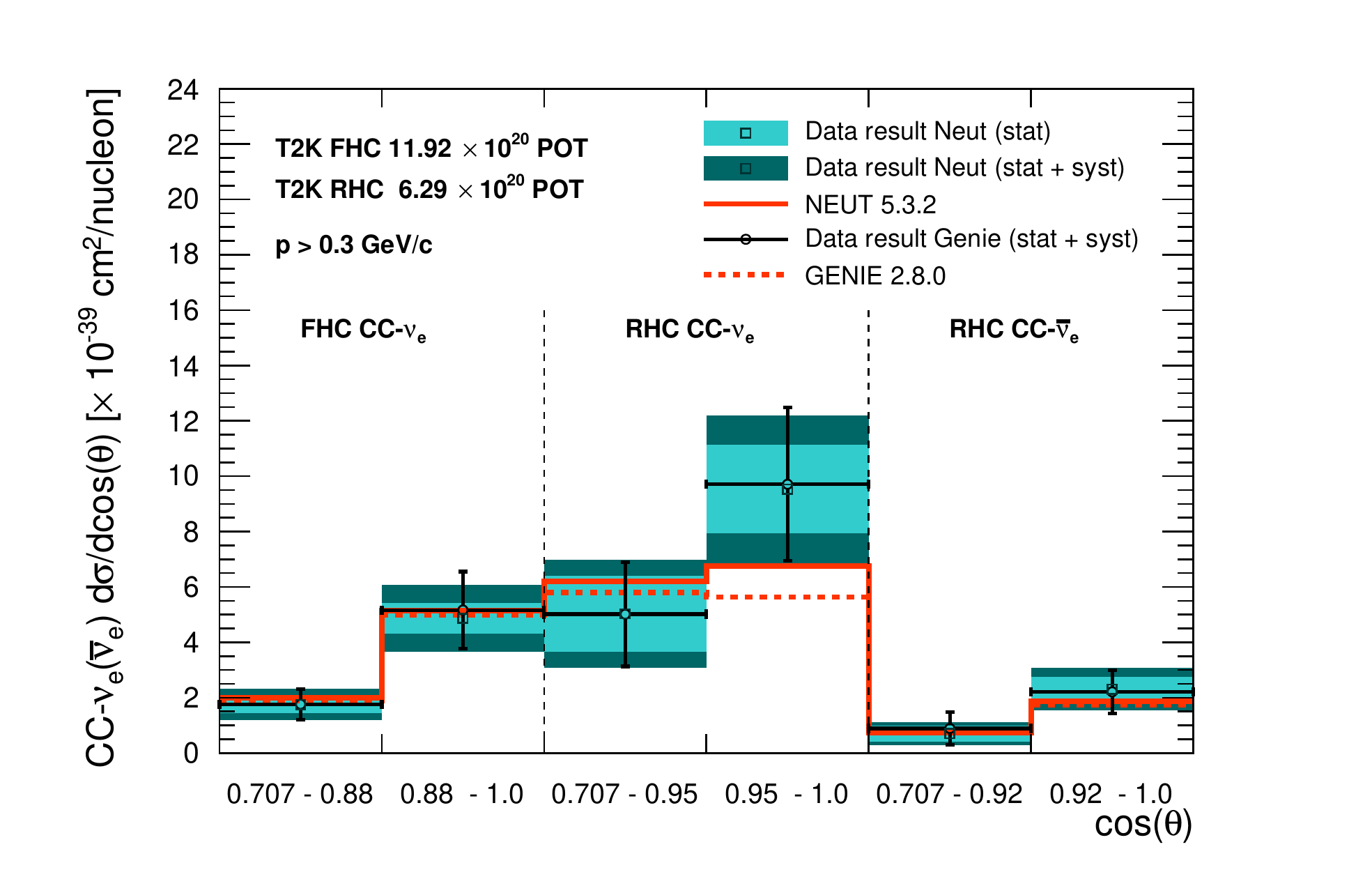}
	\caption{CC-$\nu_{e}$ and CC-$\bar\nu_{e}$ inclusive cross-section results in $d\sigma/dp$ (top) and $d\sigma/d\cos(\theta)$ (bottom) in a limited phase-space ($p > 300$~MeV/c and $\theta \leq 45^{\circ}$). The statistical uncertainty is computed by fixing all the nuisance parameters to their post-fit nominal values and redoing the fit. The systematic uncertainty is computed by subtracting in quadrature the statistical uncertainty from the total uncertainty.}
	\label{fig:dataxsresults}
\end{figure}

The differential cross-section results in electron and positron momentum, $\rm d\sigma/dp$, using NEUT (5.3.2) or GENIE (2.8.0) as input MC are shown in the top plot in Figure~\ref{fig:dataxsresults} and they are in agreement with the predictions. The CC-$\nu_{e}$ cross-sections are expected to be larger in RHC since the neutrino energy spectrum peaks at higher energy and it is much broader with larger contribution from higher energy neutrinos. Differences between the results using either NEUT or GENIE simulations are expected due to small differences in the efficiency corrections (Figure~\ref{fig:signaleff_finebin}) and small differences in the muon, proton and other backgrounds (Table~\ref{tab:genieneutEventRate}) which are kept constant in the fit. The cross-section results are dominated by the statistical uncertainty, especially in RHC. The statistical uncertainty is estimated by fixing all the nuisance parameters to their post-fit nominal values and repeating the fit.

The differential cross-sections are also calculated in electron and positron scattering angles, $\rm d\sigma/d\cos(\theta)$, for both NEUT and GENIE. They are calculated in the same angular regions defined in Table~\ref{tab:binsummary} and for $p > 300$~MeV/c. The results are shown in the bottom plot in Figure~\ref{fig:dataxsresults} and they are in agreement with the NEUT and GENIE predictions.

\begin{figure}[htbp]
 \centering
	\includegraphics[width=0.99\textwidth]{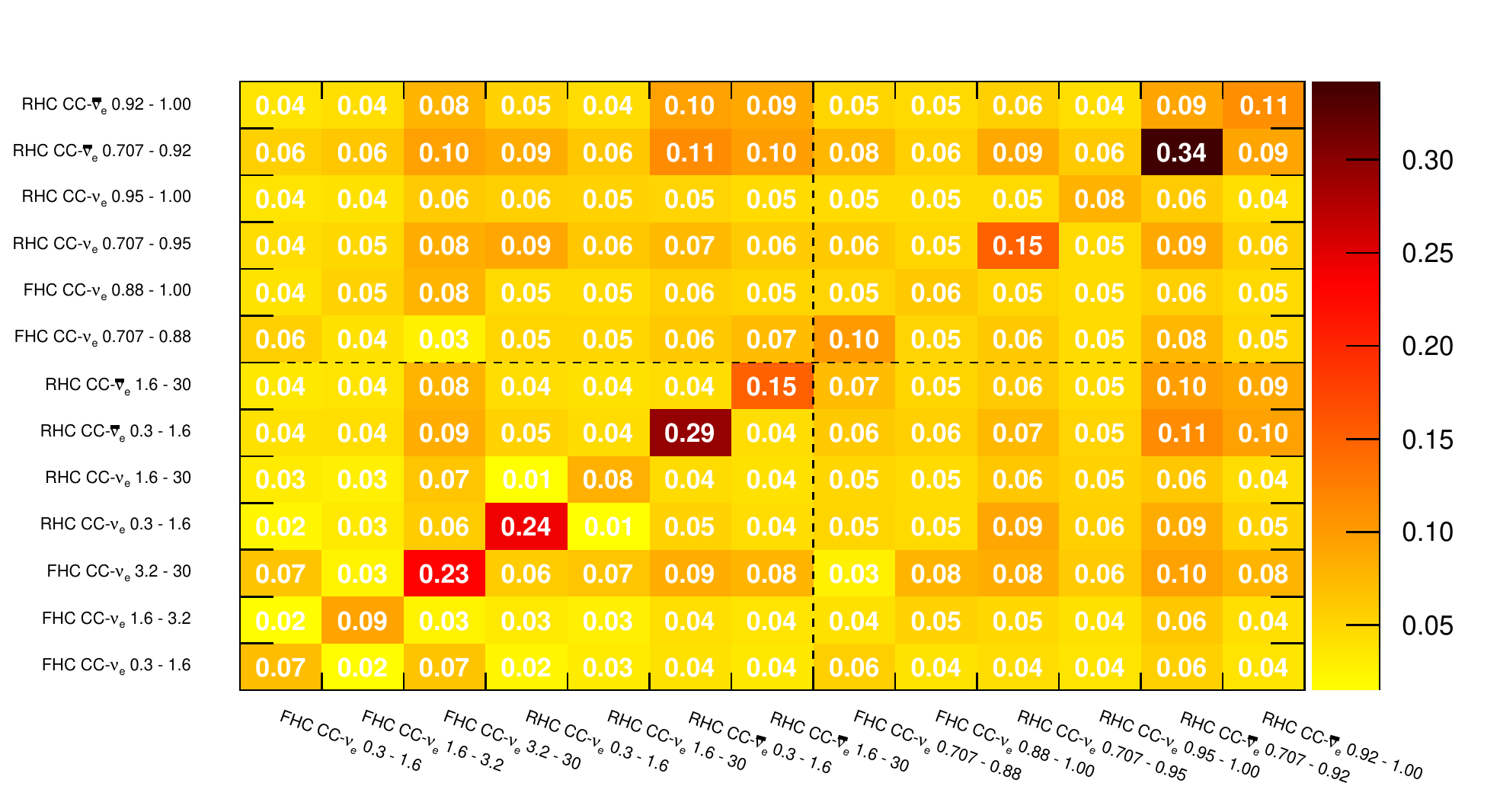}
	\caption{Cross-section fractional covariance matrix for $d\sigma/dp$ (bottom left area) and $d\sigma/d\cos(\theta)$ (top right area) measurements for NEUT (5.3.2). The top left and bottom right areas show the covariance between $d\sigma/dp$ and $d\sigma/d\cos(\theta)$ measurements.}
	\label{fig:dataxscov}
\end{figure}

The systematic uncertainties are propagated in the final cross-section measurements using toy experiments. For each toy experiment the best-fit values and the post-fit covariance are used to vary the number of signal events. Simultaneously, the flux, efficiency and the number of targets are also varied for each toy resulting in a new measurement of the cross-section using equation~\ref{eq:xs}. For N toy experiments the covariance, in a fractional form, is computed from
\begin{equation}
 V_{ij} = \frac{1}{N} \sum_{i=1}^{N}\frac{\left(\frac{d\sigma_{i}^{variation}}{dk_{i}} -  \frac{d\sigma_{i}^{meas.}}{dk_{i}}\right) \left(\frac{d\sigma_{j}^{variation}}{dk_{j}} -  \frac{d\sigma_{j}^{meas.}}{dk_{j}}\right)}{\frac{d\sigma_{i}^{meas.}}{dk_{i}}\frac{d\sigma_{j}^{meas.}}{dk_{j}}},
 \label{eq:cov}
\end{equation}
where $k$ is either $p$ or $\cos(\theta)$, $\frac{d\sigma_{i}^{meas.}}{dk_{i}}$ is the measured differential cross-section in bin $i$ and $\frac{d\sigma_{i}^{variation}}{dk_{i}}$ is the differential cross-section in bin $i$ calculated from a toy experiment variation. The single differential cross-sections in momentum and $\cos(\theta)$ are calculated using the same two dimensional fit and the covariance matrix should include the correlations between $d\sigma/dp$ and $d\sigma/d\cos(\theta)$. The full fractional covariance matrix as calculated from Equation~\ref{eq:cov} and is shown in Figure~\ref{fig:dataxscov}. 

\subsection{Total cross-sections in limited phase-space}
\label{subsec:xstotalresults}

The total cross-sections in the measured phase-space ($p > 300$~MeV/c and $\theta \leq 45^{\circ}$) using NEUT and GENIE MC are shown in Table~\ref{tab:totalxsresult_limited}. The results are compatible with the NEUT and GENIE predictions, although larger cross-sections are measured in RHC, but with large statistical uncertainties.

\begin{table}[!ht]
 \caption{Measurement of the flux integrated CC-$\nu_{e}$ and CC-$\bar\nu_{e}$ inclusive total cross-sections in a limited phase-space ($p > 300$~MeV/c and $\theta \leq 45^{\circ}$) obtained using NEUT (5.3.2) and GENIE (2.8.0) MC. The statistical uncertainty is computed by fixing all nuisance parameters to their post-fit nominal and redoing the fit. The systematic uncertainty is computed by subtracting in quadrature the statistical uncertainty from the total uncertainty. The mean of the neutrino energy, $<E>$, in each beam mode is also shown.}
 \begin{center}
 \begin{tabular}{r|c|c|c}
 \hline
 \hline
 \multicolumn{1}{c|}{Selection} & Measured $\sigma$ & Nominal $\sigma$  & $<E>$ \\
 & $[/10^{-39} \rm cm^{2}/nucleon]$ & $[/10^{-39} \rm cm^{2}/nucleon]$ & GeV \\ 
 \hline
 FHC CC-$\nu_{e}$ NEUT   & $6.62\pm1.32(\rm stat)\pm1.30(\rm syst)$     & 7.18 & 1.28 \\
                  GENIE  & $6.93\pm1.40(\rm stat)\pm1.33(\rm syst)$ & 6.87 & \\
 RHC CC-$\nu_{e}$ NEUT   & $14.56\pm4.90(\rm stat)\pm2.31(\rm syst)$        & 12.96 & 1.98 \\
                  GENIE        & $14.73\pm5.06(\rm stat)\pm2.01(\rm syst)$ & 11.44  & \\
 RHC CC-$\bar\nu_{e}$ NEUT & $3.01\pm1.36(\rm stat)\pm0.57(\rm syst)$ & 2.61 & 0.99   \\
                      GENIE & $3.10\pm1.46(\rm stat)\pm0.53(\rm syst)$ & 2.51 \\
 \hline
 \hline
 
\end{tabular}
\end{center}
       
\label{tab:totalxsresult_limited}
            
\end{table}


\subsection{Comparisons to additional models}
\label{subsec:modelcomparisons}

Using the NUISANCE framework~\cite{nuisance}, the fit results are compared to cross-section predictions from recent neutrino generator models in NEUT (5.4.0), GENIE (2.12.10) and also from NuWro (19.02)~\cite{nuwro}. NEUT 5.4.0 uses a local Fermi gas (instead of spectral function). Other interaction modelling and final state interactions are similar to NEUT 5.3.2 (detailed in section~\ref{sec:analysissamples}). GENIE 2.12.10 interaction modelling is similar to 2.8.0 (detailed in section~\ref{sec:analysissamples}), with the "empirical MEC" model for the description of multi-nucleon interactions enabled. NuWro simulates the CC quasi-elastic process with the Llewellyn-Smith model with axial mass value of 1.03~$\rm GeV/c^{2}$. The nuclear model is simulated using the relativistic Fermi gas including random phase approximation corrections~\cite{rpa}. Multi-nucleon interactions are simulated similar to NEUT using the model from~\cite{Nieves2p2h}. For pion production a single $\Delta$-model by Adler-Rarita-Schwinger~\cite{adler} is used for the hadronic mass W~$<$~1.6~$\rm GeV/c^{2}$ with axial mass value of 0.94~$\rm GeV/c^{2}$. A smooth transition to DIS processes is made for W between 1.3 and 1.6~$\rm GeV/c^{2}$. The total cross section is based on the Bodek and Yang approach~\cite{BodekYangNeut}. Similar to NEUT, final state interactions are simulated using a semi-classical cascade model.

\begin{figure}[htbp]
 \centering
	\includegraphics[width=0.99\textwidth]{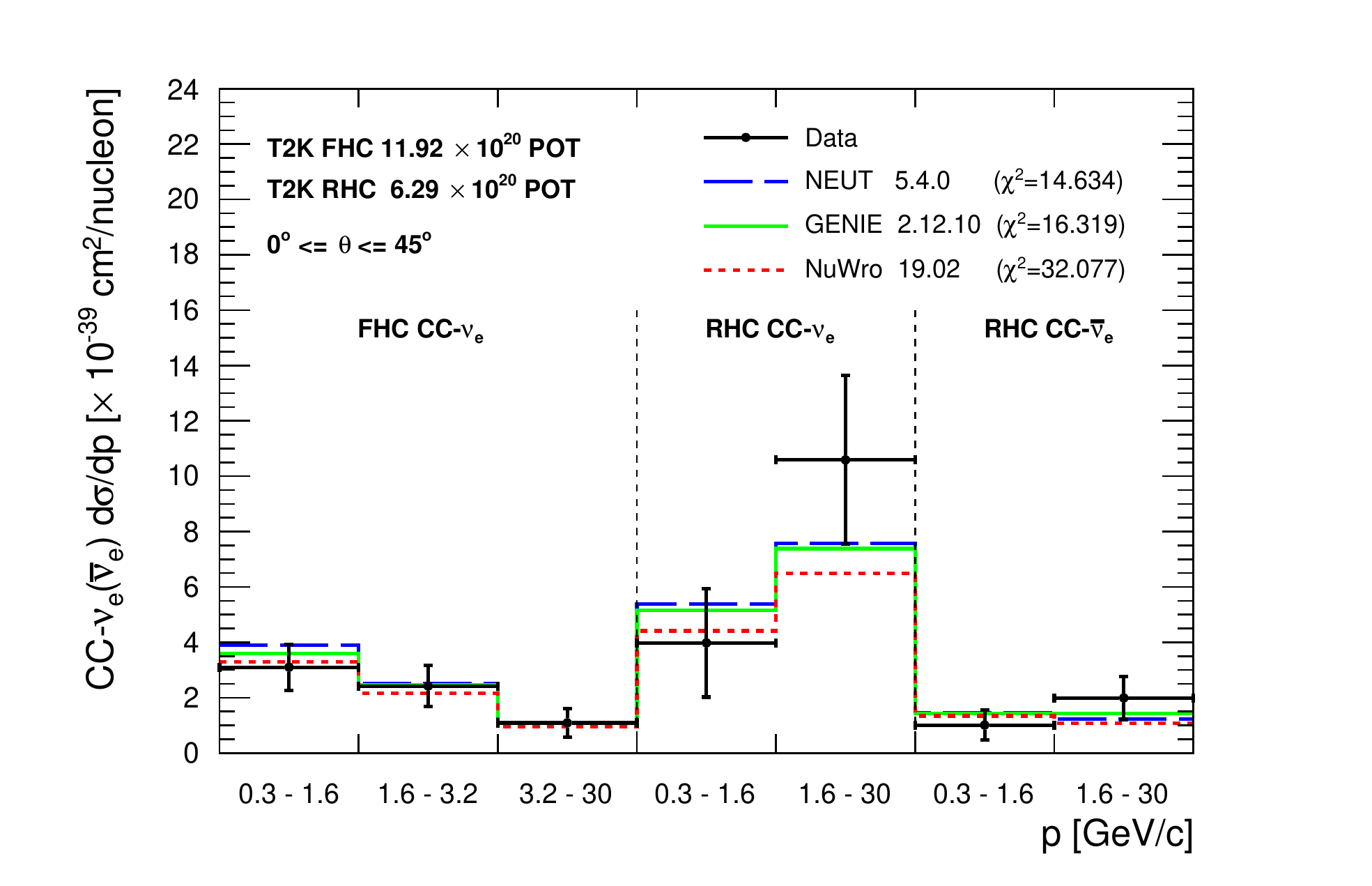}
	\includegraphics[width=0.99\textwidth]{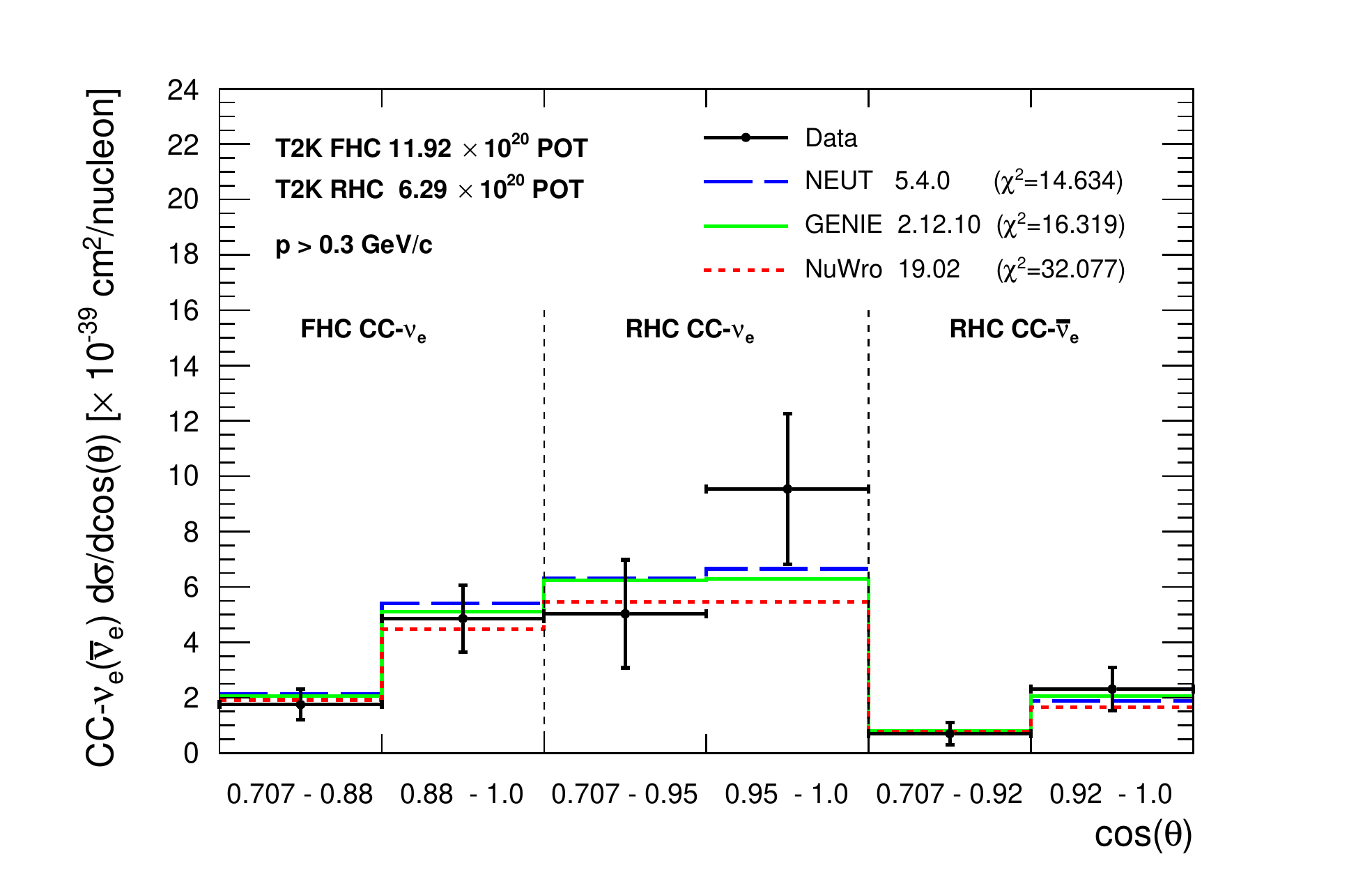}
	\caption{Flux integrated CC-$\nu_{e}$ and CC-$\bar\nu_{e}$ inclusive cross-section results in a limited phase-space ($p > 300$~MeV/c and $\theta \leq 45^{\circ}$) with comparisons to neutrino generator models from NEUT 5.4.0, GENIE 2.12.10 and NuWro 19.02 obtained using the NUISANCE framework. The top plot shows the results in momentum and the bottom plot the results in scattering angle. The $\chi^{2}$ is the total from the combined measurements in momentum and $\cos(\theta)$.}
	\label{fig:datamodelcomp}
\end{figure}

The comparisons of the data to NEUT 5.4.0, GENIE 2.12.10 and NuWro 19.02 are shown in Figure~\ref{fig:datamodelcomp}. A $\chi^{2}$ between the data measurements and each neutrino generator model predictions is defined as 
\begin{equation}
 \chi^{2} = \sum_{i} \sum_{j} \left(\frac{d\sigma_{i}^{meas.}}{dk_{i}} - \frac{d\sigma_{i}^{model}}{dk_{i}}\right) V_{ij}^{-1} \left(\frac{d\sigma_{j}^{meas.}}{dk_{j}} - \frac{d\sigma_{j}^{model}}{dk_{j}}\right),  
 \label{eq:chi2def}
\end{equation}

where $k$ is either $p$ or $\cos(\theta)$, $\frac{d\sigma_{i}^{meas.}}{dk_{i}}$ is the differential cross-section measurement in bin $i$, $\frac{d\sigma_{i}^{model}}{dk_{i}}$ is the differential cross-section model prediction in bin $i$ and $V_{ij}$ is the covariance matrix as defined in equation~\ref{eq:cov} and shown in Figure~\ref{fig:dataxscov}. The $\chi^{2}$ is measured for each neutrino generator individually and is summarised in Table~\ref{tab:chi2comp}. NEUT 5.4.0 has the lowest $\chi^{2}$ compared to our data. GENIE 2.12.10 has a slightly larger $\chi^{2}$. The $\chi^{2}$ for NuWro 19.02 is significantly larger. The $\chi^{2}$ is also calculated individually for the single differential cross-sections $d\sigma/dp$ and $d\sigma/d\cos(\theta)$. A reduced covariance is used considering only the momentum or $\cos(\theta)$ part of the full covariance in Figure~\ref{fig:dataxscov}. In these cases the $\chi^{2}$, in both momentum and $\cos(\theta)$ measurements, are smaller and similar for all neutrino generators. This highlights the importance of using the combined cross-section measurements in momentum and $\cos(\theta)$ when doing model comparisons, rather than using each cross-section measurement in momentum or $\cos(\theta)$ individually.

\begin{table}[!ht]
 \caption{The $\chi^{2}$ comparing data with neutrino generator models. The $\chi^{2}$ is calculated using Equation~\ref{eq:chi2def}. The full covariance, as shown in Figure~\ref{fig:dataxscov}, is used for $p-\cos(\theta)$ $\chi^{2}$ calculation. A reduced covariance considering only the momentum and $\cos(\theta)$ part of the full covariance is used to calculate the $p$-only and $\cos(\theta)$-only $\chi^{2}$ respectively. The number of degrees of freedom (ndof) for each $\chi^{2}$ is also shown.}
 
\begin{center}
 \begin{tabular}{l|c|c|c}
 \hline
 \hline
 Generator & $p-\cos(\theta)$ $\chi^{2}$ & $p$-only $\chi^{2}$ & $\cos(\theta)$-only $\chi^{2}$      \\
               & (ndof = 13) & (ndof = 7) & (ndof = 6) \\
 \hline
 NEUT 5.4.0    & 14.63 & 5.82 & 5.34 \\
 GENIE 2.12.10 & 16.32 & 4.16 & 4.55 \\
 NuWro 19.02   & 32.08 & 4.52 & 5.08 \\
 \hline
 \hline
 
\end{tabular}

\end{center}
\label{tab:chi2comp}
            
\end{table}

\section{Summary and conclusions}
\label{sec:summary}

Electron-like neutrino and anti-neutrino events are selected in the T2K off-axis near detector ND280, using both FHC and RHC modes. A significant amount of photon background populates the low momentum and high angle regions, constrained by an independent photon control selection. The regions dominated by the photon background also show significant data and MC discrepancies and are dominated by large systematic uncertainties. The flux integrated single differential cross-sections, as a function of momentum and scattering angle, are measured by fitting simultaneously the CC inclusive selections and their corresponding photon control selections. To minimize detector effects, the cross-sections are measured in a limited phase-space, $p > 300$~MeV/c and $\theta \leq 45^{\circ}$. The results are consistent from the two fits with both NEUT 5.3.2 and GENIE 2.8.0 predictions. The cross-section results are also compared with more recent neutrino generator models using NEUT 5.4.0, GENIE 2.12.10 and NuWro 19.02. The best agreement is observed with NEUT 5.4.0. These are the first CC-$\nu_{e}$ cross-section measurements using both FHC and RHC fluxes and the first CC-$\bar\nu_{e}$ cross-section measurement since the Gargamelle measurements in 1978. The data release from this paper can be found here~\cite{nuedata}.

\acknowledgments

We thank the J-PARC staff for superb accelerator performance. We thank the CERN NA61/SHINE Collaboration for providing valuable particle production data. We acknowledge the support of MEXT, Japan; NSERC (Grant No. SAPPJ-2014-00031), NRC and CFI, Canada; CEA and CNRS/IN2P3, France; DFG, Germany; INFN, Italy; National Science Centre (NCN) and Ministry of Science and Higher Education, Poland; RSF (Grant \#19-12-00325) and Ministry of Science and Higher Education, Russia; MICINN and ERDF funds, Spain; SNSF and SERI, Switzerland; STFC, UK; and DOE, USA. We also thank CERN for the UA1/NOMAD magnet, DESY for the HERA-B magnet mover system, NII for SINET4, the WestGrid and SciNet consortia in Compute Canada, and GridPP in the United Kingdom. In addition, participation of individual researchers and institutions has been further supported by funds from ERC (FP7), RFBR project number 20-32-70196, "la Caixa” Foundation (ID 100010434, fellowship code LCF/BQ/IN17/11620050), the European Union’s Horizon 2020 Research and Innovation Programme under the Marie Sklodowska-Curie grant agreements no. 713673 and no. 754496, and H2020 Grants No. RISE-RISE-GA822070-JENNIFER2 2020 and RISE-GA872549-SK2HK; JSPS, Japan; Royal Society, UK; French ANR Grant No. ANR-19-CE31-0001; and the DOE Early Career program, USA.




\begin{thebibliography}{99}
\bibitem{T2KExperiment} K. Abe et al. (T2K Collaboration), \emph{The T2K Experiment}, \href{https://doi.org/10.1016/j.nima.2011.06.067} {\emph{Nucl. Instrum. Meth. A} {\bf 659}, 106, 2011}.
\bibitem{Gargamelle} J. Blietschau et al. (Gargamelle Collaboration), \emph{Total Cross-Sections for electron-neutrino and anti-electron-neutrino Interactions and Search for Neutrino Oscillations and Decay}, \href{https://doi.org/10.1016/0550-3213(78)90299-7} {\emph{Nucl. Phys. B} {\bf 133}, 205, 1978}.
\bibitem{ND280NueXs} K. Abe et al. (T2K Collaboration), \emph{Measurement of the Inclusive Electron Neutrino Charged Current Cross Section on Carbon with the T2K Near Detector}, \href{https://doi.org/10.1103/PhysRevLett.113.241803} {\emph{Phys. Rev. Lett.} {\bf 113}, 241803, 2014}.
\bibitem{MinervaNue} J. Wolcott et al. (MINERvA Collaboration), \emph{Measurement of Electron Neutrino Quasielastic and Quasielasticlike Scattering on Hydrocarbon at $\left\langle E_{\nu} \right\rangle = 3.6$~GeV}, \href{https://doi.org/10.1103/PhysRevLett.116.081802} {\emph{Phys. Rev. Lett.} {\bf 116}, 081802, 2016}.
\bibitem{DUNETDR} B. Abi et al. (DUNE Collaboration), \emph{Deep Underground Neutrino Experiment (DUNE), Far Detector Technical Design Report, Volume II DUNE Physics}, \href{https://arxiv.org/pdf/2002.03005.pdf} {\emph{arXiv:2002.03005 [physics.ins-det]}, 2020}.
\bibitem{HYPERKTDR} K. Abe et al. (Hyper-K Collaboration), \emph{Hyper-Kamiokande Design Report}, \href{https://arxiv.org/pdf/1805.04163.pdf} {\emph{arXiv:1805.04163 [physics.ins-det]}, 2018}.
\bibitem{ND280cc0pi} K. Abe et al. (T2K Collaboration), \emph{Measurement of double-differential muon neutrino charged-current interactions on $C_{8}H_{8}$ without pions in the final state using the T2K off-axis detector}, \href{https://journals.aps.org/prd/abstract/10.1103/PhysRevD.93.112012} {\emph{Phys. Rev. D} {\bf 93}, 112012, 2016}.
\bibitem{ND280numucc4pi} K. Abe et al. (T2K Collaboration), \emph{Measurement of inclusive double-differential $\nu_{\mu}$ charged-current cross section with improved acceptance in the T2K off-axis near detector}, \href{https://journals.aps.org/prd/pdf/10.1103/PhysRevD.98.012004} {\emph{Phys. Rev. D} {\bf 98}, 012004, 2018}.
\bibitem{ND280nucleff} K. Abe et al. (T2K Collaboration), \emph{Characterisation of nuclear effects in muon-neutrino scattering on hydrocarbon with a measurement of final-state kinematics and correlations in charged-current pionless interactions at T2K}, \href{https://journals.aps.org/prd/abstract/10.1103/PhysRevD.98.032003} {\emph{Phys. Rev. D} {\bf 98}, 032003, 2018}.
\bibitem{dagostini} G. D'Agostini \emph{A multidimensional unfolding method based on Bayes' theorem}, \href{https://doi.org/10.1016/0168-9002(95)00274-X}{\emph{Nucl. Instrum. Methods} {\bf A362}, 487, 1995}

\bibitem{T2Kbeamline} K. Abe et al. (T2K Collaboration), \emph{T2K neutrino flux prediction}, \href{https://doi.org/10.1103/PhysRevD.87.012001} {\emph{Phys. Rev. D} {\bf 87}, 012001, 2013}.
\bibitem{fluka1} T. Bohlen, F. Cerutti, M. Chin, A. Fasso, A. Ferrari, P. Ortega, A. Mairani, P. Sala, G. Smirnov, and V. Vlachoudis, \emph{The FLUKA Code: Developments and Challenges for High Energy and Medical Applications}, \href{https://doi.org/10.1016/j.nds.2014.07.049} {\emph{Nuclear Data Sheets} {\bf 120}, 211, 2014}.
\bibitem{fluka2} A. Ferrari, P. R. Sala, A. Fasso, and J. Ranft, \emph{FLUKA : A multi-particle transport code}, \href{http://cds.cern.ch/record/898301} {\emph{CERN-2005-010, SLAC-R-773, INFN-TC-05-11}, 2005}.
\bibitem{geant3} R. Brun, F. Carminati, and S. Giani, \emph{GEANT: Detector Description and Simulation Tool}, \href{https://cds.cern.ch/record/1082634} {{\bf CERN-W5013}, 1994}.
\bibitem{gcalor} C. Zeitnitz and T. A. Gabriel, \emph{The GEANT-CALOR Interface}, \emph{Proceedings of International Conference on Calorimetry in High Energy Physics}, 1993.
\bibitem{na61shine1} N. Abgrall et al. (NA61/SHINE Collaboration), \emph{Measurements of cross sections and charged pion spectra in proton-carbon interactions at 31 GeV/c} \href{https://doi.org/10.1103/PhysRevC.84.034604} {\emph{Phys. Rev. C} {\bf 84}, 034604, 2011}.
\bibitem{na61shine2} N. Abgrall et al. (NA61/SHINE Collaboration), \emph{Measurement of production properties of positively charged kaons in proton-carbon interactions at 31 GeV/c}, \href{https://doi.org/10.1103/PhysRevC.85.035210} {\emph{Phys. Rev. C} {\bf 85}, 035210, 2012}.
\bibitem{na61shine3} N. Abgrall et al. (NA61/SHINE Collaboration), \emph{Measurements of $\pi^{\pm}$, $K^{\pm}$, $K^{\pm}$, $K^0_S$, $\Lambda$ and proton production in proton-carbon interactions at 31 GeV/c with the NA61/SHINE spectrometer at the CERN SPS}, \href{https://doi.org/10.1140/epjc/s10052-016-3898-y} {\emph{Eur. Phys. J. C} {\bf 76}, 84, 2016}.

\bibitem{ND280P0D} S. Assylbekov et al., \emph{The T2K ND280 Off-Axis Pi-Zero Detector}, \href{https://doi.org/10.1016/j.nima.2012.05.028} {\emph{Nucl. Instrum. Meth. A} {\bf 686}, 48, 2012}.
\bibitem{ND280FGD} P.-A. Amaudruz et al., \emph{The T2K Fine-Grained Detectors}, \href{https://doi.org/10.1016/j.nima.2012.08.020} {\emph{Nucl. Instrum. Meth. A} {\bf 696}, 1, 2012}.
\bibitem{ND280TPC} N. Abgrall et al., \emph{Time projection chambers for the T2K near detectors}, \href{https://doi.org/10.1016/j.nima.2011.02.036} {\emph{Nucl. Instrum. Meth. A} {\bf 637}, 25, 2011}.
\bibitem{ND280ECal} K. Allan et al, \emph{The Electromagnetic Calorimeter for the T2K Near Detector ND280}, \href{https://iopscience.iop.org/article/10.1088/1748-0221/8/10/P10019/pdf} {\emph{JINST} {\bf 8}, P10019, 2013}.
\bibitem{ND280SMRD} S. Aoki et al., \emph{The T2K Side Muon Range Detector (SMRD)}, \href{https://doi.org/10.1016/j.nima.2012.10.001} {\emph{Nucl. Instrum. Meth. A} {\bf 698}, 135, 2013}.

\bibitem{ND280NueSel} K. Abe et al. (T2K Collaboration), \emph{Measurement of the intrinsic electron neutrino component in the T2K neutrino beam with the ND280 detector}, \href{https://journals.aps.org/prd/abstract/10.1103/PhysRevD.89.092003} {\emph{Phys. Rev. D} {\bf 89}, 092003, 2014}.

\bibitem{neutmc} Y. Hayato, \emph{A neutrino interaction simulation program library NEUT}, \href{https://www.actaphys.uj.edu.pl/fulltext?series=Reg&vol=40&page=2477} {\emph{Acta Phys. Pol. B} {\bf 40}, 2477, 2009}.
\bibitem{geniemc} C. Andreopoulos et al., \emph{The GENIE Neutrino Monte Carlo Generator}, \href{https://doi.org/10.1016/j.nima.2009.12.009} {\emph{Nucl. Instrum. Meth. A}, {\bf 614}, 87, 2010}.
\bibitem{NueNumuCCQE} Melanie Day and Kevin S. McFarland, \emph{Differences in Quasi-Elastic Cross-Sections of Muon and Electron Neutrinos}, \href{https://journals.aps.org/prd/abstract/10.1103/PhysRevD.86.053003} {\emph{Phys. Rev. D} {\bf 86}, 053003, 2012}.

\bibitem{LlewellynSmith} C. H. Llewellyn Smith, \emph{Neutrino reactions at accelerator energies}, \href{https://doi.org/10.1016/0370-1573(72)90010-5} {\emph{Phys. Rep.} {\bf 3}, 261 - 379, 1972}.

\bibitem{SpectralFunction} O. Benhar, A. Fabrocini, S. Fantoni, and I. Sick, \emph{Spectral function of finite nuclei and scattering of GeV electrons}, \href{https://doi.org/10.1016/0375-9474(94)90920-2} {\emph{Nucl. Phys. A} {\bf 579}, 493 - 517, 1994}.

\bibitem{Nieves2p2h} J. Nieves, I. R. Simo, and M. V. Vacas, \emph{The nucleon axial mass and the MiniBooNE quasielastic neutrino–nucleus scattering problem} \href{https://doi.org/10.1016/j.physletb.2011.11.061} {\emph{Phys. Lett. B} {\bf 707}, 72 - 75, 2012}.

\bibitem{ReinSehgal} D. Rein and L.M. Sehgal, \emph{Neutrino-excitation of baryon resonances and single pion production} \href{https://doi.org/10.1016/0003-4916(81)90242-6} {\emph{Ann. Phys. (N.Y.)} {\bf 133}, 79 - 153, 1981}.

\bibitem{GRV98} M. Gluck, E. Reya, and A. Vogt, \emph{Dynamical parton distributions revisited} \href{https://link.springer.com/article/10.1007\%2Fs100529800978} {\emph{Eur. Phys. J. C} {\bf 5}, 461 - 470, 1998}.

\bibitem{BodekYangNeut} A. Bodek and U. K. Yang, \emph{Modeling Neutrino and Electron Scattering Cross Sections in the Few GeV Region with Effective LO PDFs} \href{https://doi.org/10.1063/1.1594324} {\emph{AIP Conf. Proc.} {\bf 670}, 110, 2003}.

\bibitem{BodekRitchie} A. Bodek and J. L. Ritchie, \emph{Further studies of Fermi-motion effects in lepton scattering from nuclear targets} \href{https://doi.org/10.1103/PhysRevD.24.1400} {\emph{Phys. Rev. D} {\bf 24}, 1400, 1981}.

\bibitem{BodekYangGenie} A. Bodek and U. K. Yang, \emph{A Unified Model for inelastic e - N and $\nu$ - N cross sections at all $Q^{2}$} \href{https://doi.org/10.1063/1.2122031} {\emph{AIP Conf. Proc.} {\bf 792}, 2005}.



\bibitem{T2KOscLong} K. Abe et al. (T2K Collaboration), \emph{Measurements of neutrino oscillation in appearance and disappearance channels by the T2K experiment with $6.6\times10^{20}$ protons on target}, \href{https://journals.aps.org/prd/pdf/10.1103/PhysRevD.91.072010} {\emph{Phys. Rev. D} {\bf 91}, 072010, 2015}.

\bibitem{geant4} S. Agostinelli et al. (GEANT4 Collaboration), \emph{GEANT4: A Simulation toolkit}, \href{https://doi.org/10.1016/S0168-9002(03)01368-8} {\emph{Nucl. Instrum. Meth. A} {\bf 506}, 250, 2003}.

\bibitem{ND280SingleGamma} K. Abe et al. (T2K Collaboration), \emph{Search for neutral-current induced single photon production at the ND280 near detector in T2K}, \href{https://iopscience.iop.org/article/10.1088/1361-6471/ab227d/pdf} {\emph{J. Phys. G: Nucl. Part. Phys.} {\bf 46}, 08LT01, 2019}.
\bibitem{t2kfluxerror} K. Abe et al. (T2K Collaboration), \emph{Measurement of neutrino and antineutrino oscillations by the T2K experiment including a new additional sample of $\nu_{e}$ interactions at the far detector}, \href{https://journals.aps.org/prd/abstract/10.1103/PhysRevD.96.092006} {\emph{Phys. Rev. D} {\bf 96}, 092006, 2017}.

\bibitem{histfactory} K.S. Cranmer, G. Lewis, L. Moneta, A. Shibata and W. Verkerke, \emph{HistFactory: A tool for creating statistical models for use with RooFit and RooStats}, \href{https://cds.cern.ch/record/1456844/files/CERN-OPEN-2012-016.pdf}{\emph{Tech. Rep. CERN-OPEN-2012-016} (2012)}.
\bibitem{roostats} L. Moneta, K. Belasco, K.S. Cranmer, S. Kreiss, A. Lazzaro, D. Piparo, G. Schott, W. Verkerke and M. Wolf, \emph{The RooStats Project, PoS ACAT2010}, 057, \href{https://arxiv.org/pdf/1009.1003.pdf} {\emph{arXiv:1009.1003 [physics.data-an]}, 2010}.
\bibitem{roofit} W. Verkerke and D. P. Kirkby, \emph{The RooFit toolkit for data modeling, eConf C0303241 MOLT007}, \href{https://arxiv.org/pdf/physics/0306116.pdf}{\emph{arXiv:physics/0306116 [physics.data-an]}, 2003}.
\bibitem{barlowbeeston} R.J. Barlow and C. Beeston, \emph{Fitting using finite MC samples}, \href{https://doi.org/10.1016/0010-4655(93)90005-W} {\emph{Computer Physics Comm. {\bf 77}}, 219 - 228, 1993}.
\bibitem{nuisance} P.Stowell et al, \emph{NUISANCE: a neutrino cross-section generator tuning and comparison framework}, \href{https://iopscience.iop.org/article/10.1088/1748-0221/12/01/P01016} {\emph{JINST {\bf 12}}, P01016, 2017}.
\bibitem{nuwro} T.Golan, J.T.Sobczyk and J.Zmuda, \emph{NuWro: the Wrocław Monte Carlo Generator of Neutrino Interactions}, \href{https://doi.org/10.1016/j.nuclphysbps.2012.09.136} {\emph{Nucl. Phys. Proc. Suppl. \bf{229-232}}, 2012}
\bibitem{rpa} J. Nieves, I. Ruiz Simo, and M. J. Vicente Vacas, \emph{Inclusive charged-current neutrino-nucleus reactions}, \href{https://doi.org/10.1103/PhysRevC.83.045501} {\emph{Phys. Rev. C} {\bf 83}, 045501, 2011}.

\bibitem{adler} K. M. Graczyk, D. Kielczewska, P. Przewlocki, J. T. Sobczyk, \emph{$C^{A}_{5}$ axial form factor from bubble chamber experiments}, \href{	https://doi.org/10.1103/PhysRevD.80.093001} {\emph{Phys. Rev. D} {\bf 80}, 093001, 2009}.

\bibitem{nuedata}\href{https://t2k-experiment.org/results/2020_nuecc} {https://t2k-experiment.org/results/2020\_nuecc}









\end{thebibliography}
\end{document}